\documentclass[10pt]{article}
\usepackage[a4paper, left=1.5cm, right=1.5cm, top=1.5cm, bottom=2.5cm]{geometry}  % Adjust the margin here
\usepackage[authoryear,round]{natbib}
\usepackage{url}
\usepackage{authblk}
\usepackage{mathtools}

\usepackage[T1]{fontenc}     % proper font encoding
\usepackage{a4wide}
\usepackage{lipsum}
\usepackage{graphicx}
\usepackage{xcolor}
\usepackage{caption}
\usepackage{subcaption}
\usepackage{amsmath}
\usepackage{amsthm}
\usepackage{amssymb}
\usepackage{upgreek} % upright Greek letters
\usepackage{algorithm} % for depicting algorithms
\usepackage{algpseudocode}
\usepackage{hyperref}
\usepackage{float}
\usepackage{tikz}
\usepackage{cleveref}
\usepackage{enumerate}% http://ctan.org/pkg/enumerate
\usepackage{enumitem}% http://ctan.org/pkg/enumitem
\usetikzlibrary{arrows.meta}

\usepackage{blkarray}
\usepackage{array}
\usepackage{makecell}
\usepackage{svg}
\usepackage{multicol}
\usetikzlibrary{matrix}
\usepackage{comment}

\usepackage{siunitx} % for si units
\usepackage{physics}
\DeclareMathOperator{\Div}{Div}

\DeclareMathOperator{\TSB}{TSB}
\DeclareMathOperator{\PSB}{PSB}
\DeclareMathOperator{\THB}{THB}
\usepackage{todonotes}

\usepackage{abstract}
\title{Influence of the geometry on the mechanical performance of tubular interlockings: A study of the Sine Block}

\author[1]{Domen Macek\thanks{Corresponding author: domen.macek@ifam.rwth-aachen.de}}
\author[2]{Meike Weiß}
\author[3]{Reymond Akpanya}
\author[1]{Tim Brepols}
\author[4]{Hagen Holthusen}
\author[2]{Alice~C.\ Niemeyer}
\affil[1]{RWTH Aachen University, Institute of Applied Mechanics, Mies-van-der-Rohe-Str.\ 1, 52074 Aachen, Germany}
\affil[2]{RWTH Aachen University, Chair of Algebra and Representation Theory, Pontdriesch 10-16, 52062 Aachen, Germany}
\affil[3]{The School of Mathematics and Statistics, The University of Sydney, Carslaw F07, Camperdown NSW 2006, Australia}
\affil[4]{University of Erlangen-Nuremberg, Institute of Applied Mechanics, 91058 Erlangen, Germany}

\date{}

\begin{document}
\maketitle

%######################################################################################
%                                 ABSTRACT
%######################################################################################
\begin{abstract}
Topological interlocking assemblies (TIA) are arrangements of blocks such that rigid‑body motions of the blocks are fully constrained by their neighbours and a fixed frame. In this work, we investigate tubular interlocking structures derived from the sine curve and parametrised by several geometric design parameters. We analyse the behaviour of these parametrised tubular interlockings under various boundary conditions and examine how our proposed parameters influence the mechanical response. For this purpose, we first develop a simplified multibody dynamics formulation that enables an efficient exploration of how the design parameters of the block influence the load transfer within the assembly. To further corroborate these results, we perform several finite element simulations, which give insights into the mechanical behaviour of our proposed TIA. Our results show that the block geometry plays a decisive role in the mechanical performance of the corresponding TIA. We additionally discuss the problem of exploding TIAs and demonstrate that the TIA resulting from our Sine Block does not exhibit this behaviour. Lastly, we provide evidence that non-exploding TIAs possess better mechanical properties than exploding ones.
\end{abstract}

%######################################################################################
%                                 SECTIONS
%######################################################################################
\section{Introduction}
The construction industry faces increasing pressure to embrace sustainable materials and innovative architectural approaches that enhance resource efficiency and reduce waste. A promising approach is modular construction, which improves recyclability by enabling components to be reused rather than monolithically manufactured and later separated.
Our work focuses on topological interlocking assemblies (TIAs), which allow mortarless modular structures and offer excellent impact resistance, high energy absorption, and strong resistance to local failure and cracking \citep{Dyskin2012,Krause2012}.

A \emph{topological interlocking assembly} (TIA) is defined as an arrangement of blocks that are in contact with each other together with a subset of blocks called the frame such that, if the frame is fixed, any non-empty finite subset of blocks of the assembly is prevented from moving. Note that the kinematic restriction of the entire assembly is enforced by the contacts of neighbouring blocks together with the restraint of the frame.
For a precise mathematical definition of topological interlocking assemblies, the reader is referred to \cite{InterlockingSymmetry}.
The concept of TIAs dates back to the 18th century, with early interlocking blocks designed by  Abeille and Truchet in \cite{abeille_memoire_1735} later generalized by \cite{frezier_theorie_1738}.
\cite{dyskin_new_2001,dyskin_toughening_2001} pioneered the modern study of TIAs, introducing the concept as a novel material design and coining the term `topological interlocking'.

We call an assembly \emph{planar} if the blocks of the assembly are arranged between two parallel planes such that each block has a non-empty intersection with both planes.
Various construction methods for planar TIAs have been established in \citep{InterlockingSymmetry,Dyskin2003,VoroNoodles,DelaunayLofts}.
By applying continuous deformations to a planar assembly it is possible to obtain an assembly that realises a cylindrical tube, assembled from the deformed blocks of the planar assembly. Note that a tube is a hollow cylinder with inner and outer radii, bounded by two base faces. If the constructed cylindrical assembly of blocks can be equipped with a frame in such a way that the assembly exhibits interlocking properties, we call it a \emph{tubular TIA}. In many cases, the top and bottom blocks are defined as the frame.
Tubular TIAs offer promising opportunities for modular construction applications such as pillars, tunnel walls, and shafts. In the literature, tubular TIA  blocks have been constructed using different approaches, see  \cite{spherical_interlocking,HexBased}. We further refer the reader to  \cite{InnovativeApproach,ImpactPerformance,ImpactBehaviourTunnel} for detailed studies investigating tubular TIA as tunnels and tubes.

The definition of a TIA requires that, once the frame is fully fixed in space, all blocks are kinematically constrained. Hence, the fixation of the frame is essential for the interlocking property to manifest. If the frame is not fixed, it may be possible to remove blocks from the assembly without causing penetrations with other blocks in the assembly. It is therefore reasonable to assume that TIAs, that do not permit certain motions of the blocks if the frame is not fixed, have better mechanical properties than TIAs that do permit such movements. This paper provides evidence for this hypothesis by analysing the so-called \emph{explosion of a tubular interlocking}. Here, the explosion of a tubular TIA is a set of motions where every block of the assembly without the frame being fixed moves simultaneously radially outwards from the central line of the tube. If these movements of the blocks do not cause any intersections with other blocks, the TIA is called exploding. The notion of exploding TIAs has first been addressed in \cite{spherical_interlocking}.

\subsection{Contribution}
In this work, we introduce a novel interlocking block, called the \emph{Sine Block} tailored to the construction of stable cylindrical assemblies.
This Sine Block is, to our knowledge, the first interlocking block that generates tubular interlocking assemblies that do not \emph{explode}. We showcase this behaviour by multibody dynamics and demonstrate the importance of studying non-exploding tubular TIAs by providing first evidence that exploding tubular TIAs do not successfully interlock for several boundary conditions.

Tough many TIAs have been proposed in the past, not many have been parametrised and examined systematically. Some previous works show that in the case of planar TIAs both block variation as well as arrangement of blocks have a strong influence on the performance \citep{InfluenceArrangement,Mirkhalaf,WANG2025121390,Weizmann}. One of the main contribution of this paper is the use of multibody dynamics (MBD), which enables rapid pre-assessment of the design space and helps to focus costly finite element method (FEM) simulations on promising geometries.
This approach is essential for systematically considering TIAs with many varying parameters.
Using the newly introduced Sine Block and an existing hexagon‑based block introduced in \cite{HexBased} as an example, the MBD model is tested and shown to be a reliable predictor.

As a first step, we introduce the Sine Block together with its design parameters, including those controlling the amplitude, shift and frequency of the underlying sine curve, as well as the number of blocks and the wall thickness of the assembly, see \Cref{sec:SinusBlock}.
We propose a systematic workflow for analysing the large parameter space of the Sine Block to demonstrate that all parameters strongly influence the mechanical behaviour of the tubular TIA.
To evaluate the mechanical performance of the Sine Block assemblies, various loading conditions are considered. Among the examined boundary value problems, the so-called \emph{pipe} boundary value problem (BVP) is of particular relevance. In this configuration, a constant internal pressure $p_0$ is applied to the inner surface of the tubular structure, thereby representing internal pressurization. This boundary condition is especially critical for tubular TIAs, as it assesses whether the assembly can maintain radial interlocking and resist outward expansion under pressure. The corresponding non-exploding behaviour has not yet been systematically investigated for such TIAs. Consequently, this case constitutes the central focus of the present study. 

Our analysis relies on two complementary approaches:
\begin{enumerate}[leftmargin=*, labelwidth=0pt, labelsep=0.5em, align=left]
    \item \emph{Multibody dynamics (MBD) with contact, formulated as force equilibrium.}  
    This method enables fast evaluation of the interlocking mechanism. It accounts for external forces, body interactions, and the unknown contact forces arising between adjacent blocks, see \Cref{subsec:NewtonEuler} for more details. Compared to finite element methods (FEM), these simulations are computationally inexpensive, allowing large-scale exploration of different block compositions. The primary outcome of such simulations is the ability to determine whether a given assembly \emph{successfully interlocks} under external forces, i.e., whether it can withstand static loads without collapsing. This constitutes the fundamental question when designing new interlocking geometries. In addition, the contact forces acting on the discretised surfaces of the bodies allow for the computation of the corresponding contact pressures. 

    \item \emph{Finite element method (FEM).}  
    For more detailed mechanical investigations, FEM simulations in Abaqus are employed. These analyses provide insight into stress distributions, deformation fields, and potential weak points within the Sine Block assemblies.  
\end{enumerate}
Our proposed workflow applies these approaches sequentially: first, MBD simulations are used to scan the parameter space and identify promising configurations. Second, FEM simulations are applied to the selected geometries for deeper mechanical analysis.
The analysis following the described workflow demonstrates that the design of a given geometry has a huge influence on its mechanical performance.
Additionally, in \Cref{sec:Hex}, we compare the Sine Block with the hexagon-based interlocking block constructed in \cite{HexBased}. 
Our results indicate that the hexagonal‑based interlocking block explodes and that it interlocks successfully only for one specific boundary condition. This instance supports our claim that non-exploding TIAs have better mechanical properties than exploding TIAs.

The scripts used to generate the blocks and the meshes for the Abaqus analysis, as well as the code for the MBD simulation, are provided in \cite{code}.
%\alice{Im paragraph davor sollte vielleicht gesagt werden, dass die FEMs mit Abaqus durchgefuehrt werden}
\section{Tubular Interlocking Blocks}\label{sec:blocks}

We begin by introducing the tubular blocks that form the central objects of our study. In particular, we present the Sine Block depending on seven design parameters based on the sine curve. Furthermore, we recall the hexagonal-based block constructed by \cite{HexBased} and equip this block in this work with three design parameters.
This enables us to compare the mechanical properties of the hexagonal-based block with the properties of our proposed block, see \Cref{sec:Hex}.

\subsection{Sine Block}\label{sec:SinusBlock}
\begin{figure}[h!]
    \centering
    \begin{subfigure}{0.31\textwidth}
        \centering
        \includegraphics[width=1\linewidth]{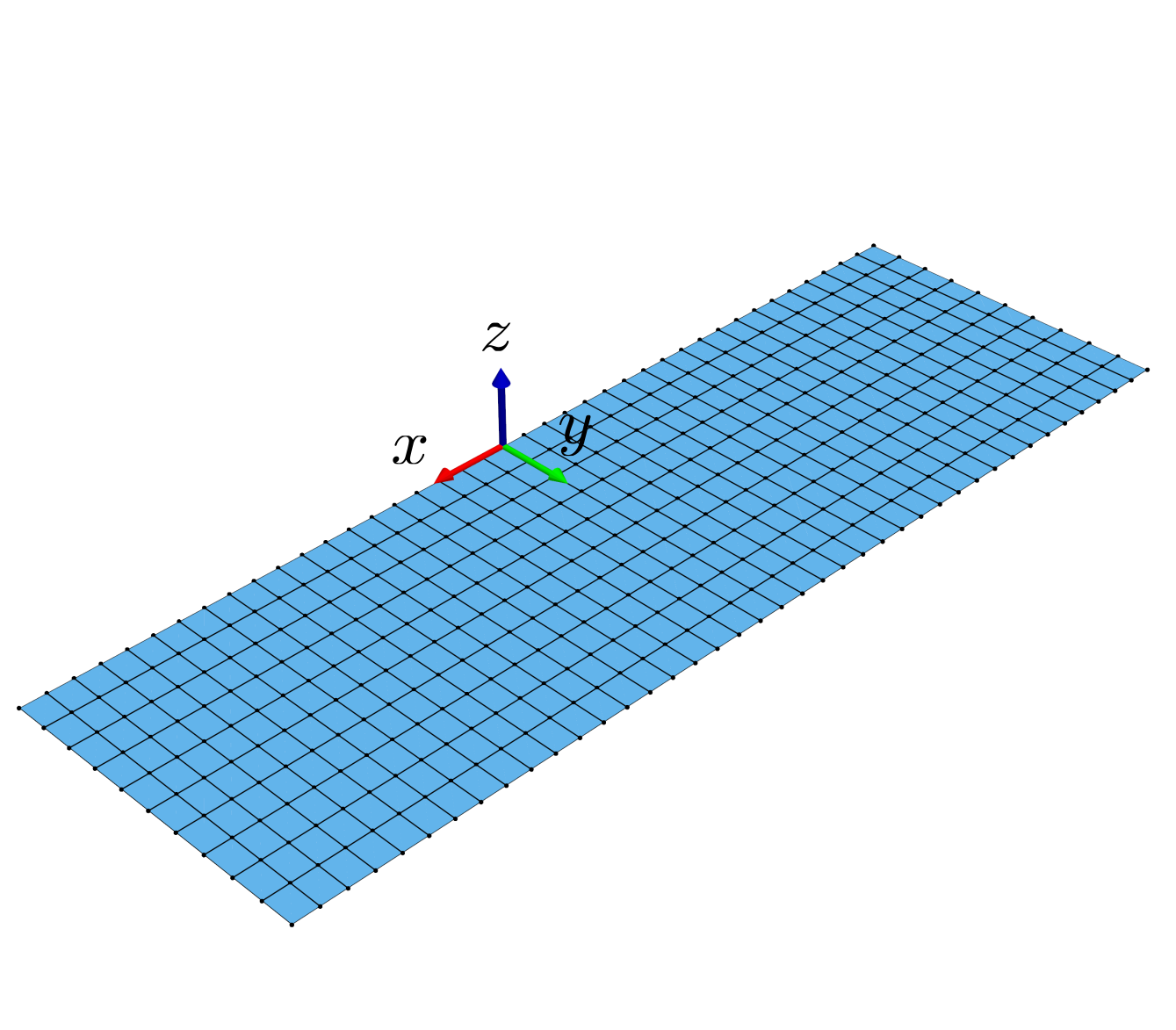}
        \caption{}
        \label{subfig:sinus_map_1}
    \end{subfigure}
    \begin{subfigure}{0.31\textwidth}
        \centering
        \includegraphics[width=1\linewidth]{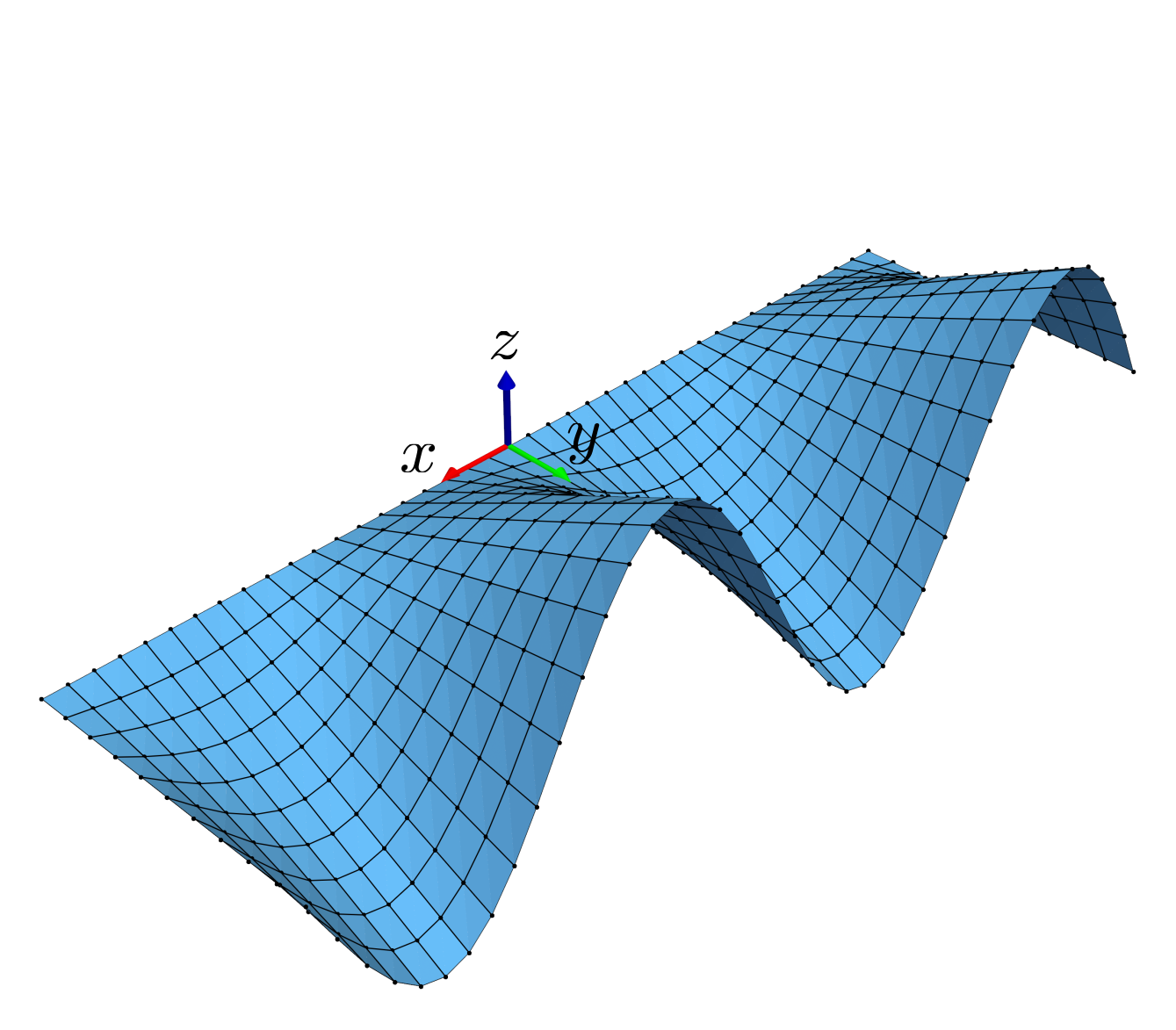}
        \caption{}
        \label{subfig:sinus_map_2}
    \end{subfigure}
    \begin{subfigure}{0.31\textwidth}
        \centering
        \includegraphics[width=1\linewidth]{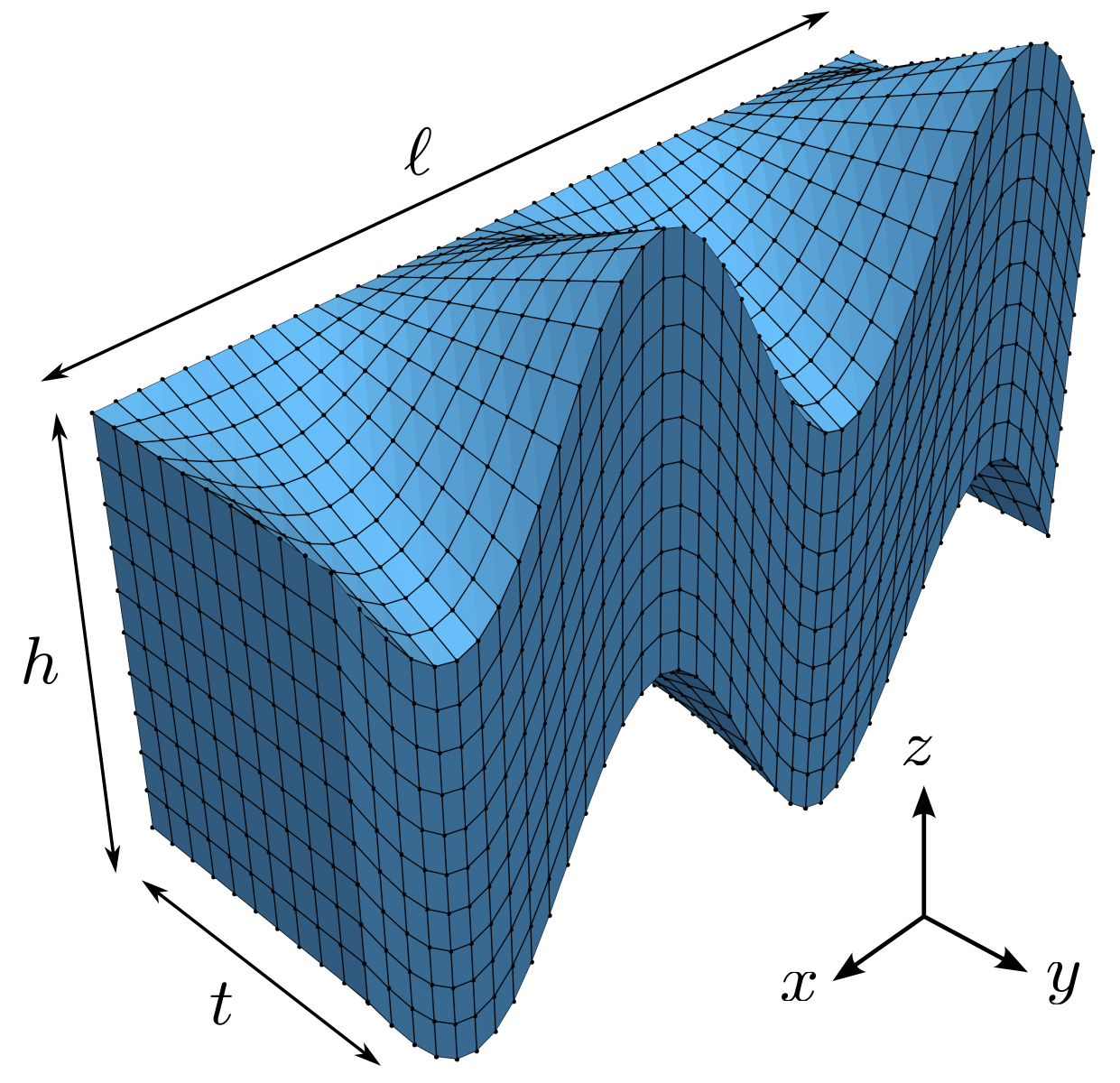}
        \caption{}
        \label{subfig:sinus_map_3}
    \end{subfigure}
    \caption{Transformation steps for constructing the planar block $\PSB(h,\ell,t,a,f=1,s=0)$.}
    \label{fig:sinus_mapping}
\end{figure}

\begin{figure}[h!]
    \centering
    \begin{subfigure}{0.48\textwidth}
        \centering
        \includegraphics[scale=0.4]{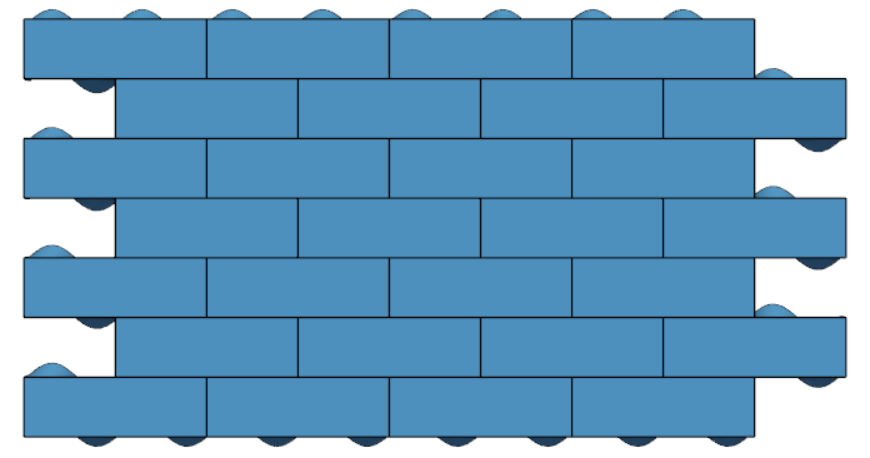}
        \caption{}
        \label{subfig:PlaneAssembly_bottom}
    \end{subfigure}
    \begin{subfigure}{0.48\textwidth}
        \centering
        \includegraphics[scale=0.4]{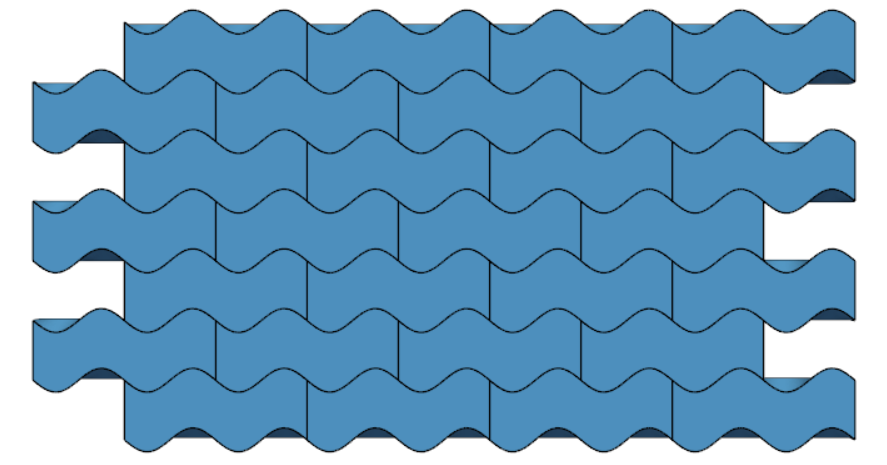}
        \caption{}
        \label{subfig:PlaneAssembly_top}
    \end{subfigure}
    \caption{Planar assembly of the planar block $\PSB(h,\ell,t,a,f=1,s=0)$: (a) View of the faces at $y=0$ and (b) view of the faces at $y=t$.}
    \label{fig:planarAss}
\end{figure}

\begin{figure}[h!]
    \centering
    \begin{subfigure}{0.48\textwidth}
        \centering
        \includegraphics[width=1.0\linewidth]{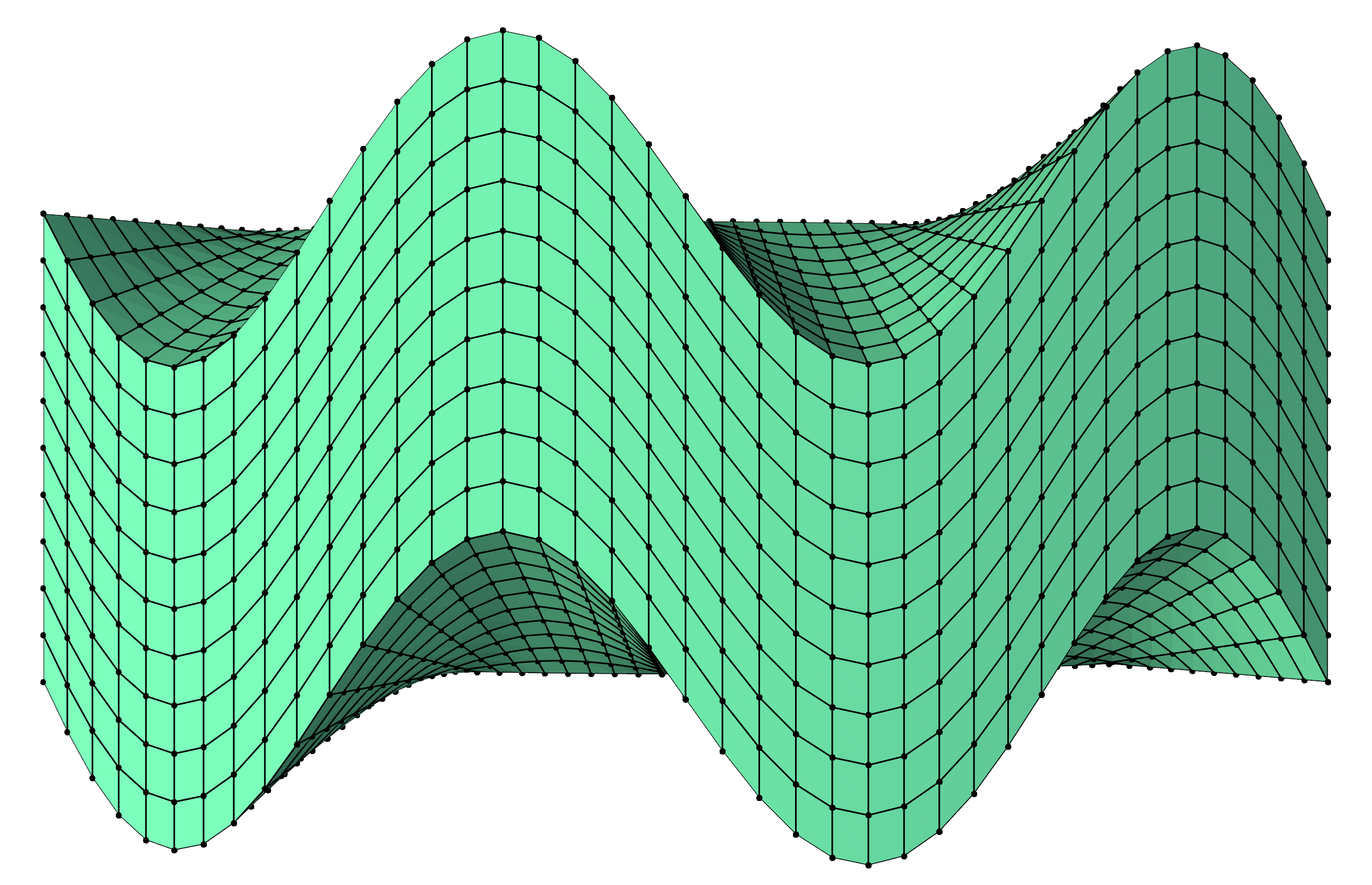}
        \caption{}
        \label{subfig:SineBlock}
    \end{subfigure}
    \begin{subfigure}{0.48\textwidth}
        \centering
        \includegraphics[width=0.5\linewidth]{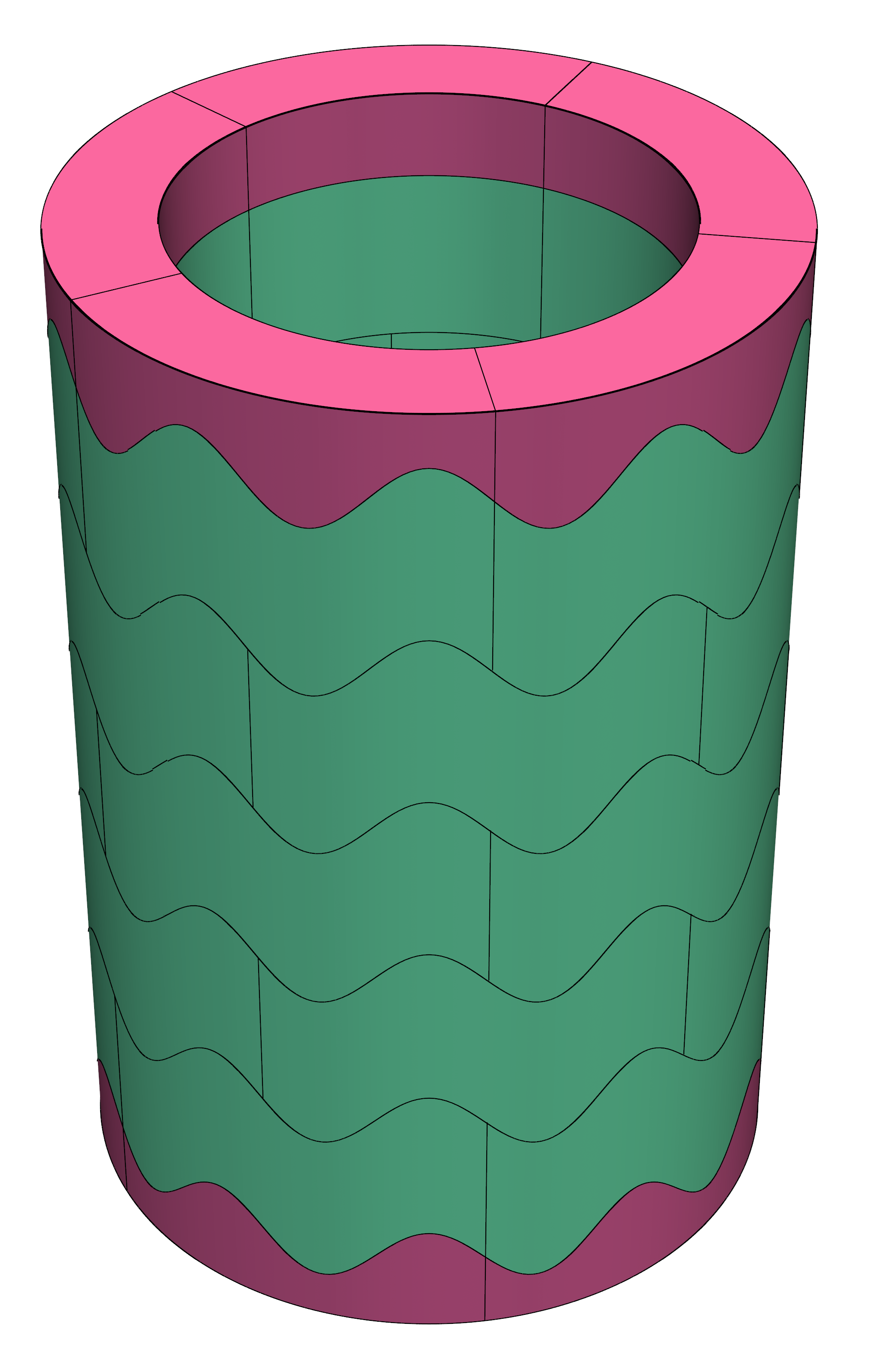}
        \caption{}
        \label{subfig:SineAss}
    \end{subfigure}
    \caption{Hexahedron mesh of $\TSB(h,a,f=1,s=0,n=5,r_i,t)$ (a) and the assembly of $\TSB(h,a,f=1,s=0,n=5,r_i,t)$ with $L=5$ and the frame depicted in red (b).}
\end{figure}

As the name indicates, the geometry of the Sine Block is based on the sine curve. This block is obtained by first constructing a planar assembly and use it to realise a tubular TIA by mapping the given assembly onto a cylinder.
To proceed, we examine a brick whose bottom and top face lie in planes that are parallel to the $xy$-plane. The bottom face is situated at $z=0$ and defined by the rectangle with corner coordinates $\left(-\frac{\ell}{2},0,0\right),\, \left(\frac{\ell}{2},0,0\right),\, \left(\frac{\ell}{2},t,0\right),\, \left(-\frac{\ell}{2},t,0\right),$
while the top face lies at $z=h$ with corners at $\left(-\frac{\ell}{2},0,h\right),\, \left(\frac{\ell}{2},0,h\right),\, \left(\frac{\ell}{2},t,h\right),\, \left(-\frac{\ell}{2},t,h\right),$ where $h,\ell,t\in\mathbb{R}_{>0}$.
We call $h$ the \emph{block height} and $t$ the \emph{block thickness}. 
The mesh corresponding to each of these faces is depicted in \Cref{subfig:sinus_map_1}.
Next, the bottom face is deformed using a sine curve whose amplitude continuously diminishes to zero as the $y$-coordinate decreases.
In particular, the edge at $y=t$ is deformed according to the function $a \cdot \sin(2\pi f x + s)$, where $a$ is the \emph{amplitude}, $f$ the \emph{frequency} and $s$ the \emph{phase shift} of the sine curve. The deformed mesh of the face can be seen in \Cref{subfig:sinus_map_2}.
Subsequently, all planes within the original brick that are parallel to the bottom face are deformed in an analogous manner.
The constructed planar block based on the sine curve is denoted by $\PSB(h,\ell,t,a,f,s)$ and a discretization of this block is shown in \Cref{subfig:sinus_map_3}.
We shift every second row horizontally depending on the chosen frequency, as illustrated in \Cref{fig:planarAss}, to obtain a planar assembly that will lead to a tubular interlocking.
In the final step, we map the planar blocks defined by $\PSB(h,\ell,t,a,f,s)$ onto a cylinder (see \Cref{fig:cylindrical_mapping})  with inner radius $r_i\in\mathbb{R}_{>0}$ and outer radius $r_o:=r_i+t$ by taking $n\in\mathbb{N}$ blocks in one layer. Note that the parameters $n$ and $r_i$ uniquely determine the value of $\ell$ via the relation $$\ell=r_i\cdot \sqrt{2\cdot\left(1-\cos(\frac{\pi}{n})\right)}.$$ 
Thus, we obtain tubular interlocking blocks based on the sine curve and denoted by $\TSB(h,a,f,s,n,r_i,t)$. We call the resulting block for $s=0$ the \emph{Sine Block} and for $s=0.5$ the \emph{Cosine Block}. The resulting Sine Block for $f=1,\, s=0$ and $n=5$ is depicted in \Cref{subfig:SineBlock} together with a hexahedron mesh as a discretization for the FEM simulation. These blocks can be arranged analogously to the planar assembly by taking $L$ layers and shift each second layer by the half of the block.
In these tubular assemblies, we add one layer at both the top and the bottom of the tube. The blocks in these layers are defined as the frame and are not included in the total number $L$ of layers. For the intended application, the frame blocks are trimmed along the sides that do not touch other blocks, ensuring that these exposed faces are planar. An example of this tubular interlocking is shown in \Cref{subfig:SineAss}.

The eight parameters of a tubular TIA consisting of Sine Blocks $\TSB(h,a,f,s,n,r_i,t)$ are summarised in \Cref{tab:sine_param}.

\begin{table}[h!]
    \centering
    \begin{tabular}{|
        >{\centering\arraybackslash}m{1.6cm}|
        >{\centering\arraybackslash}m{1.6cm}|
        >{\centering\arraybackslash}m{1.6cm}|
        >{\centering\arraybackslash}m{1.6cm}|
        >{\centering\arraybackslash}m{1.6cm}|
        >{\centering\arraybackslash}m{1.6cm}|
        >{\centering\arraybackslash}m{1.6cm}|
        >{\centering\arraybackslash}m{1.6cm}|
    }
        \hline
        \makecell{$h$} &
        \makecell{$a$} &
        \makecell{$f$} &
        \makecell{$s$} &
        \makecell{$n$} &
        \makecell{$r_i$} &
        \makecell{$t$} &
        \makecell{$L$} \\
        \hline
        height of the block & amplitude of the sine curve & frequency of the sine curve & shift of the sine curve & number of blocks in one layer & inner radius of the cylinder & thickness of the blocks & number of layers in the assembly \\ 
        \hline
    \end{tabular}
    \caption{Parameters defining the tubular TIA composed of Sine Blocks.}
    \label{tab:sine_param}
\end{table}

\subsection{Hexagon-based Block}
\begin{figure}[h!]
    \centering
    \begin{subfigure}{0.31\textwidth}
        \centering
        \includegraphics[width=1\linewidth]{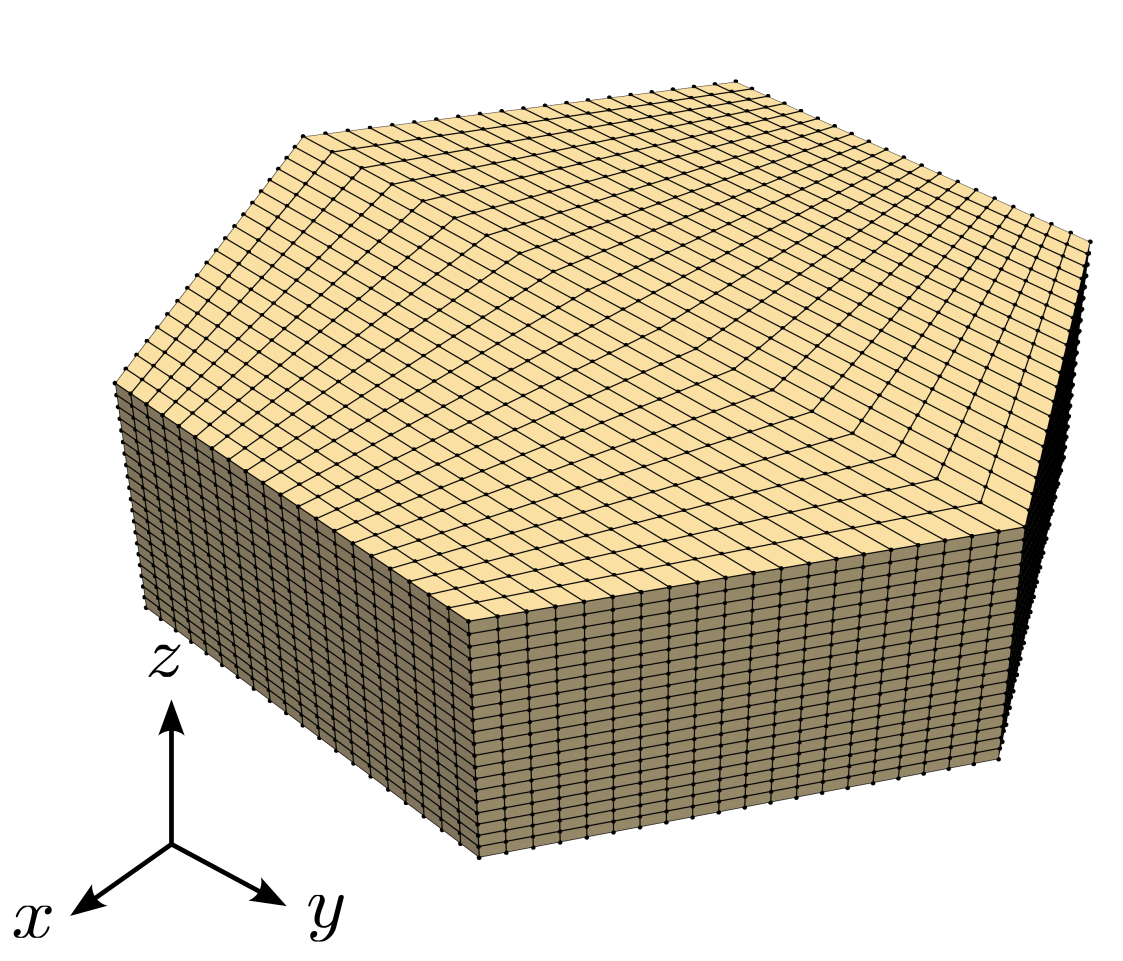}
        \caption{}
        \label{subfig:hex_map_1}
    \end{subfigure}
    \begin{subfigure}{0.31\textwidth}
        \centering
        \includegraphics[width=0.8\linewidth]{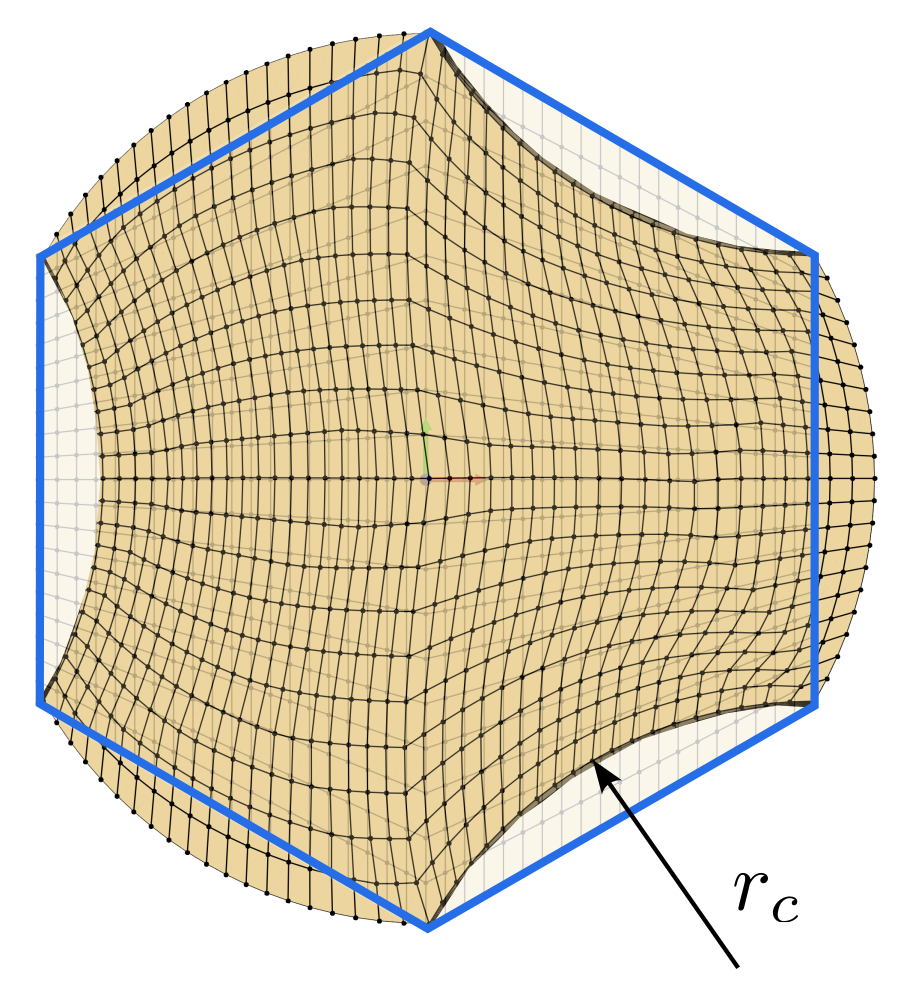}
        \caption{}
        \label{subfig:hex_map_2}
    \end{subfigure}
    \begin{subfigure}{0.31\textwidth}
        \centering
        \includegraphics[width=1\linewidth]{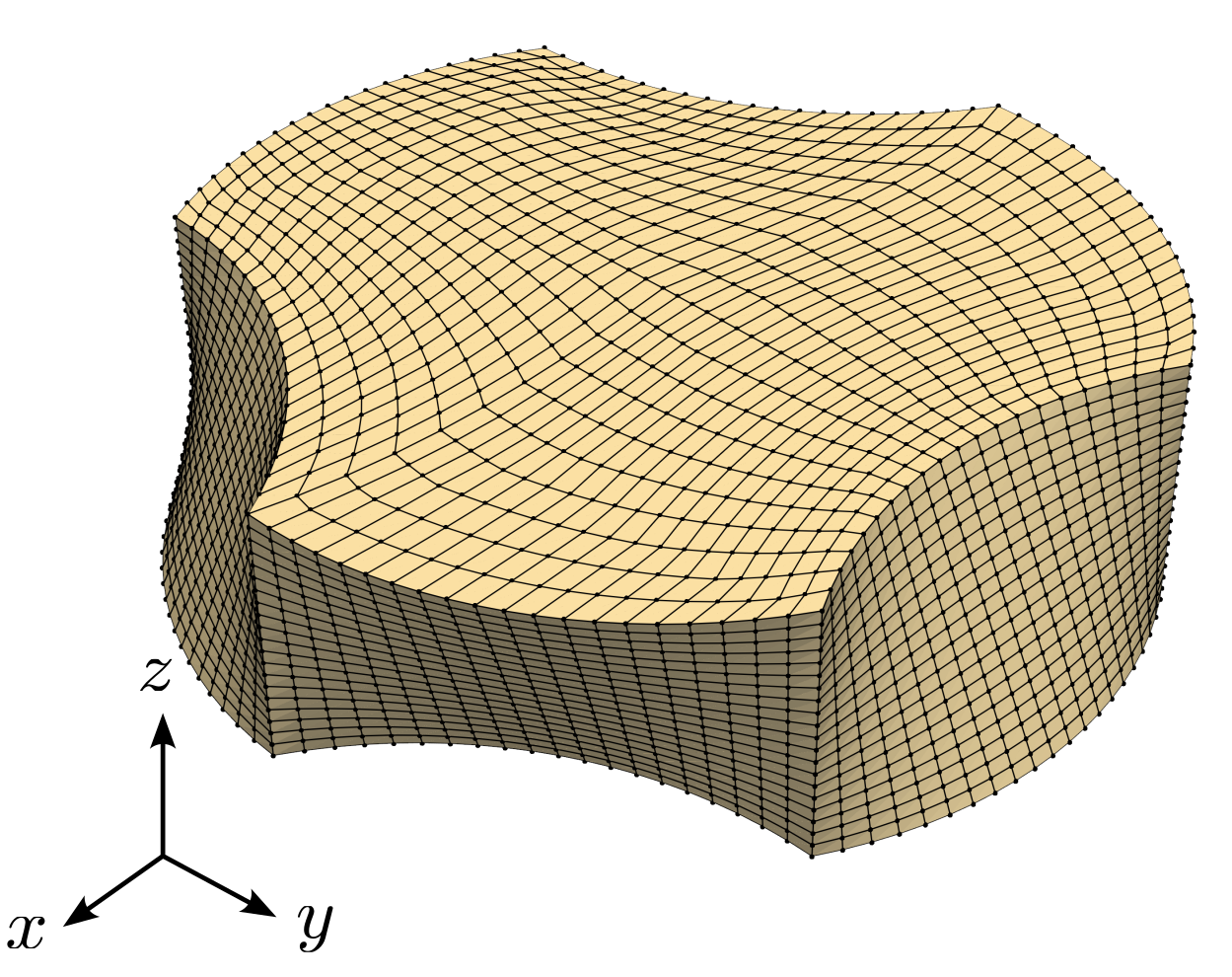}
        \caption{}
        \label{subfig:hex_map_3}
    \end{subfigure}
    \caption{Transformation of extruded hexagon (a) via the circumscribed circle (b) to the planar hexahedral block (c).}
    \label{fig:hex_mapping}
\end{figure}

\begin{figure}[h!]
    \centering
    \begin{subfigure}{0.48\textwidth}
       \centering
        \includegraphics[width=0.7\linewidth]{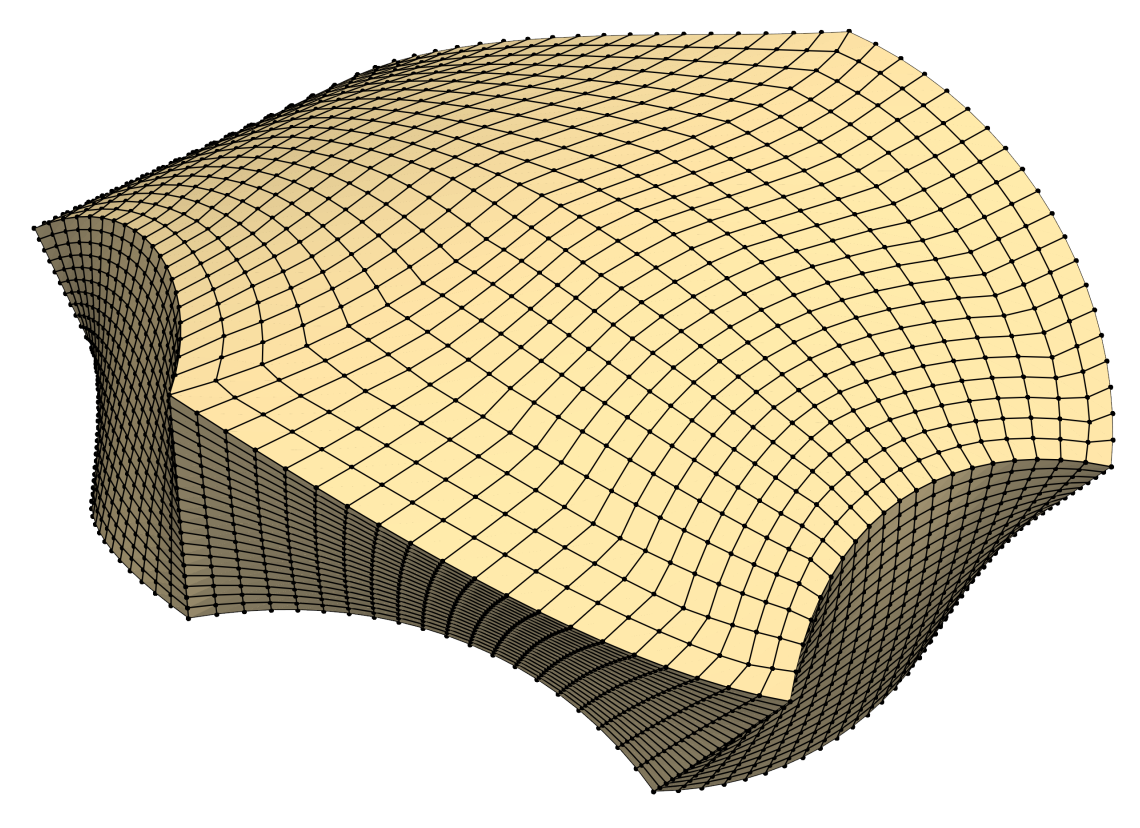}
       \caption{}
       \label{fig:hexagon-based-block}
    \end{subfigure}
    \begin{subfigure}{0.48\textwidth}
        \centering
        \includegraphics[width=0.5\linewidth]{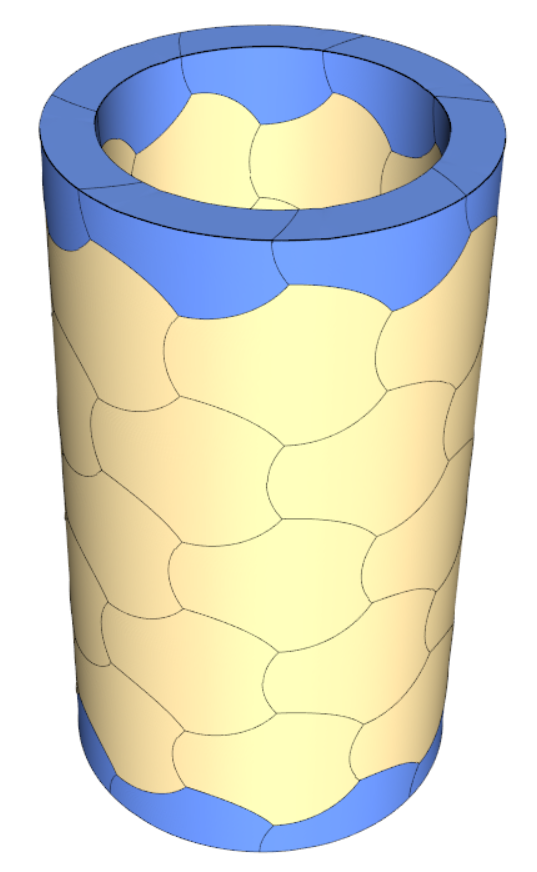}
        \caption{}
        \label{fig:assembly_hexagon}
    \end{subfigure}
    \caption{Hexahedron mesh of $\THB(n=7, r_i, r_o)$ (a) and the assembly of the hexagon-based block $\THB(n=7, r_i, r_o)$ with four layers $L=5$ (b).}
\end{figure}

Another tubular TIA block is the hexagon-based block which is introduced in \cite{HexBased}. The corresponding planar block is constructed by extruding a hexagon and deforming it using the circumscribed circle $r_c$, see \Cref{fig:hex_mapping}. Then, this block is mapped onto the tube, which has an inner radius of $r_i$ and an  outer radius of $r_o$. The resulting block is depicted in \Cref{fig:hexagon-based-block}.

The blocks are arranged in $L$ layers of $n$ blocks each. Thus, we define the hexagon-based block as $\THB(n,r_i,r_o)$. The assembly of $\THB(n = 7,r_i,r_o)$ with $L=5$ is shown in \Cref{fig:assembly_hexagon}. For a more detailed description of the hexagon-based block, the reader is referred to the mentioned publication.

\section{Problem formulation}

From a mechanical point of view, \emph{interlocking} denotes the ability of an assembly to suppress relative motion among its components and thus attain an immobilized state under prescribed boundary conditions. In this sense, interlocking emerges when geometry and contact interactions jointly block admissible motion directions. 

To analyze this, we describe this mechanical problem using two complementary theories based on different physical assumptions. In the first stage, we employ a simplified multibody dynamics (MBD) formulation, which enables a rapid and reliable assessment of force equilibrium together with kinematic immobilization. This stage evaluates whether geometry alone provides sufficient constraints to ensure interlocking and, in addition, delivers accurate estimates of the contact pressure distribution between interacting blocks.
In the second stage, we reformulate the problem within the framework of continuum mechanics. The blocks comprising the assemblies are modelled as deformable continua, leading to a finite element (FE) formulation that allows for a more detailed mechanical analysis. This continuum-based approach captures the deformation and stress fields within quasi-static assemblies, thereby providing deeper insights into their mechanical behaviour.

%#####################################################################################
\subsection{Newton--Euler Dynamics with Contacts}\label{subsec:NewtonEuler}

We focus on quasi-static problems and therefore neglect inertia. In the absence of inertia, the equations of motion reduce to a \emph{static complementarity} problem: contact forces must balance external loads while respecting non-penetration and the unilateral, non-negative nature of contact reactions \citep{Moreau1966,Loetstedt1982,Pfeiffer1996,Anitescu1997,Trinkle1997,Stewart2000,Gilardi2002}. 

\begin{figure}[h!]
    \centering
    \includegraphics[scale=1.0]{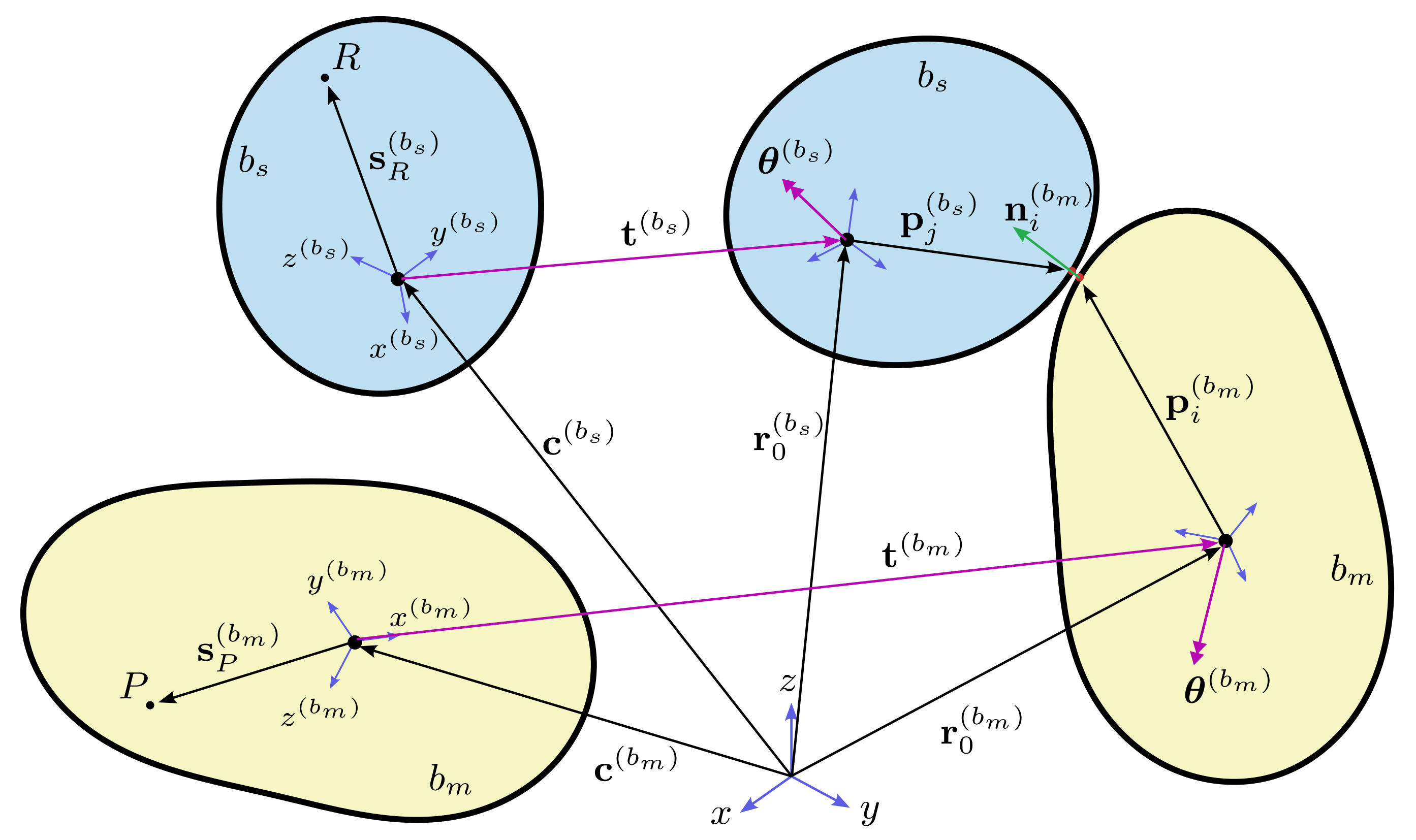}
    \caption{Kinematic description of body motion: reference frame, translation, and rotation vectors.}
    \label{fig:mbd_definitions}
\end{figure}

Let us consider a collection of $N_b$ rigid bodies in $\mathbb{R}^n$, whose configurations are described by global generalized coordinates \citep{Nikravesh1988}
\begin{equation*}
    \vb{q} = \big( \vb{q}^{(1)},\dots,\vb{q}^{(N_b)} \big) \in \mathbb{R}^{6 N_b},
\end{equation*}
where \(\vb{q}^{(b)}\) are the generalized coordinates of body \(b \in\{1,\dots,N_b\}\) with
\begin{equation*}
    \vb{q}^{(b)} 
    = \left( x, y, z, \theta_x, \theta_y, \theta_z \right)
    = \left(\vb{t}^{(b)},\vb*{\theta}^{(b)} \right)\in \mathbb{R}^6 .
\end{equation*}
Here, vector $\vb{t}^{(b)}$ denotes the translation of a chosen reference point $\vb{c}^{(b)}$ of a body $b$ (e.g.\ the centroid), and $\vb*{\theta}^{(b)}$ is the body rotation vector. 
The spatial position vector of material point $P$ rigidly attached to body $b$ is given by
\begin{equation*}
    \vb{r}_P^{(b)}(\vb{q}^{(b)}) 
    = \vb{r}_0^{(b)}(\vb{t}^{(b)}) + \vb{p}_P^{(b)}(\vb*{\theta}^{(b)}).
\end{equation*}
The vector $\vb{r}_0^{(b)}$ represents the translational motion of the body reference frame, while $\vb{p}_P^{(b)}$ accounts for the rotational contribution of point $P$ relative to this frame (see \Cref{fig:mbd_definitions}). These quantities are defined as
\begin{equation*}
    \vb{r}_0^{(b)}(\vb{t}^{(b)}) = \vb{c}^{(b)} + \vb{t}^{(b)}, 
    \qquad
    \vb{p}_P^{(b)}(\vb*{\theta}^{(b)}) = \vb{R}^{(b)}(\vb*{\theta}^{(b)})\,\vb{s}_P^{(b)},
\end{equation*}
where $\vb{R}^{(b)} \in \mathrm{SO}(3)$ is the rotation matrix parametrized by $\vb*{\theta}^{(b)}$, and $\vb{s}_P^{(b)}$ is the position vector of the point $P$ relative to the body-fixed reference frame located at $\vb{c}^{(b)}$.
%$\vb{R}^{(b)} = \mathrm{exp} \left( [\vb*{\theta}]_\times\right) \in \mathrm{SO}(3)$ is a body rotation and $\vb{s}_P^{(b)}$ is the body-fixed position of $P$ relative to the reference point.
The rotation matrix can be represented using the exponential map of the \emph{skew-symmetric matrix} corresponding to the rotation vector $[\vb*{\theta}^{(b)}]_\times$. Specifically, we have
\begin{equation*}
    \vb{R}^{(b)}(\vb*{\theta}^{(b)}) = \mathrm{exp} \left( [\vb*{\theta}^{(b)}]_\times\right),
    \quad
    [\vb*{\theta}^{(b)}]_\times =
    \begin{bmatrix}
        0 & -\theta_z & \theta_y \\
        \theta_z & 0 & -\theta_x \\
        -\theta_y & \theta_x & 0
    \end{bmatrix}.
\end{equation*}
This representation concisely and rigorously relates rotation vectors to rotation matrices. %and is widely used in multibody dynamics.
Using $\delta\vb{R}^{(b)} = \vb{R}^{(b)} \, [\delta\vb*{\theta}^{(b)}]_\times$ and the property $\vb{R}^{(b)} \, \delta\vb*{\theta}^{(b)} = \delta\vb*{\theta}^{(b)}$, the infinitesimal position variation is
\begin{equation}\label{eq:position-variation}
    \delta \vb{r}_P^{(b)} = 
    \delta \vb{r}_0^{(b)} + \delta(\vb{R}^{(b)}\vb{s}_P^{(b)}) = 
    \delta \vb{t}^{(b)} + \vb{R}^{(b)} [\delta\vb*{\theta}^{(b)}]_\times \, \vb{s}_P^{(b)}=  
    \delta \vb{t}^{(b)} + \delta\vb*{\theta}^{(b)} \times \vb{p}_P^{(b)},
\end{equation}
where $\delta\vb*{\theta}^{(b)} \in \mathbb{R}^3$ is the infinitesimal rotation vector.

For an assembly of $N_b$ rigid bodies, contact pairs are defined for every unordered body pair $(m,s)$ with $1 \le m < s \le N_b$. The local contact set for pair $(b_m,b_s)$ is
\begin{equation*}
    \mathcal{I}^{(m,s)} \coloneqq \{\, (i,j) \mid i \in \mathcal{C}_{b_m},\; j \in \mathcal{C}_{b_s},\; j \text{ is the unique opposing point associated with } i \,\},
\end{equation*}
where $\mathcal{C}_{b_m}$ and $\mathcal{C}_{b_s}$ denote the sets of candidate contact points on bodies $b_m$ and $b_s$, respectively. Under the one-to-one pairing assumption in \(\mathcal{I}^{(m,s)}\), the points on the master side can be indexed as
\begin{equation*}
    \mathcal{C}_{b_m} = \{1,\ldots,n_c^{(m,s)}\},
\end{equation*}
with $n_c^{(m,s)} = |\mathcal{C}_{b_m}| = |\mathcal{I}^{(m,s)}|$. The global active contact index set is then written in expanded form as
\begin{equation*}
    \mathcal{I} \coloneqq \left\{(m,s,i,j)\ \middle|\ 1 \le m < s \le N_b,\ (i,j) \in \mathcal{I}^{(m,s)}\right\}.
\end{equation*}
Each contact tuple $(m,s,i,j) \in \mathcal{I}$ is defined by a material point on body $b_m$ with global position $\vb{r}_i^{(b_m)}\!\left(\vb{q}^{(b_m)}\right)$, its associated opposing point $\vb{r}_j^{(b_s)}\!\left(\vb{q}^{(b_s)}\right)$ on body $b_s$, and a unit normal vector $\vb{n}_i$, oriented from the surface of body $b_m$ toward body $b_s$. We restrict attention to frictionless unilateral contact between rigid bodies. Under this assumption, the contact constraint is characterized by a scalar \emph{gap function}, which quantifies the signed normal separation -- or, when negative, the penetration -- between the two opposing surface points \citep{Anitescu1997,Trinkle1997,Pfeiffer1996,Gilardi2002}, defined as
\begin{equation}\label{eq:gap-mbd}
    g_{m s}^{(i,j)}(\vb{q}) \coloneqq \vb{n}_i \cdot \left( \vb{r}_j^{(b_s)}(\vb{q}^{(b_s)}) - \vb{r}_i^{(b_m)}(\vb{q}^{(b_m)}) \right) \geq 0 .
\end{equation} 

With these definitions, the governing equations follow from the principle of virtual work applied to a constrained system. The Lagrange multiplier method augments the virtual work functional by enforcing the active constraints \(g_{m s}^{(i,j)}(\vb{q}) = 0\) through scalar multipliers \(\uplambda_{m s}^{(i,j)}\). In this constrained variational setting, the contribution $\delta W_{\mathrm{c}}$ of all active contacts \((m,s,i,j)\in\mathcal{I}\) to the total virtual work $\delta W$ is
\begin{equation}\label{eq:Wc_mbd}
    \delta W_{\mathrm{c}}
    = \sum_{(m,s,i,j) \in \mathcal{I}}
    \uplambda_{m s}^{(i,j)}\, \delta g_{m s}^{(i,j)} (\vb{q}) .
\end{equation}
Because the multipliers enter as coefficients of the virtual variations \(\delta g_{m s}^{(i,j)}\), the Lagrange multiplier \(\uplambda_{m s}^{(i,j)}\) is, by construction, the generalized force conjugate to the constraint \(g_{m s}^{(i,j)}\).
% Consequently, the associated reaction force acting on the bodies is of the form
% \begin{equation}
%     \uplambda_{(i,j)}\, \vb{n}_{(i,j)},
% \end{equation}
% where $\vb{n}_{(i,j)}$ is the outward normal of the gap function.
For an arbitrary virtual displacement $\delta\vb{q}$ and any contact tuple $(m,s,i,j)\in\mathcal{I}$, let
\(\delta\vb{q}^{(m,s)} \coloneqq (\delta\vb{q}^{(b_m)},\,\delta\vb{q}^{(b_s)})^{\intercal}\)
denote the subvector of generalized variations associated with bodies $b_m$ and $b_s$. The virtual variation of \eqref{eq:gap-mbd} is
\begin{align*}
    \delta g_{m s}^{(i,j)}
    &= 
    \frac{\partial g_{m s}^{(i,j)}}{\partial \vb{q}^{(m,s)}} \cdot \delta \vb{q}^{(m,s)} \\
    &= \vb{n}_i \cdot \left( 
        \delta \vb{t}^{(b_s)} 
        + \delta\vb*{\theta}^{(b_s)} \times \vb{p}_j^{(b_s)}
        - \delta \vb{t}^{(b_m)} 
        - \delta\vb*{\theta}^{(b_m)} \times \vb{p}_i^{(b_m)}
    \right) \\
    &= {\delta \vb{t}^{(b_s)}} \cdot \vb{n}_i  
       + {\delta\vb*{\theta}^{(b_s)}} \cdot (\vb{p}_j^{(b_s)} \times \vb{n}_i) 
       - {\delta \vb{t}^{(b_m)}} \cdot \vb{n}_i 
       - {\delta\vb*{\theta}^{(b_m)}} \cdot (\vb{p}_i^{(b_m)} \times \vb{n}_i) .
\end{align*}
In matrix notation, this can be expressed as
\begin{equation}\label{eq:gap-variation}
    \delta g_{m s}^{(i,j)}= 
    \begin{pmatrix}
        \delta \vb{t}^{(b_m)}\\
        \delta\vb*{\theta}^{(b_m)}\\
        \delta \vb{t}^{(b_s)}\\
        \delta\vb*{\theta}^{(b_s)}
    \end{pmatrix}^{\!\intercal}
    \underbrace{
    \begin{pmatrix}
        -\vb{n}_i\\
        -(\vb{p}_i^{(b_m)} \times \vb{n}_i) \\
        \vb{n}_i\\
        (\vb{p}_j^{(b_s)} \times \vb{n}_i)
    \end{pmatrix}
    }_{\displaystyle \coloneqq \vb{G}_{m s}^{(i,j)}}
    = 
    \delta {\vb{q}^{(m,s)}}^{\intercal} \vb{G}_{m s}^{(i,j)} ,
\end{equation}
where $\vb{G}_{m s}^{(i,j)}$ denotes the \emph{Jacobian matrix} of the local gap function \eqref{eq:gap-mbd}, also referred to as the \emph{contact Jacobian}.
% \begin{equation}
%     \vb{G}^{(i,j)} \coloneqq 
%     \begin{pmatrix}
%         -\vb{n}_i\\
%         -(\vb{p}_i^{(b_m)} \times \vb{n}_i) \\
%         \vb{n}_i\\
%         (\vb{p}_j^{(b_s)} \times \vb{n}_i)
%     \end{pmatrix}.
% \end{equation}
This yields the contribution \eqref{eq:Wc_mbd} to the total virtual work $\delta W$, expressed in compact global form as
\begin{equation*}
    \delta W_{\mathrm c}
    =
    \sum_{(m,s,i,j)\in\mathcal{I}} \uplambda_{m s}^{(i,j)}\,\delta g_{m s}^{(i,j)}(\vb{q})
    =
    \delta \vb{q}^{\intercal}(\vb{G}\vb*{\uplambda})
    =
    \delta \vb{q}^{\intercal} \vb{f}_c ,
\end{equation*}
where $\vb{f}_c \coloneqq \vb{G}\vb*{\uplambda}$ denotes the global generalized contact force, $\vb{G}$ is the \emph{global contact Jacobian} obtained by assembling all local contact Jacobians $\vb{G}_{m s}^{(i,j)}$, and $\vb*{\uplambda} = \left(\uplambda_{m s}^{(i,j)}\right)_{(m,s,i,j)\in\mathcal{I}}$ is the global vector of contact Lagrange multipliers.  

% An external force--couple pair $(\vb{F}_k^{(b_s)},\vb{M}_k^{(b_s)})$ applied at point $P_k$ on body $b$ contributes to the external virtual work
An external force--couple pair $(\vb{F}_k^{(b_s)}, \vb{M}_k^{(b_s)})$, where $\vb{F}_k^{(b_s)}$ denotes the applied force and $\vb{M}_k^{(b_s)}$ the associated moment acting on body $b_s$ at point $P_k$, contributes to the external virtual work
\begin{align*}
    \delta W_{\text{ext},k}^{(b_s)}
    &= \vb{F}_k^{(b_s)} \cdot \delta\vb{r}_{P_k}^{(b_s)} + \vb{M}_k^{(b_s)} \cdot \delta\vb*{\theta}_{P_k}^{(b_s)} \\
    &= \vb{F}_k^{(b_s)} \cdot \left(\delta\vb{t}^{(b_s)} + \delta\vb*{\theta}^{(b_s)}\times\vb{p}_{P_k}^{(b_s)}\right)
       + \vb{M}_k^{(b_s)} \cdot \delta\vb*{\theta}^{(b_s)} \\
    &= \vb{F}_k^{(b_s)} \cdot \delta\vb{t}^{(b_s)}
       + \left(\vb{M}_k^{(b_s)} + \vb{p}_{P_k}^{(b_s)}\times\vb{F}_k^{(b_s)}\right)^{} \cdot \delta\vb*{\theta}^{(b_s)}.
\end{align*}
Summing over all external loads on body $b_s$ gives
\begin{equation*}
    \delta W_{\text{ext}}^{(b_s)}
    = {\delta\vb{t}^{(b_s)}} \cdot \left(\sum_{k} \vb{F}_k^{(b_s)}\right)
      + {\delta\vb*{\theta}^{(b_s)}} \cdot \left(\sum_{k} \left(\vb{M}_k^{(b_s)} + \vb{p}_{P_k}^{(b_s)}\times\vb{F}_k^{(b_s)}\right)\right).
\end{equation*}
Thus, the generalized external-force vector for body $b_s$ is
\begin{equation*}
    \vb{f}_{\text{ext}}^{(b_s)}
    =
    \begin{pmatrix}
        \displaystyle \sum_{k} \vb{F}_k^{(b_s)} \\
        \displaystyle \sum_{k} \left(\vb{M}_k^{(b_s)} + \vb{p}_{P_k}^{(b_s)}\times\vb{F}_k^{(b_s)}\right)
    \end{pmatrix}.
\end{equation*}
An analogous expression holds for every body $b\in\{1,\ldots,N_b\}$. Collecting the contributions of all bodies yields
\begin{equation*}
    \vb{f}_{\text{ext}} = 
    \begin{pmatrix} 
    \vb{f}_{\text{ext}}^{(1)} \\ \vdots \\ \vb{f}_{\text{ext}}^{(N_b)}
    \end{pmatrix},
    \qquad\text{and}\qquad
    \delta W_{\text{ext}} = \delta\vb{q}^{\intercal}\,\vb{f}_{\text{ext}}.
\end{equation*}

Expressing the total virtual work $\delta W$ as the sum of the generalized external contribution $\delta W_{\mathrm{ext}}$ (e.g., gravity and applied loads) and the contact contribution $\delta W_{\mathrm{c}}$, the principle of virtual work yields
\begin{equation*}
    \delta W
    = \delta W_{\text{ext}} + \delta W_{\mathrm c}
    = \delta\vb{q}^{\intercal}\left( \vb{f}_{\text{ext}} + \vb{f}_{\mathrm{c}} \right)
    = 0.
\end{equation*}
Since the above virtual work identity must hold for all admissible $\delta\vb{q}$, we obtain the static force balance
\begin{equation}\label{eq:force-balance}
    \vb{f}_{\text{ext}} + \vb{f}_{\mathrm c} = \vb{0}.
\end{equation}
Each contact tuple $(m,s,i,j)\in\mathcal{I}$ is governed by complementarity conditions that enforce non-penetration and exclude tensile (pulling) contact forces, so that reactions arise only at active contacts \citep{Loetstedt1982,Pfeiffer1996,Anitescu1997,Stewart2000}. Together with~\eqref{eq:force-balance}, these conditions define the static complementarity problem
\begin{subequations}\label{eq:SCP}
\renewcommand{\theequation}{\theparentequation.\alph{equation}}
\begin{align}
    &\vb{f}_{\mathrm{ext}}(\vb{q}) + \vb{G}(\vb{q})\,\vb*{\uplambda} = \vb{0},\\[4pt]
    &g_{m s}^{(i,j)}(\vb{q}) \ge 0,\qquad 
    \uplambda_{m s}^{(i,j)} \ge 0,\qquad 
    \uplambda_{m s}^{(i,j)} \, g_{m s}^{(i,j)} (\vb{q}) = 0,
    \quad \forall (m,s,i,j)\in\mathcal{I} .
\end{align}
\end{subequations}

When the configuration $\vb{q}$ is unknown, it must be solved simultaneously with contact activation. In this case, the gap functions $g_{m s}^{(i,j)}(\vb{q})$ depend on $\vb{q}$, and the active set evolves consistently with equilibrium. The resulting formulation is a nonlinear complementarity problem (NCP), or equivalently, a mathematical program with complementarity constraints (MPCC), a viewpoint consistent with complementarity-based rigid-body formulations \citep{Moreau1966,Anitescu2004}.

However, for a given configuration $\vb{q}$ (i.e., fixed geometry), the quantities $\vb{f}_{\text{ext}}(\vb{q})$, $\vb{G}(\vb{q})$, and $g_{m s}^{(i,j)}(\vb{q})$ are known. In this setting, the only unknowns are the contact reactions $\vb*{\uplambda}$, which must balance the system at that configuration. Consequently, the governing equations reduce to a linear system with fixed active set and zero gap, which is then solved using established complementarity and contact-dynamics algorithms \citep{Baraff1989,Coutinho2001,Kaufman2008,Anitescu2004}:
\begin{align} \label{eq:SCP-linear}
    \vb{f}_{\text{ext}} + \vb{G} \vb*{\uplambda} = \vb{0}
    \quad \text{s.t.}
    \quad g_{m s}^{(i,j)} = 0,\quad 
    \uplambda_{m s}^{(i,j)} \ge 0,
    \quad \forall (m,s,i,j)\in\mathcal{I}.
\end{align}

In the present setting, this is a purely \emph{static} (force-equilibrium) problem. It asks whether the given set of contact normals (the columns of $\vb{G}$) can generate the required reaction $-\vb{f}_{\text{ext}}$ through a nonnegative linear combination. If so, an equilibrium exists and the corresponding multipliers $\vb*{\uplambda}$ define the contact reactions; if not, the structure cannot sustain the applied load in this configuration. 
This is a necessary condition for static support, but it is not sufficient to guarantee that the configuration is motionless. In unilateral contact, a system may admit balanced contact forces and still possess a feasible infinitesimal motion, which motivates combining equilibrium checks with mechanism analysis \citep{Wang2018,Wang2019}.

\emph{Kinematic feasibility} addresses this missing condition. Instead of asking whether forces can balance, we ask whether a nonzero infinitesimal generalized displacement direction $\delta \vb{q}$ can satisfy contact admissibility.
For frictionless unilateral contact, admissibility requires $g_{m s}^{(i,j)}(\vb{q}) \ge 0$ for all $(m,s,i,j)\in\mathcal{I}$. Using the first-order gap variation \eqref{eq:gap-variation}, we obtain (see, for instance, Section~3.1 in \cite{Wang2019})
\begin{equation*}
    \delta g_{m s}^{(i,j)}(\vb{q}) = {\vb{G}_{m s}^{(i,j)}}^{\intercal}\delta\vb{q}^{(m,s)} \ge 0,
    \qquad \forall (m,s,i,j)\in\mathcal{I}.
\end{equation*}
At active contacts ($g_{m s}^{(i,j)} = 0$), the sign of $\delta g_{m s}^{(i,j)}$ determines whether the motion opens the contact ($>0$), preserves it ($=0$), or immediately violates non-penetration ($<0$). 
% Higher-order terms become relevant only when $\delta g_{m s}^{(i,j)}=0$; these correspond to second-order effects and are outside first-order feasibility analysis.
Collecting the first-order admissibility conditions over all active contacts yields the global kinematic feasibility condition
\begin{equation}\label{eq:kinematic-feasibility}
    \vb{G}^{\intercal}\delta\vb{q} \ge \vb{0},
\end{equation}
which defines the cone of admissible infinitesimal generalized displacement directions, i.e., \emph{mechanisms},
\begin{equation*}
    \mathcal{M} = \{ \delta \vb{q} \mid \vb{G}^{\intercal} \delta \vb{q} \ge \vb{0} \}.
\end{equation*}
If this cone contains a nonzero direction, the contact state admits a mechanism and is therefore \emph{kinematically underconstrained}. Conversely, if the only admissible direction is $\delta\vb{q}=\vb{0}$, all first-order motions are blocked and the assembly is \emph{kinematically immobilized}.

This kinematic criterion alone does not determine whether the mechanism will actually evolve under the applied loads. For that, we evaluate the external work rate along an admissible direction:
\begin{equation*}
    \vb{f}_{\text{ext}}^{\intercal} \, \delta \vb{q}.
\end{equation*}
Its sign distinguishes the physically relevant cases: zero corresponds to a \emph{neutral (undriven)} mechanism, a positive value identifies a \emph{load-activated} (first-order unstable) mechanism, and a negative value indicates that the applied \emph{loads oppose the admissible motion}. Hence, a load-activated mechanism requires
\begin{equation*}
    \vb{f}_{\text{ext}}^{\intercal} \, \delta \vb{q} > 0.
\end{equation*}

Together, \eqref{eq:SCP-linear} and \eqref{eq:kinematic-feasibility} provide complementary views of the same contact state: the former enforces static equilibrium in force space, while the latter tests whether the contact geometry blocks first-order motion in configuration space.

%######################################################################################
% Continuum mechanics
%######################################################################################

\subsection{Continuum mechanics framework}
In the framework of continuum mechanics, the interlocking problem can be formulated as follows. The reference configuration of body $b$, denoted by $\Omega_0^{(b)} \subset \mathbb{R}^3$, is defined as the set of all material points $P$ of the body at time $t = 0$. Each material point is uniquely identified by its position vector $\vb{X}^{(b)}$, measured with respect to a fixed origin $O$. The current configuration $\Omega_t^{(b)} \subset \mathbb{R}^3$ describes the positions $\vb{x}^{(b)}$ of these points at time $t$. On every domain $\Omega_0^{(b)}$ of body $b$, the displacement field is defined as
\begin{equation*}
    \vb{u}^{(b)}(\vb{X}^{(b)},t) = \vb{x}^{(b)}(\vb{X}^{(b)},t) - \vb{X}^{(b)}.
\end{equation*}
The boundary of each body in the reference configuration $\partial \Omega_0^{(b)}$, is partitioned into three disjoint subsets $\partial \Omega_0^{(b)} = \Gamma_u^{(b)} \cup \Gamma_\sigma^{(b)} \cup \Gamma_c^{(b)}$. Here, $\Gamma_{\sigma}^{(b)}$ denotes the surface on which tractions $\tilde{\vb{t}}_0^{(b)}$ are prescribed (Neumann boundary); $\Gamma_{u}^{(b)}$ denotes the surface on which displacements $\tilde{\vb{u}}^{(b)}$ are imposed (Dirichlet boundary); and $\Gamma_{c}^{(b)}$ denotes the contact surface. The contact interaction is mathematically expressed as a set of non-linear boundary conditions (see, e.g., \cite{laursen2003computational, wriggers2006computational}).
The initial boundary value problem governing finite deformation elastodynamics for each body can be stated as
\begin{subequations}\label{eq:inital-BVP}
\renewcommand{\theequation}{\theparentequation.\alph{equation}}
\begin{align}
    \Div \vb{P}^{(b)} + \tilde{\vb{b}}_0^{(b)} &= \rho_0^{(b)} \ddot{\vb{u}}^{(b)} && \text{in } \Omega_0^{(b)} \times [0,t],\\
    \vb{u}^{(b)} &= \tilde{\vb{u}}^{(b)} && \text{on } \Gamma_u^{(b)} \times [0,t],\\
    \vb{P}^{(b)} \vb{N}^{(b)} &= \tilde{\vb{t}}_0^{(b)} && \text{on } \Gamma_{\sigma}^{(b)} \times [0,t],\\
    \vb{u}^{(b)} (\vb{X}^{(b)}, 0) &= \tilde{\vb{u}}^{(b)} (\vb{X}^{(b)}) && \text{in } \Omega_0^{(b)},\\
    \dot{\vb{u}}^{(b)} (\vb{X}^{(b)}, 0) &= \dot{\tilde{\vb{u}}}^{(b)} (\vb{X}^{(b)}) && \text{in } \Omega_0^{(b)},\\
    g_n^{(b_m,b_s)}(\vb{X}^{(b_m)},t) &\geq 0 && \text{on } \Gamma_c^{(b_m,b_s)} \times [0,t] \label{eq:cont-1}\\
    p_n^{(b_m,b_s)}(\vb{X}^{(b_m)},t) &\leq 0 && \text{on } \Gamma_c^{(b_m,b_s)} \times [0,t] \label{eq:cont-2}\\
    p_n^{(b_m,b_s)}(\vb{X}^{(b_m)},t) \, g_n^{(b_m,b_s)}(\vb{X}^{(b_m)},t) &= 0 && \text{on } \Gamma_c^{(b_m,b_s)} \times [0,t] , \label{eq:cont-3}
\end{align}
\end{subequations}
where $\vb{P}^{(b)}$ denotes the first Piola-Kirchhoff stress tensor, $\Div(\cdot)$ is the Lagrangian divergence operator, and $\vb{N}^{(b)}$ is the outward unit normal vector to the reference configuration. The frictionless contact conditions in the normal direction~\eqref{eq:cont-1}, \eqref{eq:cont-2} and \eqref{eq:cont-3} must hold on $\Gamma_c^{(b_m, b_s)}$ at all times $t$, where $g_n^{(b_m,b_s)}$ represents the normal gap function between a pair of bodies $b_m$ and $b_s$
\begin{equation*}
    g_{n}^{(b_m,b_s)}(\vb{X}, t)=
    -\vb{N}^{(b_m)}_{\mathrm{c}} \cdot\left[
    \vb{x}^{(b_m)}\left(\vb{X}^{(b_m)}, t\right)
    -{\vb{x}}^{(b_s)}\left({\vb{X}}^{(b_s)}, t\right)\right]
    \quad \text{on } \Gamma_c^{(b_m,b_s)},
\end{equation*}
and $p_n^{(b_m, b_s)}$ the contact pressure. The vector $\vb{N}^{(b_m)}_c$ denotes the contact normal, defined as the outward unit normal to the current master surface. The true (Cauchy) stress tensor $\vb*{\sigma}^{(b)}$ is related to the first Piola--Kirchhoff stress by
\begin{equation*}
    \vb*{\sigma}^{(b)} = \frac{1}{\det(\vb{F}^{(b)})} \, \vb{P}^{(b)} \left(\vb{F}^{(b)}\right)^T,
\end{equation*}
with $\vb{F}^{(b)} = \partial \vb{x}^{(b)} / \partial \vb{X}^{(b)}$ denoting the deformation gradient.
%#########################################################################################

\section{Implementation and simulation set-up}

\subsection{Numerical implementation of MBD framework}

For the investigations presented in the following sections, the static problem~\eqref{eq:SCP-linear} provides a sufficient level of accuracy. All components required for this study -- including geometry and assembly generation, geometric discretization, and the numerical solution of the static and kinematic feasibility problems~\eqref{eq:SCP-linear} and~\eqref{eq:kinematic-feasibility} -- have been implemented in a stand-alone code, available at \cite{code}.

Considering the static problem~\eqref{eq:SCP-linear} for a fixed configuration $\vb{q}$, the number of contact tuples $(m,s,i,j)\in\mathcal{I}$ generally exceeds the number of degrees of freedom (DOFs). Consequently, the system becomes \emph{statically indeterminate}, in the sense that multiple contact force vectors $\vb*{\uplambda}$ may satisfy~\eqref{eq:SCP-linear}. Several strategies can be employed to resolve this indeterminacy. One approach introduces a small elastic compliance (penalty), where contact forces result from small deformations, thereby yielding a unique solution. Alternatively, the problem can be formulated as an optimization problem, for instance as the quadratic program
\begin{equation}\label{eq:QP}
    \min_{\vb*{\uplambda}} \ \tfrac{1}{2} \| \vb*{\uplambda} \|^2
    \quad \text{s.t.} \quad
    \vb{f}_{\text{ext}} + \vb{G}\vb*{\uplambda} = \vb{0},
    \quad \uplambda_{m s}^{(i,j)} \ge 0,
    \quad \forall (m,s,i,j)\in\mathcal{I}.
\end{equation}
This formulation can be interpreted as selecting, among all admissible contact force distributions that satisfy equilibrium and unilateral contact constraints, the minimum-norm solution. This selection admits a clear physical interpretation as the rigid-contact limit of a compliant contact model with quadratic elastic energy, yielding a force distribution that suppresses unnecessary force amplification by distributing loads as uniformly and gently as possible across contacts, in close analogy to minimum-energy principles in elasticity.

For the kinematic feasibility condition~\eqref{eq:kinematic-feasibility}, the numerical task is to determine whether a nonzero generalized displacement direction $\delta \vb{q}$ exists such that $\vb{G}^{\intercal}\delta\vb{q} \ge \vb{0}$. To avoid the trivial solution $\delta\vb{q}=\vb{0}$, we solve an auxiliary linear program. Introducing a vector $\vb{s}\in\mathbb{R}^{6N_b}_{\ge 0}$, we write
\begin{equation}\label{eq:kinematic-feasibility-lp}
  \max_{\delta\vb{q},\vb{s}}\; \vb{1}^{\intercal}\vb{s} \quad 
  \text{s.t.} \quad 
  \vb{G}^{\intercal}\delta\vb{q} \ge \vb{0},\; 
  -\vb{s} \le \delta\vb{q} \le \vb{s},\; 
  \vb{0} \le \vb{s} \le \vb{1}.
\end{equation}
Let $z^{\star}$ denote the optimal value of~\eqref{eq:kinematic-feasibility-lp}. If $z^{\star}=0$, the contact arrangement is kinematically immobilized; if $z^{\star}>0$, the current contact state admits a nonzero first-order admissible motion direction. In the latter case, we then evaluate whether there exists an admissible mechanism that is activated by the applied loads by maximizing the external virtual work over the admissible cone with linear normalization:
\begin{equation}\label{eq:kinematic-feasibility-mechanism}
  \max_{\delta \vb{q}}\; \vb{f}_{\text{ext}}^{\intercal}\delta \vb{q}
  \quad \text{s.t.} \quad
  \vb{G}^{\intercal}\delta \vb{q} \ge \vb{0},\;
  \|\delta \vb{q}\|_1 \le 1.
\end{equation}
To explicitly assess force suppression, we also solve the companion minimization problem over the same admissible set:
\begin{equation}\label{eq:kinematic-feasibility-mechanism-min}
  \min_{\delta \vb{q}}\; \vb{f}_{\text{ext}}^{\intercal}\delta \vb{q}
  \quad \text{s.t.} \quad
  \vb{G}^{\intercal}\delta \vb{q} \ge \vb{0},\;
  \|\delta \vb{q}\|_1 \le 1.
\end{equation}
Let $\gamma_{\max}^{\star}$ and $\gamma_{\min}^{\star}$ denote the optimal values of \eqref{eq:kinematic-feasibility-mechanism} and \eqref{eq:kinematic-feasibility-mechanism-min}, respectively. Then $\gamma_{\max}^{\star}>0$ indicates the existence of a load-activated mechanism direction, whereas $\gamma_{\max}^{\star}=0$ means that the applied force does not activate any admissible mechanism direction. Likewise, $\gamma_{\min}^{\star}<0$ quantifies force suppression via the strongest load-opposing admissible mechanism direction, while $\gamma_{\min}^{\star}=0$ means that the applied force does not suppress any admissible mechanism direction. In particular, if both $\gamma_{\max}^{\star}=0$ and $\gamma_{\min}^{\star}=0$, all admissible mechanisms are energetically neutral under the applied load, i.e., $\vb{f}_{\text{ext}}^{\intercal}\delta\vb{q}=0$ for every admissible $\delta\vb{q}$. 

The numerical implementation is based on specific modeling and discretization choices, which define the algorithmic setup shown in Algorithm~\ref{alg:static-contact}. The geometries introduced above are discretized using standard hexahedral volume elements for the bulk and quadrilateral surface elements for the boundaries. Since the multibody formulation requires only surface information for contact detection and force transmission, the MBD computations operate exclusively on the surface discretization.

\begin{algorithm}[h]
\caption{Static contact force computation for a rigid block assembly}
\label{alg:static-contact}
\begin{algorithmic}[1]
\State Perform contact detection over the assembly and identify all active contact faces
\For{each block \(b = 1,\dots,N_b\)}
    \State Compute the mass centroid of block \(b\): $\vb{c}^{(b)}$
    \For{each contact face \(i = \{1,\dots,k^{(b)}\}\) of block \(b\)}
        \State Compute the face centroid $\vb{r}_i^{(b)}$
        \State Compute the face normal \(\vb{n}_i^{(b)}\)
        \State Compute the face lever arm $\vb{p}_i^{(b)} = \vb{r}_i^{(b)} - \vb{c}^{(b)}$
        \State Compute the face torque vector
        $\vb*{\tau}_i^{(b)} = \vb{p}_i^{(b)} \times \vb{n}_i^{(b)}$
    \EndFor
    \State Construct the contact Jacobian \(\vb{G}^{(b)}\)
    \State Construct the generalized external force vector
          \(\vb{f}_{\text{ext}}^{(b)}\)
\EndFor
\State Assemble the global contact Jacobian \(\vb{G}\) and the global external force vector \(\vb{f}_{\text{ext}}\)
\State Solve the linear program~\eqref{eq:kinematic-feasibility-lp} to evaluate kinematic feasibility
\If{$z^{\star}>0$ (admissible mechanisms exist)}
    \State Solve the linear program~\eqref{eq:kinematic-feasibility-mechanism}
          to evaluate load activation of admissible mechanisms
    \State Solve the linear program~\eqref{eq:kinematic-feasibility-mechanism-min}
          to evaluate force suppression of admissible mechanisms
\EndIf
\State Solve the quadratic program~\eqref{eq:QP} to obtain the contact forces \(\vb*{\uplambda}\)
\end{algorithmic}
\end{algorithm}

\begin{figure}[h!]
    \centering
    \includegraphics[scale=1.0]{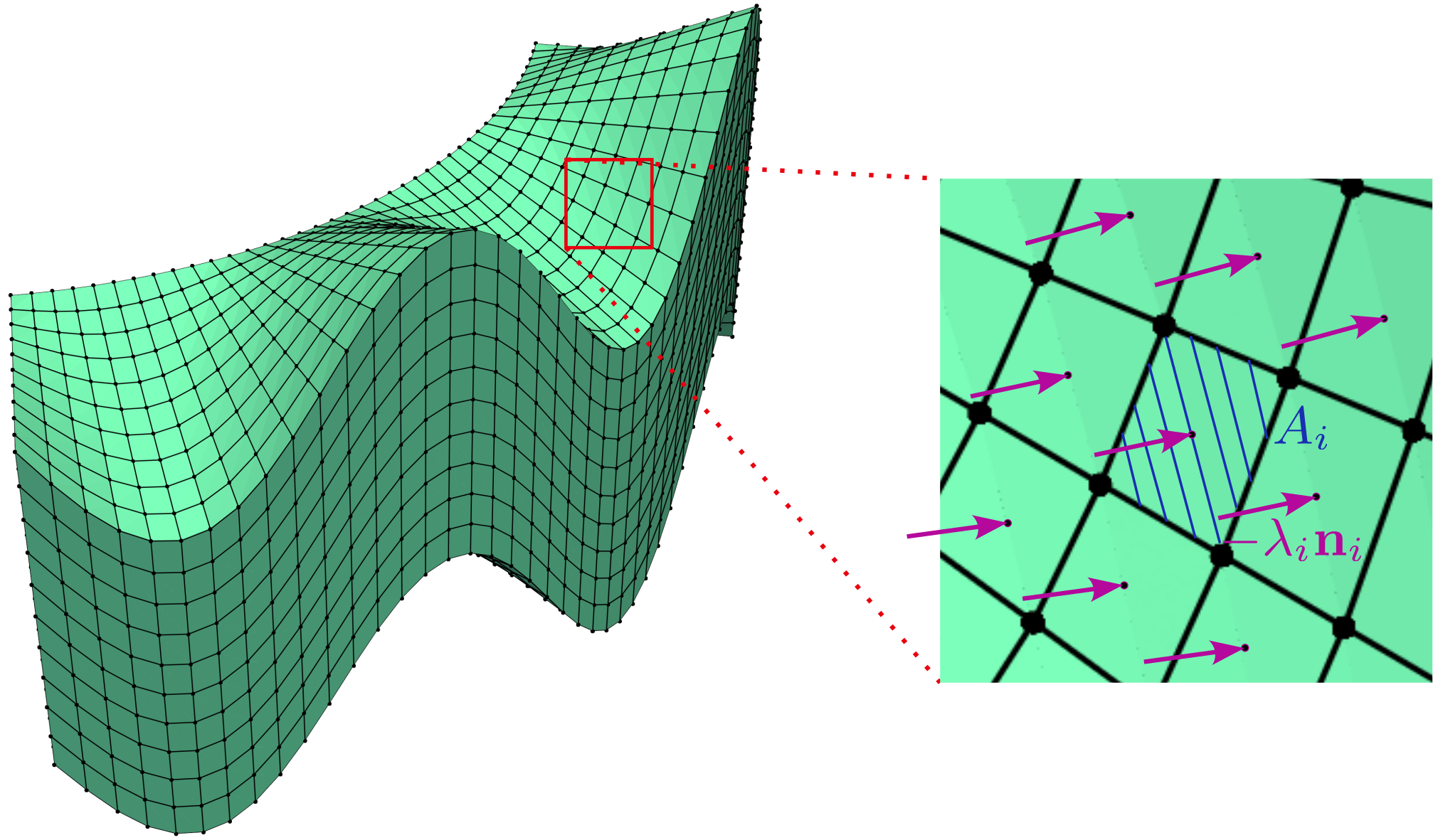}
    \caption{Contact reaction forces on the Sine Block, assumed to act at the centroids of the surface elements.}
    \label{fig:element-contact-forces}
\end{figure}

The surface meshes are constructed in accordance with the parametric mappings presented in the preceding section, ensuring a consistent evaluation of surface normals and gap functions. To simplify contact pairing and algorithmic implementation, we restrict the analysis to conforming meshes with matching nodes across potential contact interfaces. In addition, contact forces are assumed to act at the centroid of each surface element identified as being in contact, as illustrated in \Cref{fig:element-contact-forces}.
The contact pressure $p_c$ acting on a surface element $i$ is defined as the ratio between the contact force associated with that element and its corresponding area. In the present formulation, the contact force is represented by the Lagrange multiplier $\lambda_i$, leading to
\begin{equation}
   p_c^i = \frac{\lambda_i}{A_i},
\end{equation}
where $A_i$ denotes the area of the surface element $i$. This definition allows for a direct and physically consistent evaluation of contact pressures within the MBD framework. Moreover, it provides a common measure that facilitates a quantitative comparison between MBD and FEM results, which will be carried out in subsequent sections.

\subsection{FEM simulation set-up}

The finite element analyses presented in the following sections were conducted using the commercial finite element software Abaqus/CAE 2022.HF1.

Since the focus of this study is on the static response of the assemblies in different boundary value problems, quasi-static analyses are performed. The geometries of both the Sine Block and the hexagon-based block are non-convex and geometrically complex, leading to highly intricate contact interactions between individual blocks. Although the deformations are assumed to be small and are therefore modelled within the framework of linear strain theory, the blocks may undergo finite rotations. Owing to this combination of geometric complexity, i.e. large rotations and general contact conditions, the overall mechanical system exhibits a high level of nonlinearity. For this reason, an explicit dynamics FEM solver (Abaqus/EXPLICIT) is used to improve computational efficiency and to robustly handle the complex contact behaviour.

All bodies are modelled as isotropic, linear elastic materials, with concrete parameters as summarized in Table~\ref{table:material_parameters}. The contact between the blocks is defined as a soft contact using an exponential pressure--overclosure relationship. Following the Abaqus definitions, a clearance of $c_0 = 1 \cdot 10^{-3}$~mm (measured in the contact normal direction) and a contact pressure at zero clearance of $p_{c_0} = 1 \cdot 10^{-4}$~MPa are prescribed. Each block is meshed individually using 8-node hexahedral elements (see Figure~\ref{fig:sinus_mapping}), reflecting the treatment of each body as a separate entity. Hexahedral elements are selected due to their superior numerical performance and accuracy compared to tetrahedral elements.

Displacement boundary conditions are applied by fully constraining all nodes of the frame blocks, thereby fixing them in space (see Figure~\ref{fig:bvps_sinus}). Depending on the specific boundary value problem, a pressure load of $p_0 = \SI{0.01}{\mega\pascal}$ is applied to the blocks using the built-in \emph{smooth step} amplitude function. This function ensures a gradual load application and prevents abrupt accelerations that could generate stress waves and lead to noisy or inaccurate solutions.

To control dynamic effects further, both linear and quadratic volumetric strain-rate damping terms are included. Material damping is introduced to attenuate low-frequency (mass-dependent) as well as high-frequency (stiffness-dependent) responses, as summarized in Table~\ref{table:material_parameters}. No mass scaling is applied, ensuring that the physical dynamics of the system are accurately represented throughout the simulations. All Abaqus input files used in this study can be generated using the code available at~\cite{code}.

\renewcommand{\arraystretch}{1.2}
\begin{table}[h!]
  \centering
  \caption{FEM simulation parameters.}
  \begin{tabular}[h!]{c| c | c}
    \hline\hline
    \textbf{Parameter} &  \textbf{Value} & \textbf{Description} \\[5pt]
    \hline
    $\rho$ [\SI{}{\kilogram\per\cubic\milli\metre}] & $2400 \cdot 10^{-12}$ & Density
    \\
    \hline
    $E$ [\SI{}{\mega\pascal}] & $33 \cdot 10^{3}$ & Young's modulus
    \\
    \hline
    $\nu$ [\SI{}{-}]& $0.2$ & Poisson's ratio
    \\
    \hline
    $\alpha$ [\SI{}{-}] & $2.0$ & Mass proportional damping
    \\
    \hline
    $\beta$ [\SI{}{-}] & $1.0 \cdot 10^{-8}$ & Stiffness proportional damping
    \\
    \hline
    $\mu$ [\SI{}{-}] & $0.01$ & Friction coefficient
    \\ 
    \hline \hline
  \end{tabular}
  \label{table:material_parameters}
\end{table}
\renewcommand{\arraystretch}{1.0}

% ############################################################################   
\section{Evaluation criteria and measures}
% ############################################################################
A key component of the parameter space analysis is the formulation of appropriate criteria and quantitative measures to evaluate the MBD results. The following criteria and measures are introduced to characterize the mechanical performance of the assemblies:

\begin{enumerate}[leftmargin=*]
    \item \emph{Kinematic mechanism classification test:}
    For a fixed configuration, this test first determines whether the assembly admits a nonzero infinitesimal mechanism. Using the criterion in \eqref{eq:kinematic-feasibility-lp}, the configuration is classified as kinematically feasible if such a mechanism exists, and as kinematically \emph{locked} otherwise. For feasible configurations, we then evaluate \eqref{eq:kinematic-feasibility-mechanism} and \eqref{eq:kinematic-feasibility-mechanism-min} to assess whether the mechanism is admissible under the applied loads: $\gamma^{\star}>0$ corresponds to an \emph{activated} mechanism, $\gamma^{\star}=0$ to a \emph{neutral} mechanism, and $\gamma^{\star}<0$ to a \emph{suppressed} mechanism.
    
    \item \emph{Force equilibrium:}
    For \emph{locked} and \emph{suppressed} configurations, force equilibrium~\eqref{eq:SCP-linear} is evaluated. Equilibrium is considered satisfied when the relative force-balance residual is below a prescribed tolerance:
    \begin{equation*}
        \frac{\lVert \vb{G}\vb*{\uplambda} + \vb{f}_{\text{ext}} \rVert}
        {\lVert \vb{f}_{\text{ext}} \rVert} \le \mathrm{tol},
        \quad \text{with } \mathrm{tol} = 10^{-6}.
    \end{equation*}
    Here, $\vb*{\uplambda}$ denotes the vector of contact reactions, and $\vb{f}_{\text{ext}}$ denotes the generalized external load vector.
    
    \item \emph{Effective contact area:} 
    This metric quantifies the extent of the surface that effectively participates in contact interactions within the assembly (a larger value indicates a more uniformly distributed contact). 
    Let $\vb{p}_c = (p_c^{ij}) \in \mathbb{R}^{k_b \times N_b}$ denote the matrix of contact pressure values, where each column $j \in \{1, \dots, N_b\}$ corresponds to a block in the assembly, and each row $i \in \{1, \dots, k_b\}$ represents a surface element of the discretized block surface that is eligible for contact. 
    To quantify the extent of contact for each block within an assembly, we introduce the \emph{participation ratio}, which represents the effective number of contact surface elements actively engaged in contact. For a given block $j$, the participation ratio is defined as
    \begin{equation*}
        N_{\mathrm{eff}, j} \coloneqq
        \frac{\left( \sum_{i=1}^{k_b} p_c^{ij} \right)^{2}}
        {\sum_{i=1}^{k_b} (p_c^{ij})^{2}},
    \end{equation*}
    where $p_c^{ij} \geq 0$ denotes the contact pressure on the $i$-th contact surface element of block $j$, and $k_b$ is the total number of contact surface elements of the block. Intuitively, $N_{\mathrm{eff}, j}$ measures the number of elements of the contact surface that contribute significantly to the contact, providing a more informative metric than simply counting elements above a fixed threshold.
    To compute the corresponding effective contact area, the contributing surface elements must first be identified. This is achieved by sorting the contact pressures of block $j$ in descending order,
    \begin{equation*}
        p_c^{1j} \geq p_c^{2j} \geq \cdots \geq p_c^{k_b j},
    \end{equation*}
    where $k_b$ denotes the total number of surface elements considered for the block. Let $K_j = \left\lfloor N_{\mathrm{eff}, j} \right\rfloor$ be the integer part of the participation ratio. Using this value, we define a binary matrix $\vb{M} \in \{0,1\}^{k_b \times n_b}$, where $n_b$ is the total number of blocks in the assembly:
    \begin{equation*}
        M_{ij} \coloneqq 
        \begin{cases}
    1, & \text{if } p_c^{ij} \text{ is among the } K_j \text{ largest values in block } j, \\
    0, & \text{otherwise}.
    \end{cases}
    \end{equation*}
    The entries of $\vb{M}$ thus indicate which surface elements are part of the \emph{effective contact area} of each block according to the participation ratio. Summing the areas of the surface elements corresponding to $M_{ij}=1$ yields the effective contact area for block $j$, and collectively, these elements define the effective contact region of the entire assembly.
    
    \item \emph{Effective mean contact pressure:} 
    For each block $j$, the mean contact pressure over its effective contact area is defined as
    \begin{equation*}
        \bar{p}_{\mathrm{eff}, j} = \frac{\sum_{i} M_{ij} \, p_c^{ij}}{K_j},
    \end{equation*}
    where $K_j$ is the normalization factor corresponding to the effective contact area of block $j$ (as defined above).  
    The overall effective mean contact pressure of the assembly is then obtained as the arithmetic mean of all block-specific values:
    \begin{equation*}
        \bar{p}_{\mathrm{eff}} = \frac{1}{N_b} \sum_{j=1}^{N_b} \bar{p}_{\mathrm{eff}, j}.
    \end{equation*}
    This scalar quantity provides an aggregate measure characterizing the global contact behaviour of the assembly.

    \item \emph{Max–mean contact pressure:} 
    For each block $j \in \{1, \dots, m\}$, the maximum contact pressure is defined as
    \begin{equation*}
        p_{\mathrm{max}, j} = \max_{1 \leq i \leq n} p_{ij},
    \end{equation*}
    where $p_{ij}$ denotes the local contact pressure value at surface element $i$ of block $j$. 
    The overall average of these block-wise maxima is then given by
    \begin{equation*}
        \bar{p}_{\mathrm{max}} = \frac{1}{m} \sum_{j=1}^{m} p_{\mathrm{max}, j}.
    \end{equation*}
    The quantity $\bar{p}_{\mathrm{max}}$, referred to as the \emph{max–mean contact pressure}, provides a global indicator of the intensity and spatial concentration of contact pressures within the assembly.

\end{enumerate}

This set of quantitative measures enables a systematic assessment of mechanically robust assemblies. 

\section{Numerical mechanical analysis}\label{sec:analyse}

In this section, the mechanical performance of the assemblies is investigated under the prescribed boundary conditions. The equilibrium state is first evaluated using a multibody dynamics (MBD) framework to assess whether the assemblies can sustain the applied loads. Subsequently, a parametric analysis is conducted for a representative boundary value problem (BVP) -- the pipe configuration -- from which an appropriate set of parameters is selected for subsequent simulations based on finite element method (FEM).

% ############################################################################
\subsection{Sine Block}
% ############################################################################
%\paragraph{Parameter Space}
The Sine Block, $\TSB(h,a,f,s,n,r_i,t)$, as defined in \Cref{sec:SinusBlock}, is characterized by seven independent parameters. To construct a relatively slender tubular structure that is representative of a column or pillar, the inner radius $r_i$ and the block height $h$ are fixed at $r_i = 120\,\mathrm{mm}$ and $h = 200\,\mathrm{mm}$, respectively. The remaining five parameters define the parameter space, which is summarized in Table~\ref{tab:block-parameter-space}. A constant number of layers, $L = 12$, is selected, resulting in a total structure length (including the frame) of $2.6$~m.
\renewcommand{\arraystretch}{1.5}
\begin{table}[h!]
    \centering
    \begin{tabular}{|
        >{\centering\arraybackslash}m{3cm}|
        >{\centering\arraybackslash}m{2cm}|
        >{\centering\arraybackslash}m{2cm}|
        >{\centering\arraybackslash}m{3cm}|
        >{\centering\arraybackslash}m{3cm}|
    }
        \hline
        \makecell{\textbf{amplitude:}\\ $a$ [mm]} &
        \makecell{\textbf{shift:}\\ $s$ [-]} &
        \makecell{\textbf{frequency:}\\ $f$ [-]} &
        \makecell{\textbf{number of blocks} \\ \textbf{per layer:} $n$ [-]} &
        \makecell{\textbf{wall thickness:}\\ $t$ [mm]} \\
        \hline
        $\{10, 15, 20, 25\}$ & $\{0, 0.5\}$ & $\{1, 1.5, 2\}$ & $\{2,3,4,5,6\}$ & $\{30, 40, 50, 60\}$ \\ 
        \hline
    \end{tabular}
    \caption{Chosen parameter space defining the Sine Block configuration.}
    \label{tab:block-parameter-space}
\end{table}
\renewcommand{\arraystretch}{1.0}

\begin{figure}[h!]
    \centering
    \includegraphics[width=0.8\linewidth]{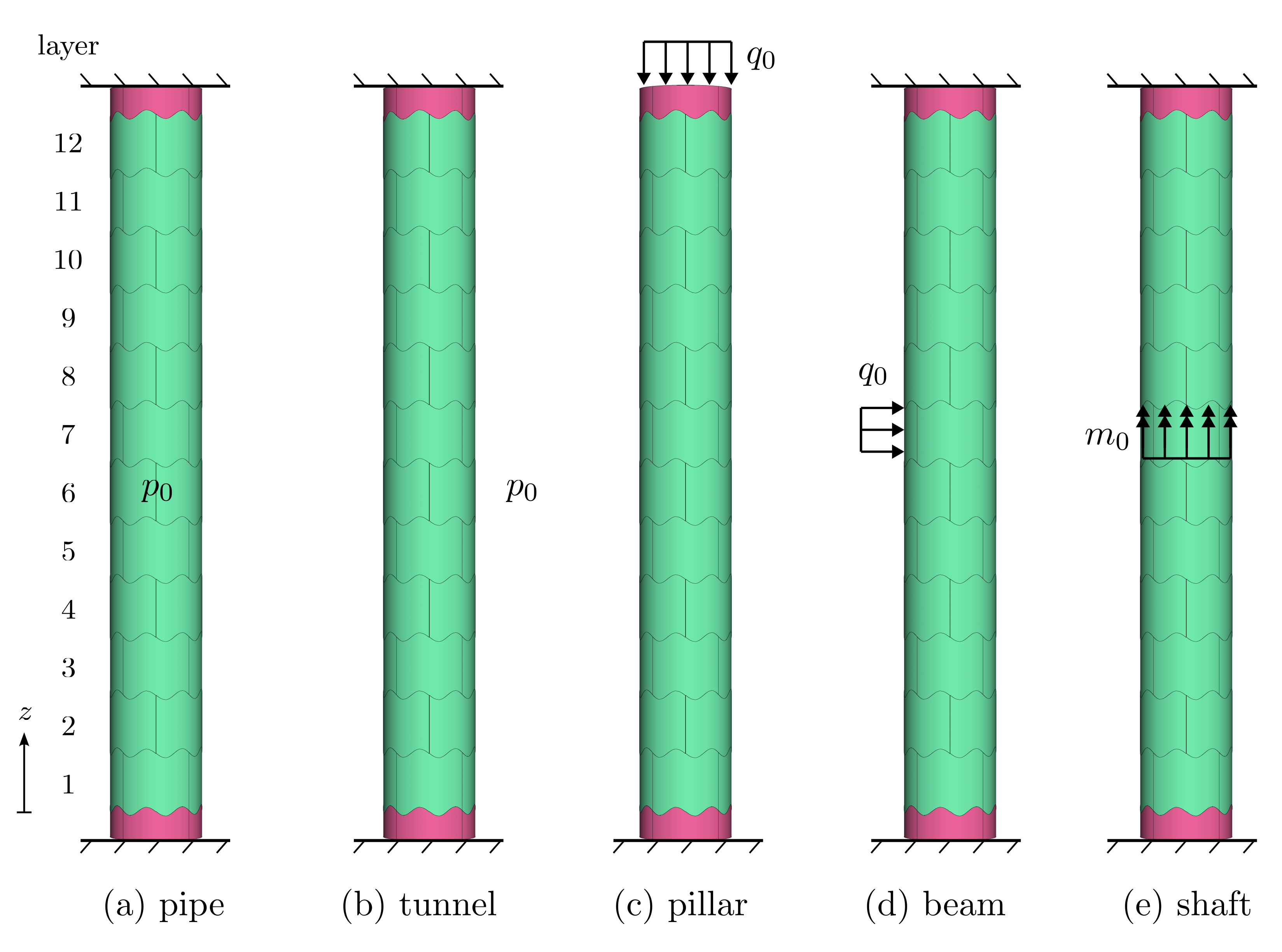}
    \caption{Five representative boundary value problems (BVPs) considered in this study of a 12-layer tubular interlocking: \textit{pipe}, \textit{tunnel}, \textit{pillar}, \textit{beam}, and \textit{shaft}. The configurations represent internal pressure, external pressure, axial compression, bending, and torsion, respectively. A constant load $p_0$ is applied either as surface pressure or tangential traction, depending on the BVP: inner pressure for the pipe, outer pressure for the tunnel, axial load on the top frame for the pillar, localized surface load on a block in the seventh layer for the beam, and tangential surface loads on all blocks in the seventh layer for the shaft, inducing a global moment.
}
    \label{fig:bvps_sinus}
\end{figure}

Five representative boundary value problems (BVPs) are investigated, as illustrated in \Cref{fig:bvps_sinus}: \textit{pipe}, \textit{tunnel}, \textit{pillar}, \textit{beam}, and \textit{shaft}. These configurations are chosen to systematically cover a broad range of loading scenarios for cylindrical systems.

In the \textit{pipe} BVP, a constant pressure $p_0$ is applied to the inner surface of the tubular structure, representing internal pressurization. Conversely, the \textit{tunnel} BVP is subjected to a constant external pressure $p_0$ acting on the outer surface, mimicking ground loading. The \textit{pillar} configuration corresponds to axial compression, where a constant surface load $p_0$ is applied to the top frame of the column. In the \textit{beam} BVP, bending-dominated behaviour is induced by applying a constant surface load $p_0$ to the outer surface of a single block located in the middle of the assembly. Since the assemblies here consist of 12 layers, we select the seventh layer. Finally, the \textit{shaft} problem represents torsional loading: all outer surfaces of the blocks in the seventh layer are subjected to a tangential surface load $p_0$, generating a global moment on the structure.
The selected BVPs therefore encompass internal and external pressure, axial compression, bending, and torsion, which are among the most common and mechanically relevant loading conditions encountered in structural mechanics of cylindrical systems.

\subsubsection{MBD simulations:}\label{subsec:MBD_results}

We begin by employing the multibody dynamics (MBD) approach as an efficient preliminary tool to assess whether a constructed block assembly, corresponding to an arbitrary set of block parameters drawn from the predefined parameter space (see Table~\ref{tab:block-parameter-space}), can achieve kinematic immobilization for all of the boundary value problems (BVPs) introduced earlier.
We first determine whether assemblies with the block parameters defined in \Cref{tab:block-parameter-space} admit a nonzero kinematic mechanism. The outcomes of this initial evaluation are summarized in Table~\ref{tab:kinematic-mechanism} where we distinguish only between $s=0$ and $s=0.5$, as these parameters influence the block’s geometry beyond scaling.
The results show that the Sine Block configuration $\TSB(h,a,f,s,n,r_i,t)$ has no kinematic mechanism for all $a>0$. This indicates that the proposed block can withstand all possible loading scenarios.
\begin{table}[h!]
\centering
\begin{tabular}{|c|>{\centering\arraybackslash}p{3cm}|>{\centering\arraybackslash}p{3cm}|}
    \hline
     &  $\boldsymbol{s=0}$ &  $\boldsymbol{s=0.5}$  \\ 
    \hline
    \textbf{Kinematic mechanism} &  locked &  locked \\ 
    \hline
\end{tabular}
\caption{Classification of the kinematic mechanism for block configurations with shift $s=0$ and $s=0.5$.}
\label{tab:kinematic-mechanism}
\end{table}

\begin{figure}[h!]
    \centering
    \includegraphics[height=0.8\textheight]{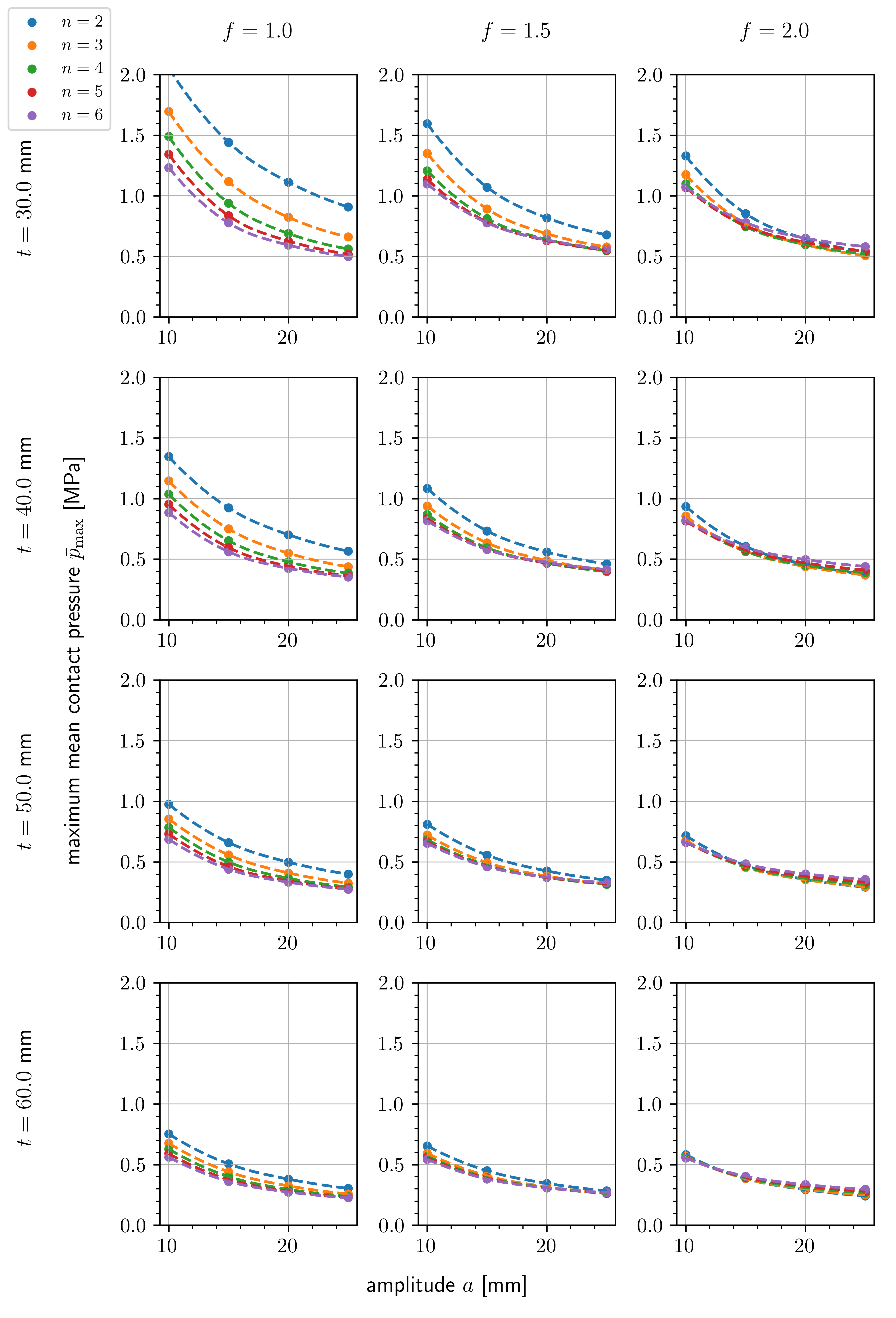}
    \caption{Parameter study of the maximum mean contact pressure $\bar{p}_{\mathrm{max}}$ for tubular \emph{Sine} $\TSB(200,a,f,s=0,n,120,t)$ block assemblies in the representative \emph{pipe} boundary value problem. The graphs illustrate the evolution of the max--mean contact pressure in the assembly over the investigated block parameter space (see Table~\ref{tab:block-parameter-space}), obtained using the multibody dynamics approach. This quantity characterizes the intensity and localization of contact forces within the assembly and serves as an indicator of potential stress concentrations. Results enable a comparative assessment of parameter configurations.}
    \label{fig:mbd_anal_sinus}
\end{figure}

\begin{figure}[h!]
    \centering
    \includegraphics[height=0.8\textheight]{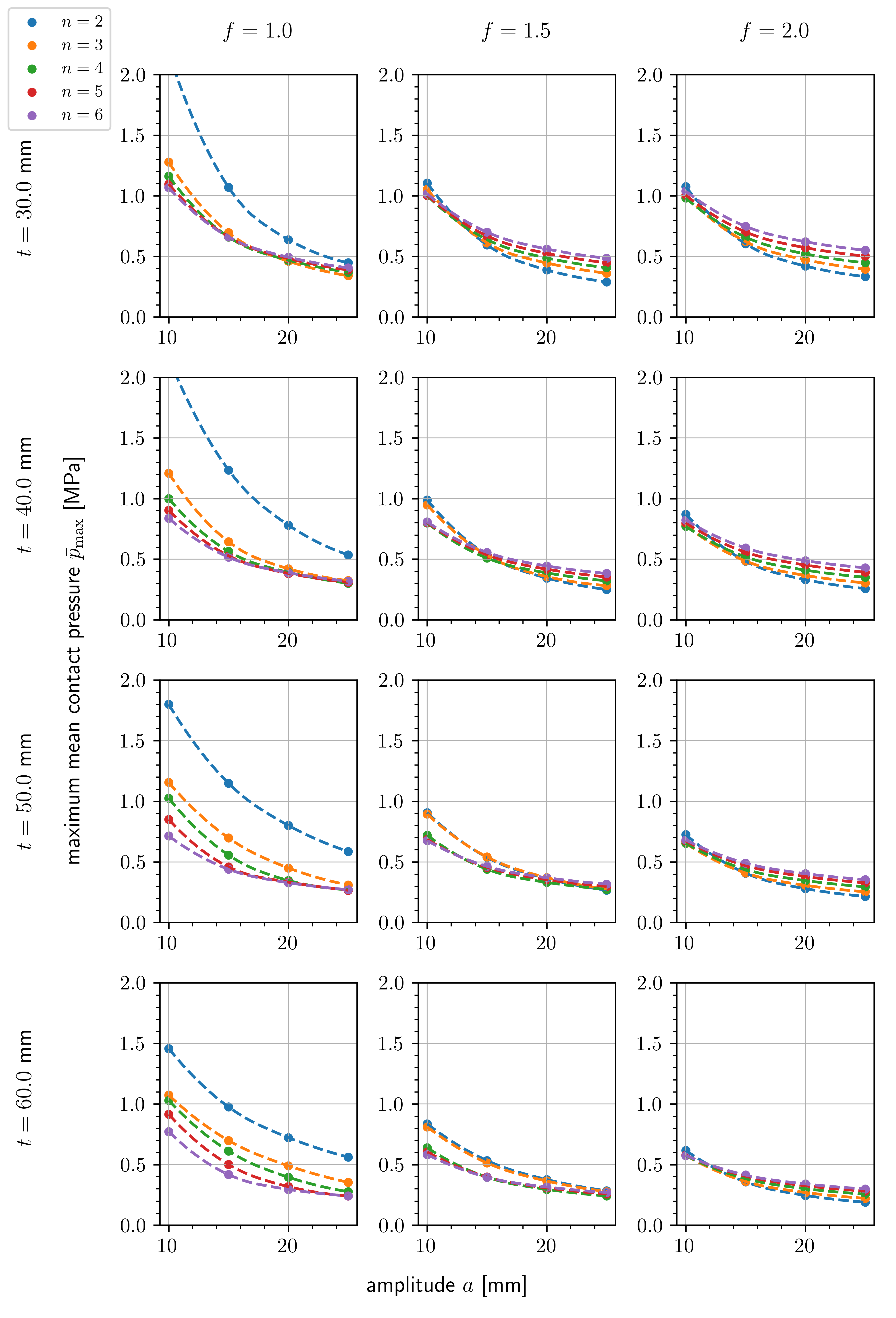}
    \caption{Parameter study of the maximum mean contact pressure $\bar{p}_{\mathrm{max}}$ for tubular \emph{Cosine} $\TSB(200,a,f,s=0.5,n,120,t)$ block assemblies in the representative \emph{pipe} boundary value problem. The graphs illustrate the evolution of the max--mean contact pressure in the assembly over the investigated block parameter space (see Table~\ref{tab:block-parameter-space}), obtained using the multibody dynamics approach. This quantity characterizes the intensity and localization of contact forces within the assembly and serves as an indicator of potential stress concentrations. Results enable a comparative assessment of parameter configurations.}
    \label{fig:mbd_anal_cosinus}
\end{figure}

Building on these findings, we conduct an extensive parameter study to explore how variations in the block parameters affect the interactions between the blocks in the assembled structure. Such a comprehensive exploration of the parameter space is feasible only with the MBD approach, as it is significantly less computationally demanding than the finite element methods (FEM). To ensure a focused and interpretable analysis, this parameter study is performed for a single representative boundary value problem, namely the \emph{pipe} configuration as this BVP can be interpreted as an equivalent mechanical condition for kinematical explosion.
Within this study, we examine the evolution of the max-mean contact pressure across the parameter space. This quantity serves as an effective indicator of the intensity and localization of contact forces within the assembly. From a mechanical design perspective, excessive localization of contact pressure is undesirable, as it may lead to stress concentrations, material damage, or premature failure. Consequently, our objective is to identify parameter configurations that promote a more uniform distribution of contact pressure.
The results of this parameter investigation are presented in Figures~\ref{fig:mbd_anal_sinus} and \ref{fig:mbd_anal_cosinus} for the $\TSB(h,a,f,s=0,n,r_i,t)$ and $\TSB(h,a,f,s=0.5,n,r_i,t)$ block geometries, corresponding to the Sine and Cosine Blocks, respectively. Based on these numerical investigations, several key insights into the mechanical behaviour of the proposed block assemblies are discussed in the following.

The amplitude $a$ of the sine curve has an influence on the maximum averaged contact pressure $\bar p_{\max}$ for all investigated values of number of blocks $n$, frequency $f$, and thickness $t$. For all configurations, $\bar p_{\max}$ decreases monotonically with increasing amplitude. The reduction is most pronounced for small amplitudes, whereas the curves progressively flatten for larger values of $a$, indicating a diminishing sensitivity to further amplitude increases. For sufficiently large amplitudes, the pressure levels obtained for different combinations of $n$ and $f$ approach a common trend, suggesting the existence of an amplitude-driven saturation regime. Within the investigated parameter range, this regime is reached at approximately $a \approx 30$~mm, beyond which further increases in amplitude lead to only marginal changes in the contact pressure response.

The block thickness $t$ strongly affects the magnitude and convergence behaviour of the maximum averaged contact pressure $\bar p_{\max}$. For all values of $a$, $n$, and $f$, increasing the thickness results in a systematic reduction of $\bar p_{\max}$. In addition, thicker blocks exhibit a faster convergence with respect to both the number $n$ of blocks and the frequency $f$. For large thicknesses, the dependence of $\bar p_{\max}$ on the remaining parameters becomes weaker.

The results further reveal a clear interdependence between the number of blocks $n$, the frequency $f$, and the amplitude $a$. For a fixed block thickness $t$, an increase in frequency combined with a higher partitioning of the layer (larger $n$) and larger amplitudes generally leads to increased maximal average contact pressures. This trend is particularly pronounced for configurations such as $\TSB(h,a=25,f,s,n=6,r_i,t)$, where the combined effect of high frequency and fine partitioning results in elevated pressure levels.

These observations indicate that the total number of sine waves within the layer,
\begin{equation*}
    n_w = 2 f n,
\end{equation*}
plays a decisive role in the contact pressure response. Specifically, there appears to be a threshold value of $n_w$ above which amplitudes of approximately $10$~mm or greater lead to a deterioration in performance, manifested by higher maximal average contact pressures. This suggests that excessively fine waviness combined with large amplitudes may induce unfavorable local stress concentrations.

Moreover, when comparing configurations with the same total number of sine waves $n_w$, systems with a smaller number $n$ of blocks consistently exhibit lower contact pressures. For example, the configuration $\TSB(h,a,f=2,s=0,n=2,r_i,t)$ yields lower pressure values than $\TSB(h,a,f=1,s=0,n=4,r_i,t)$, despite both have $n_w=8$. This indicates that fewer, larger blocks distribute contact forces more effectively than a larger number of smaller blocks for the same overall waviness, highlighting the importance of block partitioning as an independent design parameter.

\begin{figure}[H]
    \centering
    \begin{subfigure}{0.48\textwidth}
        \centering
        \includegraphics[width=0.5\linewidth]{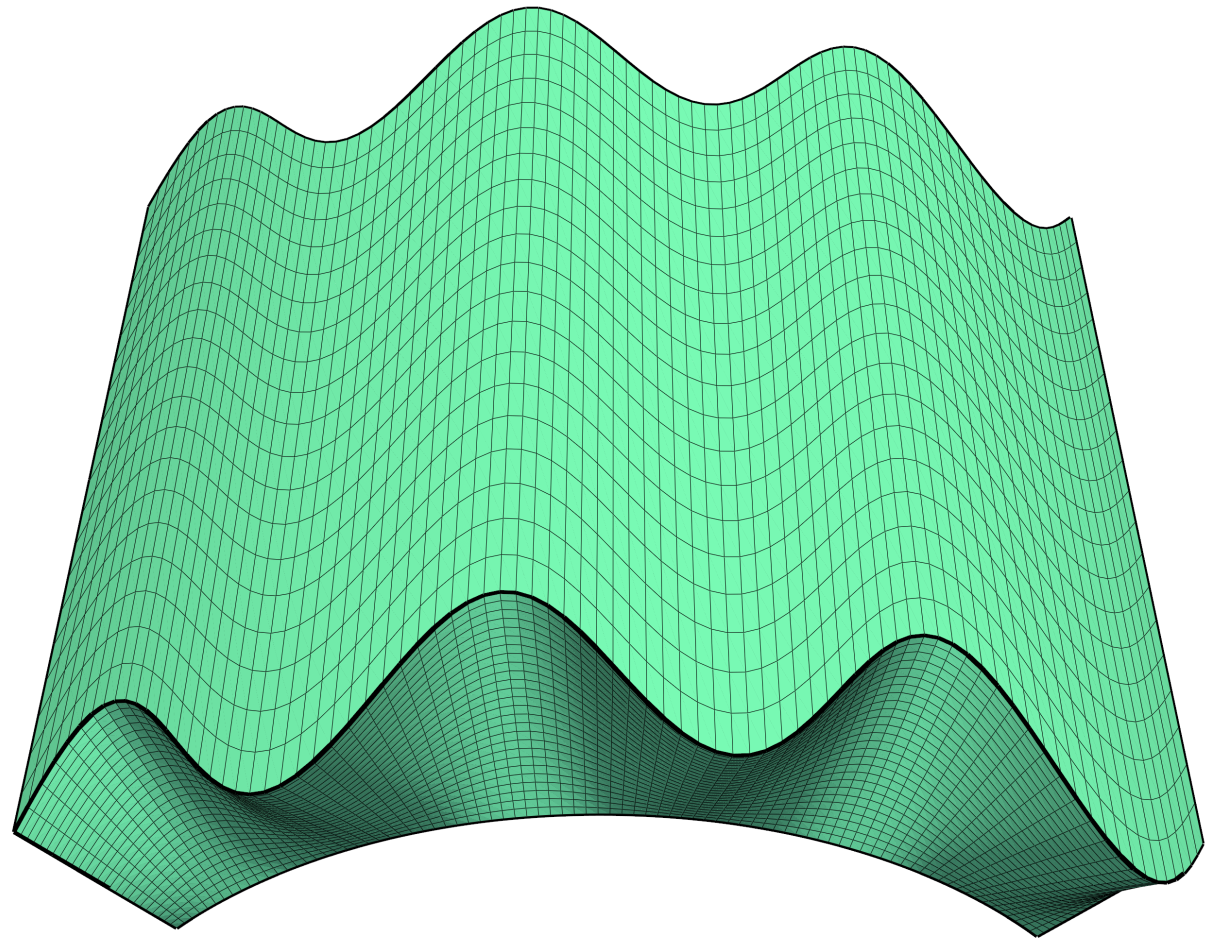}
        \caption{$\mathrm{TSB}\, {1}$}
        \label{subfig:TSB1}
    \end{subfigure}
    \begin{subfigure}{0.48\textwidth}
        \centering
        \includegraphics[width=0.5\linewidth]{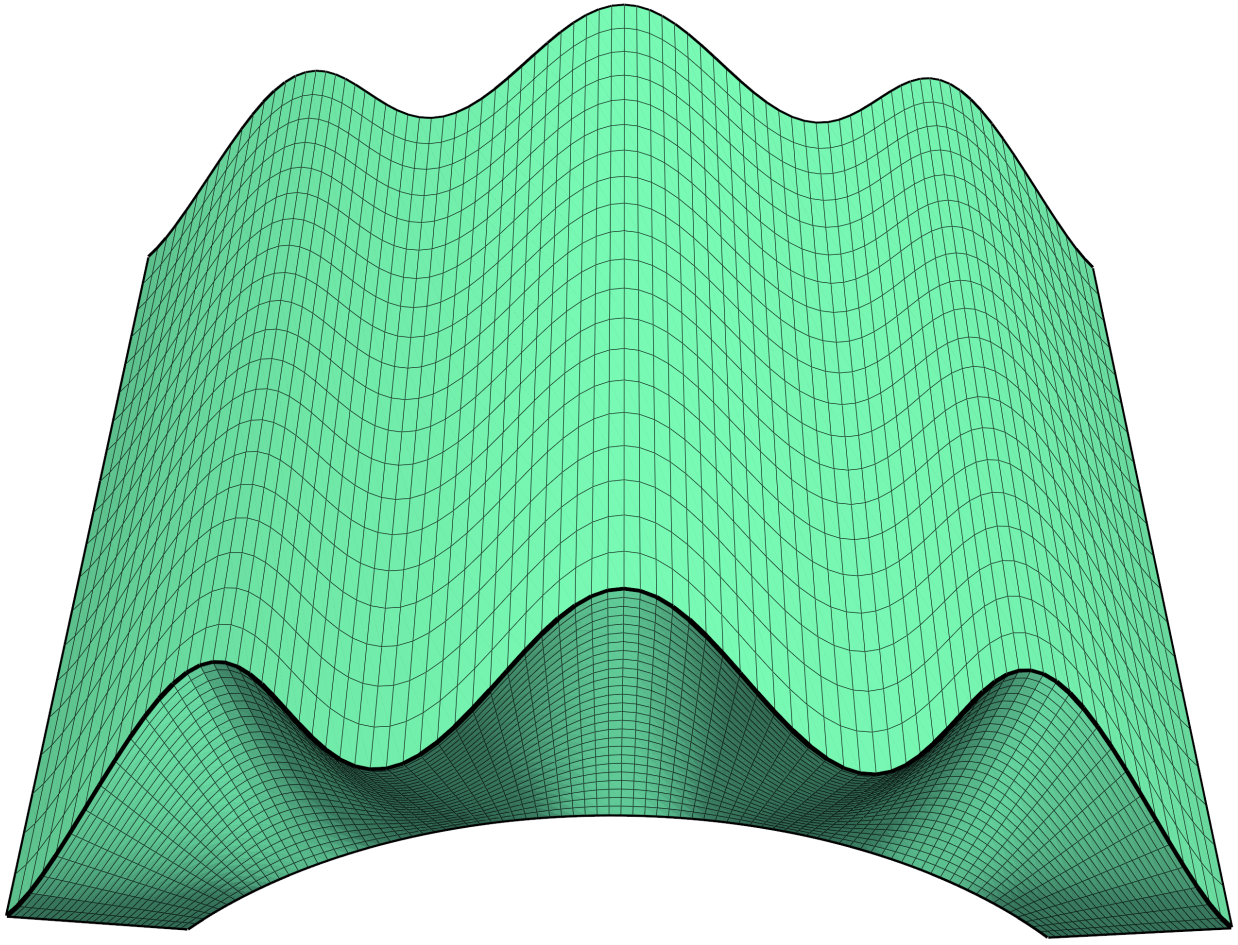}
        \caption{$\mathrm{TSB}\, {2}$}
        \label{subfig:TSB2}
    \end{subfigure}
    \caption{Block configurations selected for analysis.}
\end{figure}

Based on the results presented above, the configuration 
\[
\TSB(h,a=25,f=2,s=0.5,n=2,r_i,t=60)
\]
provides the best performance in terms of maximal average contact pressure. However, the design requirements also include modularity, low material usage, and the mitigation of stress peaks. To accommodate these considerations, we selected an assembly with higher modularity, leading to an increased number of blocks $n=4$.
Furthermore, guided by the previous insights, we avoided the highest frequency for higher layer partitioning, and therefore set the frequency to $f=1.5$. The configuration with $s=0.5$ exhibits slightly lower maximal average contact pressures compared to the $s=0$ configuration, suggesting that $s=0.5$ is the preferred setting for minimizing contact stresses. Nevertheless, we include both $s=0.5$ and $s=0$ in our further analysis to enable a direct comparison of the resulting stress distributions with finite element simulations. This decision is justified because, although the $s=0$ configuration shows higher average maximum contact pressures, it may still exhibit advantageous stress distribution characteristics within the block that are relevant for structural performance assessment.
Regarding block thickness, the differences in contact pressure among the configurations 
\[
\TSB(h,a=20,f=1.5,s=0.5,n=4,r_i,t=\{40,50,60\})
\] 
are minor. Consequently, the smallest thickness $t=40$~mm was chosen to reduce material consumption without significantly affecting performance.
Following these design considerations, we finalise two block configurations
\begin{align*}
    \mathrm{TSB}\, {1} \coloneqq &\TSB(h=200,a=20,f=1.5,s=0,n=4,r_i=120,t=40)  
    \quad \text{(see \Cref{subfig:TSB1})} %\label{eq:sin-block-1}
    \\
    \mathrm{TSB}\, {2} \coloneqq &\TSB(h=200,a=20,f=1.5,s=0.5,n=4,r_i=120,t=40) 
    \quad \text{(see \Cref{subfig:TSB2})}, %\label{eq:sin-block-2}
\end{align*}
which will be further analysed in greater detail.

For the selected block configurations $\mathrm{TSB}\, {1}$ and $\mathrm{TSB}\, {2}$, the four evaluation measures introduced above -- namely the kinematic mechanism, the effective contact area, the effective mean contact pressure, and the maximal mean contact pressure -- are computed for all five boundary value problems (BVPs). The corresponding results are summarized in Tables~\ref{tab:sin-force-equilibrium} and~\ref{tab:cosin-force-equilibrium}.
Together, these measures establish a consistent framework for systematically comparing the mechanical performance of the considered assemblies under different loading scenarios.

From a mechanical perspective, a larger effective contact area is advantageous, as it promotes a more distributed transmission of forces and thereby reduces the contact pressure levels. Correspondingly, a lower effective mean contact pressure indicates a more favourable and uniform stress state within the activated contact zones. In contrast, the maximal mean contact pressure serves as a global indicator of the intensity and spatial concentration of contact forces, highlighting potential local stress accumulations within the assembly.

A comparison of the results for $\mathrm{TSB}\,1$ and $\mathrm{TSB}\,2$ in the pipe problem illustrates the relevance of these metrics. Configuration $\mathrm{TSB}\,2$ activates a larger portion of the available contact surface. Although both block configurations exhibit a comparable effective mean contact pressure, $\mathrm{TSB}\,2$ attains a lower maximal mean contact pressure. This indicates a reduced tendency for localized stress concentrations and therefore identifies $\mathrm{TSB}\,2$ as the mechanically preferable configuration.
\begin{table}[h!]
\centering
\begin{subtable}{1.0\textwidth}
\centering
\begin{tabular}{|c|>{\centering\arraybackslash}m{3cm}|>{\centering\arraybackslash}m{3cm}|>{\centering\arraybackslash}m{3cm}|>{\centering\arraybackslash}m{3cm}|}
    \hline
    \textbf{BVP} & \textbf{Kinematic mechanism} & \textbf{Effective contact area} [\%] & $\bar{p}_{\mathrm{eff}}$ [MPa] & $\bar{p}_{\mathrm{max}}$ [MPa] \\ 
    \hline
    pipe & locked & $28.3$ & $0.189$ & $0.467$\\ 
    \hline
    tunnel & locked & $46.8$ & $0.029$ & $0.050$\\
    \hline
    pillar & locked & $73.6$ & $0.011$ & $0.019$\\
    \hline
    beam & locked & $36.5$ & $0.071$ & $0.170$\\
    \hline
    shaft & locked & $38.4$ & $0.112$ & $0.214$\\
    \hline
\end{tabular}
\caption{$\TSB \, 1$}
\label{tab:sin-force-equilibrium}
\end{subtable}

\par\medskip

\begin{subtable}{1.0\textwidth}
\centering
\begin{tabular}{|c|>{\centering\arraybackslash}m{3cm}|>{\centering\arraybackslash}m{3cm}|>{\centering\arraybackslash}m{3cm}|>{\centering\arraybackslash}m{3cm}|}
    \hline
    \textbf{BVP} & \textbf{Kinematic mechanism} & \textbf{Effective contact area} [\%] & $\bar{p}_{\mathrm{eff}}$ [\SI{}{\mega\pascal}]  & $\bar{p}_{\mathrm{max}}$ [\SI{}{\mega\pascal}] \\ 
    \hline
    pipe & locked & $32.0$ & $0.191$ & $0.387$\\ 
    \hline
    tunnel & locked & $49.1$ & $0.030$ & $0.049$\\
    \hline
    pillar & locked & $72.1$ & $0.011$ & $0.019$\\
    \hline
    beam & locked & $35.0$ & $0.071$ & $0.162$\\
    \hline
    shaft & locked & $38.2$ & $0.117$ & $0.230$\\
    \hline
\end{tabular}
\caption{$\TSB \, 2$}
\label{tab:cosin-force-equilibrium}
\end{subtable}
\caption{Overview of the kinematic mechanism, effective contact area, effective mean contact pressure, and maximal mean contact pressure for the $\mathrm{TSB}\, 1$ and $\TSB \, 2$ assemblies across all five boundary value problems (BVPs). The metrics indicate whether force equilibrium is achieved, quantify the extent of the effective contact area within the assembly, provide the mean contact pressure restricted to this area, and report the average of the maximum contact pressure per block.}
\end{table}

\begin{figure}[h!]
    \centering
    \includegraphics[scale=1.0]{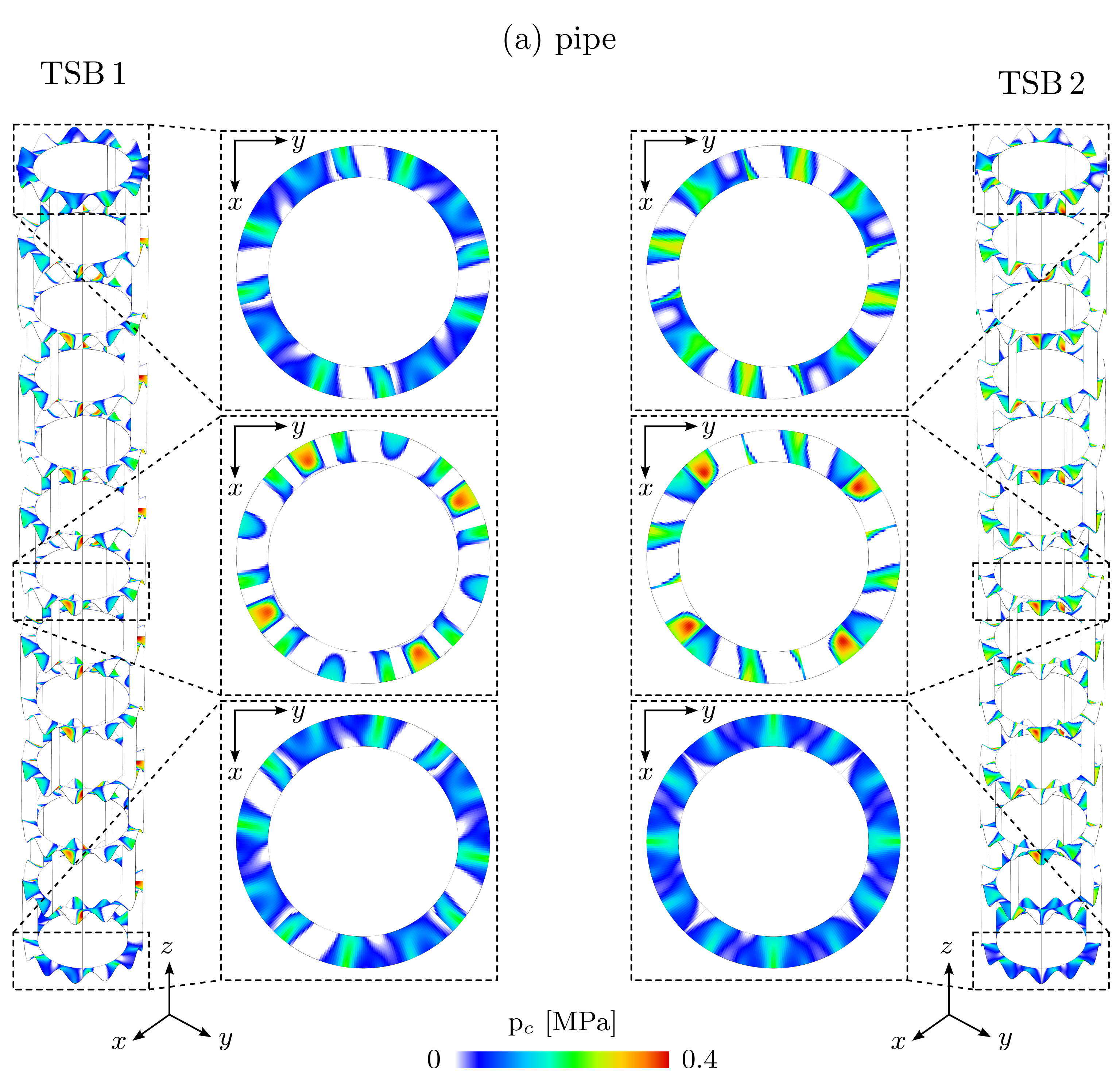}
    \caption{Contact pressure distributions obtained using the MBD approach for the \emph{pipe} BVP, illustrating the transfer of the externally applied load through the assembled block system to the bounding frame. The $\TSB\,1$ assembly is shown on the left, while the $\TSB\,2$ assembly is shown on the right. Only block surfaces that can come into contact with adjacent blocks or with the frame are displayed. The top and bottom rows present cross-sectional views highlighting the load transfer from the assembly to the bounding frame, whereas the middle row shows the internal load transfer between blocks in the sixth and seventh layers of the assembly.}
    \label{fig:pipe_MBD_pc}
\end{figure}

\begin{figure}[h!]
    \centering
    \includegraphics[scale=1.0]{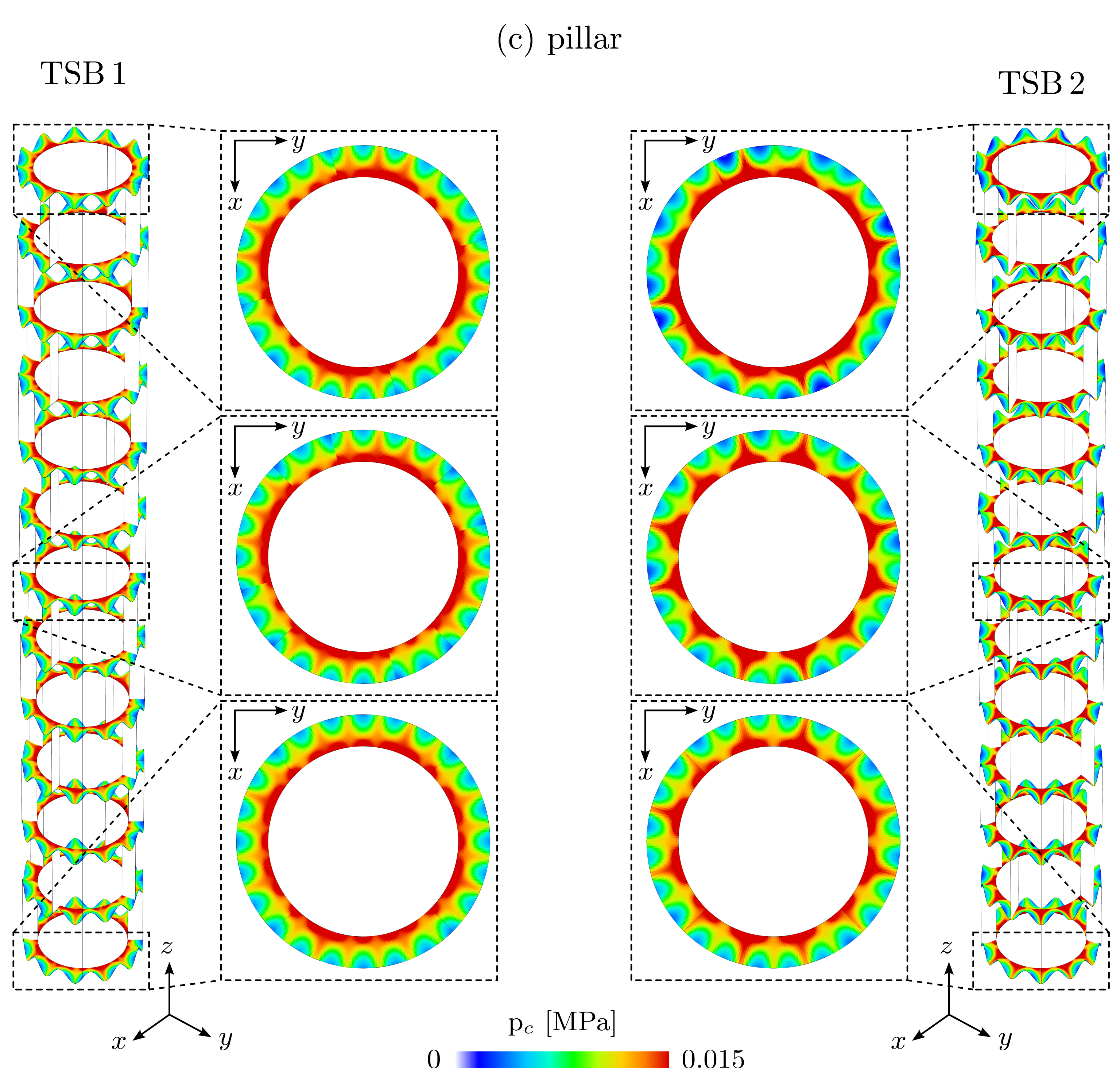}
    \caption{Contact pressure distributions obtained using the MBD approach for the \emph{pillar} BVP, illustrating the transfer of the externally applied load through the assembled block system to the bounding frame. The $\TSB\,1$ assembly is shown on the left, while the $\TSB\,2$ assembly is shown on the right. Only block surfaces that can come into contact with adjacent blocks or with the frame are displayed. The top and bottom rows present cross-sectional views highlighting the load transfer from the assembly to the bounding frame, whereas the middle row shows the internal load transfer between blocks in the sixth and seventh layers of the assembly.}
    \label{fig:pillar_MBD_pc}
\end{figure}

One of the central questions in the analysis of TIAs is how an externally applied load is transferred through the block system to the surrounding bounding frame. This load transfer mechanism can be investigated by examining the contact pressure distribution within the assembly, as obtained from the force equilibrium computation using the multibody dynamics (MBD) approach. Figures~\ref{fig:pipe_MBD_pc} and~\ref{fig:pillar_MBD_pc} present the resulting contact pressure distributions for the \emph{pipe} and \emph{pillar} configurations, respectively.
%
%The contact pressure distributions obtained using the MBD approach provide direct insight into the transfer of the externally applied load in the BVP configuration through the assembled block system to the bounding frame. 
In these two cases this load transfer mechanism is compared for two different block configurations, with the $\TSB\,1$ assembly shown on the left and the $\TSB\,2$ assembly shown on the right. 
The illustrated contact pressure distributions are evaluated over the entire assembly and are restricted to those block surfaces that are geometrically capable of coming into contact with adjacent blocks or with the bounding frame. This selective visualization ensures that only mechanically relevant interactions are depicted, thereby improving the interpretability of the load transmission paths within the structure.
A more detailed view of the load transfer mechanisms is provided by the cross-sectional representations. The top and bottom rows highlight the transmission of forces from the outermost blocks to the bounding frame, revealing how the global load is ultimately supported by the frame. In contrast, the middle row focuses on the internal force transfer between neighbouring blocks, specifically between the sixth and seventh layers of the assembly, thereby elucidating the interlocking behaviour and internal load redistribution within the block system.
%
% From the presented contact pressure distributions, the interaction between individual blocks becomes evident, as well as the specific regions of the block surfaces that are responsible for transferring forces across the assembly. The top and bottom cross-sections of the contact pressure fields in Figures~\ref{fig:pipe_MBD_pc} and~\ref{fig:pillar_MBD_pc} directly illustrate how the load is ultimately transmitted to the bounding frame. 
Moreover, these distributions highlight localized regions of elevated contact pressure that may indicate potential weak points or failure-prone areas on the block surfaces.

These two representative cases are shown primarily to illustrate the capability of the MBD approach to provide a first-order evaluation of the interlocking mechanism. In particular, the results demonstrate that MBD yields physically meaningful contact pressure patterns that capture the essential force transmission pathways within the assembly, which will be further validated by simulation based on the finite element method (FEM) in the subsequent section.
%
% \begin{figure}[h!]
%     \centering
%     \includegraphics[scale=1.0]{figures/tunnel_MBD_pc.png}
%     \caption{\color{red}Contact pressure distributions of \emph{tunnel} configuration.}
%     \label{fig:tunnel_MBD_pc}
% \end{figure}
%

Furthermore, the contact pressure patterns observed for the \emph{pipe} and \emph{pillar} configurations in Figures~\ref{fig:pipe_MBD_pc} and ~\ref{fig:pillar_MBD_pc}  are consistent with, and further support, the quantitative force equilibrium results summarized in Tables~\ref{tab:sin-force-equilibrium} and~\ref{tab:cosin-force-equilibrium}.

% #############################################################################################
%                                  CONTACT FORCES
% #############################################################################################
\subsubsection{Comparison of MBD and FEM Contact Pressure Distributions}

\begin{figure}[h!]
    \centering
    \includegraphics[width=0.8\textwidth]{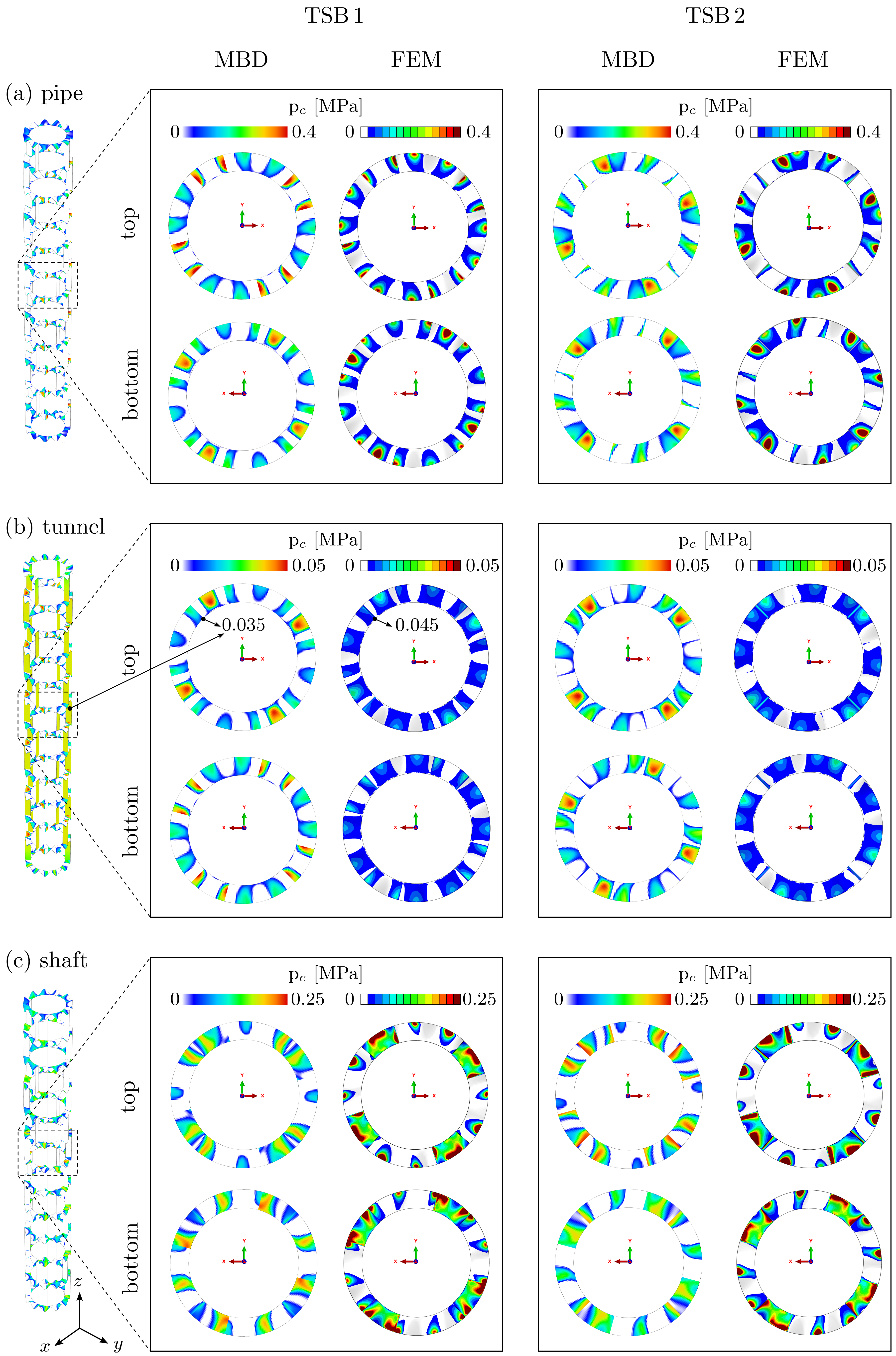}
    \caption{Comparison of contact pressure distributions obtained using the MBD and FEM approaches for layer~7 in the pipe, tunnel and shaft configurations. Results are shown for both $\TSB\,1$ ($s=0$) and $\TSB\,2$ ($s=0.5$) blocks, focusing on the top and bottom surfaces. The MBD method qualitatively reproduces the FEM contact patterns, including effective contact areas, symmetry properties, and locations of peak contact pressures, demonstrating its suitability for rapid preliminary assessment of block assemblies.}
    \label{fig:pc_MBD_FEM}
\end{figure}

We begin by comparing the contact pressure distributions between blocks determined using the multibody dynamics (MBD) approach and the finite element method (FEM). The primary objective of this comparison is to assess the accuracy of the simplified MBD simulations and to evaluate whether they can serve as a reliable and efficient substitute for FEM when fast, preliminary assessments of assemblies are required. The results of this comparison are presented in \Cref{fig:pc_MBD_FEM}.

The analysis is carried out for three boundary value problems (BVPs), namely the pipe, tunnel, and shaft configurations. In all cases, the comparison is restricted to a single representative layer, specifically layer~7. This choice is justified by the characteristics of the individual problems. For the pipe and tunnel configurations, the contact pressure distribution between blocks is periodic and repeats identically across layers; therefore, analyzing additional layers would not provide further insight. In contrast, for the shaft configuration, layer~7 corresponds to the location where the external loading is applied and where the highest contact forces are observed, making it the most critical layer for investigation.

The comparison between the MBD and FEM results is conducted on a qualitative basis. In total, 12 individual evaluations are performed. For each case, the contact pressure is examined on both the top and bottom surfaces of the selected layer, where the sinusoidal interfaces are located. These regions are of particular importance because they govern the interlocking mechanism of the blocks and are responsible for transmitting the majority of the contact forces, which tend to localise in these areas.

Furthermore, the influence of block geometry is investigated by comparing $\TSB\,1$ ($s = 0$) and $\TSB\,2$ ($s = 0.5$). The obtained contact pressure distributions corroborate the findings summarised in Tables~\ref{tab:sin-force-equilibrium} and~\ref{tab:cosin-force-equilibrium}. Specifically, for both $\TSB\,1$ and $\TSB\,2$, approximately 30\% of the surface area in the pipe configuration and about 50\% in the tunnel configuration constitute the effective contact area actively contributing to load transfer.

Overall, the MBD results for all six assemblies -- corresponding to the three BVPs and the two block geometries -- show very good agreement with the FEM simulations, as illustrated in \Cref{fig:pc_MBD_FEM}. The MBD framework reliably identifies the regions responsible for load transfer, and the locations of the contact pressure peaks predicted by FEM coincide closely with those obtained from the MBD simulations. Although discrepancies in the absolute magnitude of the contact pressures are observed, the effective contact areas and the distribution of peak pressures are captured consistently across all boundary value problems. The remaining differences arise from the fundamentally different mechanical and numerical assumptions of the two approaches: FEM employs a soft-contact formulation that allows controlled overclosure to ensure numerical stability, which can reduce peak pressures and slightly alter their distribution, and it accounts for material deformation and stress–strain behaviour that are not represented in rigid-body MBD analyses. In addition, the small initial gap between adjacent blocks introduced in the FEM model for stabilization purposes further influences the global mechanical response. This effect is particularly evident in the tunnel example (see \Cref{fig:pc_MBD_FEM}), where the rotational freedom of the blocks leads to increased stresses along the flat contact surfaces in the FEM simulation compared to the MBD results. Together, these aspects explain the observed discrepancies and their impact on the overall mechanical response of the system.

% #############################################################################################
%                                  STRESSES
% #############################################################################################
\subsubsection{Distribution of stresses}
\begin{figure}[h!]
    \centering
    \includegraphics[width=0.9\textwidth]{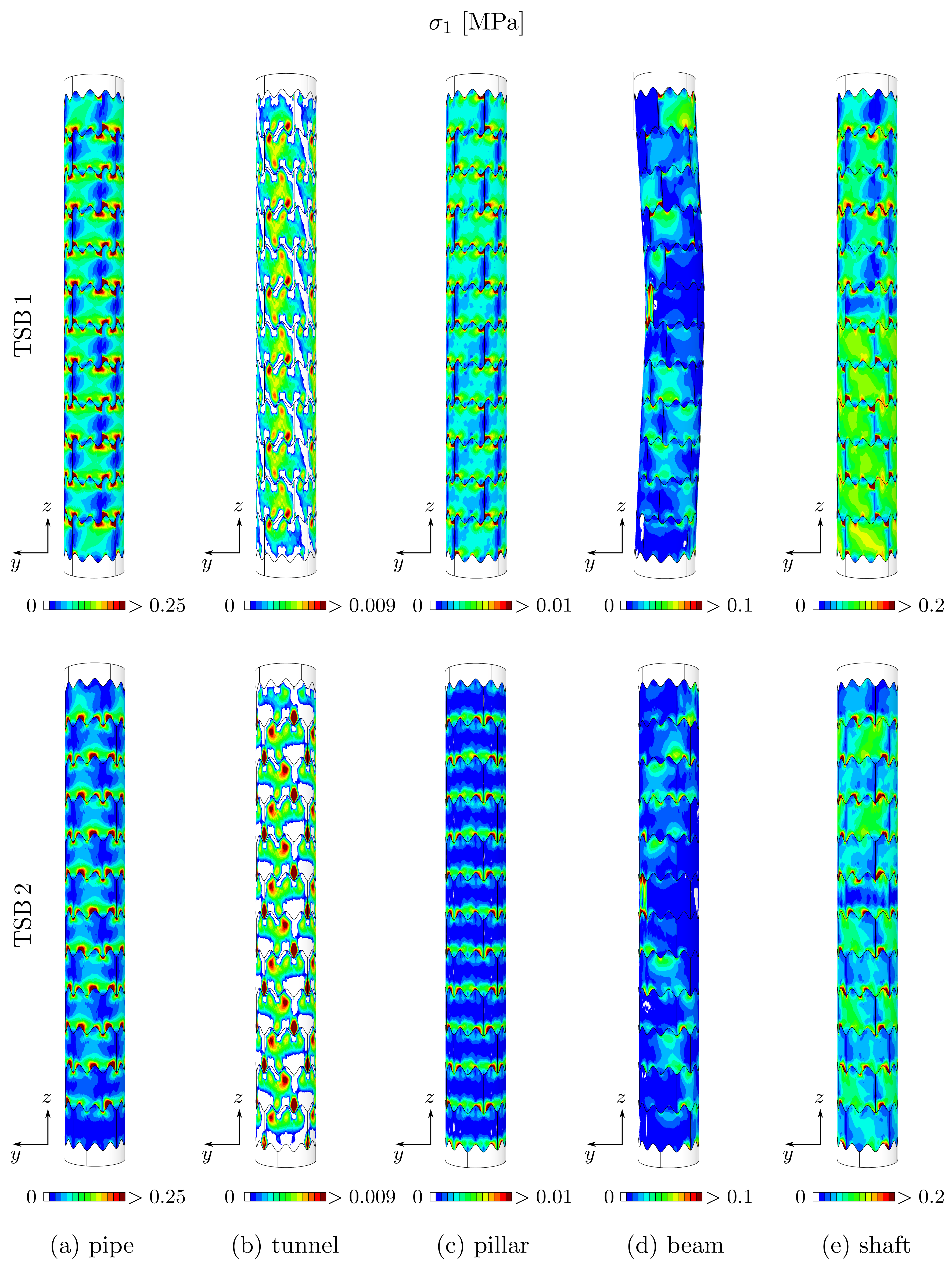}
    \caption{Distributions of the maximum principal stress $\sigma_1$ for the five BVP configurations: pipe, tunnel, pillar, beam, and shaft. The upper row shows assemblies composed of $\TSB\,1$ blocks, while the lower row presents the corresponding assemblies composed of $\TSB\,2$ blocks.}
    \label{fig:sigma1}
\end{figure}

\begin{figure}[h!]
    \centering
    \includegraphics[width=0.9\textwidth]{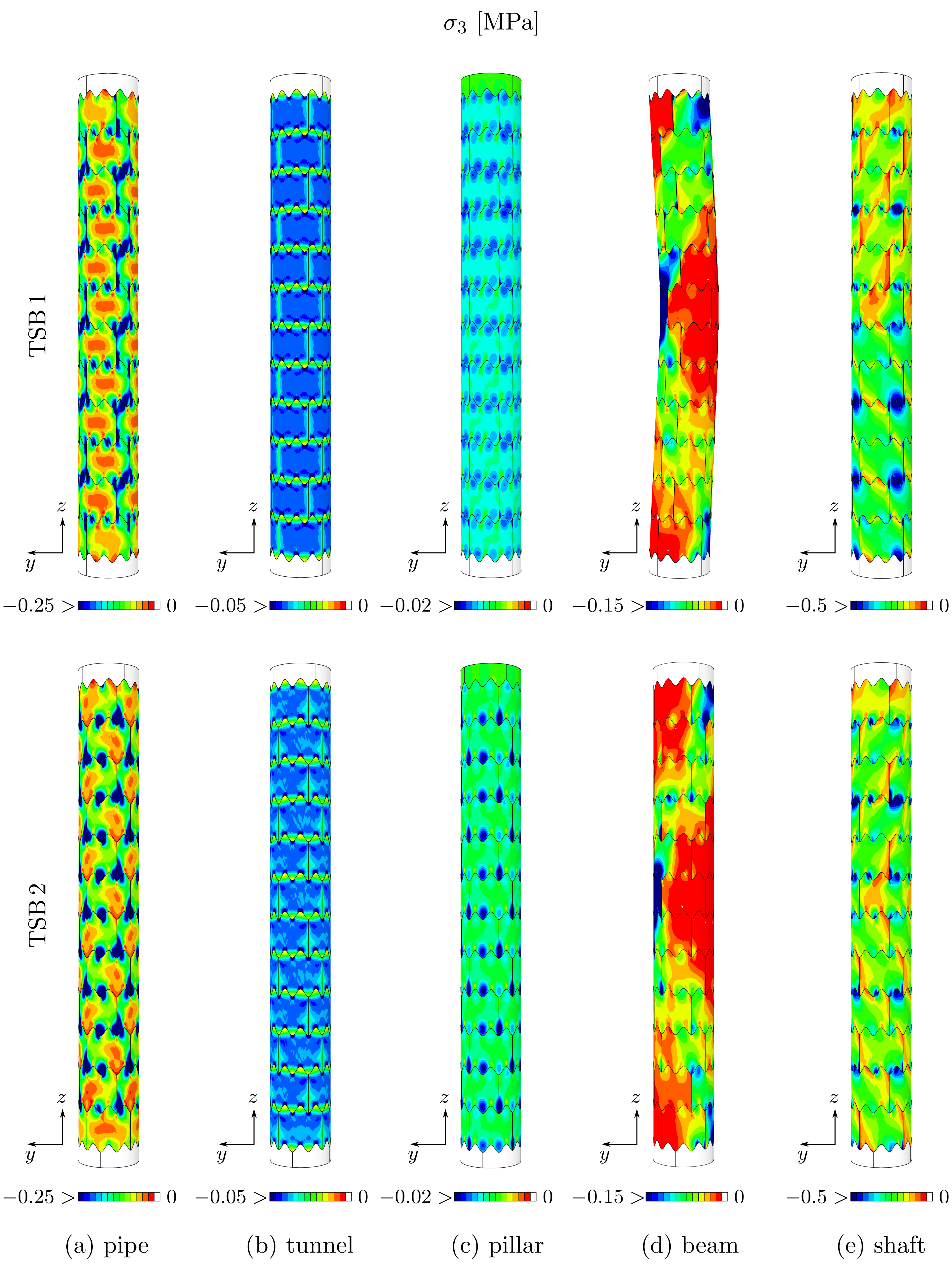}
    \caption{Distributions of the minimal principal stress $\sigma_2$ for the five BVP configurations: pipe, tunnel, pillar, beam, and shaft. The upper row shows assemblies composed of $\TSB\,1$ blocks, while the lower row presents the corresponding assemblies composed of $\TSB\,2$ blocks.}
    \label{fig:sigma3}
\end{figure}

Figures~\ref{fig:sigma1} and \ref{fig:sigma3} present the distributions of the maximum and minimum principal stresses within the assembly of the five boundary value problems (BVPs) obtained from the FEM simulations. The maximum principal stress, which corresponds to tensile stress, is a particularly relevant indicator for the prediction of cracking, given the very limited tensile capacity of concrete. Conversely, the minimum principal stress represents compressive stress and, together with the contact forces, provides valuable insight into the load transfer mechanisms within the topological interlocking assembly (TIA). The stress distributions are highly complex, largely due to the non-convex geometry of the blocks.
For clarity, only the outer side of the tube is shown in the figures. This representation is nevertheless sufficient, as it adequately captures the qualitative changes and trends in stress distributions throughout the entire assembly.

For the first three BVPs, a repeating pattern of both maximum and minimum principal stresses can be observed across the individual layers of the assembly (see Figures~\ref{fig:sigma1} and \ref{fig:sigma3}). This layer-wise similarity is a direct consequence of the homogeneous boundary conditions applied in these cases. In contrast, the fourth and fifth BVPs exhibit significantly more complex stress patterns, reflecting the increased complexity of their respective boundary and loading conditions.
Across all BVP configurations, the most critical stress concentrations are found beneath the sinusoidal contact surfaces of the blocks. The maximum principal stress distributions clearly indicate that the critical points of the assembly are located in the vicinity of the sinusoidal surfaces for both the $\TSB\,1$ and $\TSB\,2$ block geometries. Consequently, these regions emerge as prime candidates for fibre reinforcement.

A comparison between the $\TSB\,1$ and $\TSB\,2$ blocks reveals clear differences in their respective stress distributions. However, based solely on the results presented, it is not possible to conclusively~determine which block geometry leads to a more favourable overall stress state.

Finally, in the beam configuration, it is observed that the right side of the assembly exhibits practically zero maximum principal stress (tension) (see \Cref{fig:sigma1} (d)). This behaviour differs markedly from the classical case of a cylindrical beam clamped at both ends and loaded from one side, where tensile stresses would typically be expected in this region. The observed effect can be attributed to the fact that the $\TSB\,1$ block does not provide interlocking in the axial ($z$-) direction of the tube, but only in the radial and angular directions.
%\meike{wird irgendwo über die shaft Ergebnisse gesprochen?}

\begin{figure}[h!]
    \centering
    \includegraphics[width=0.9\textwidth]{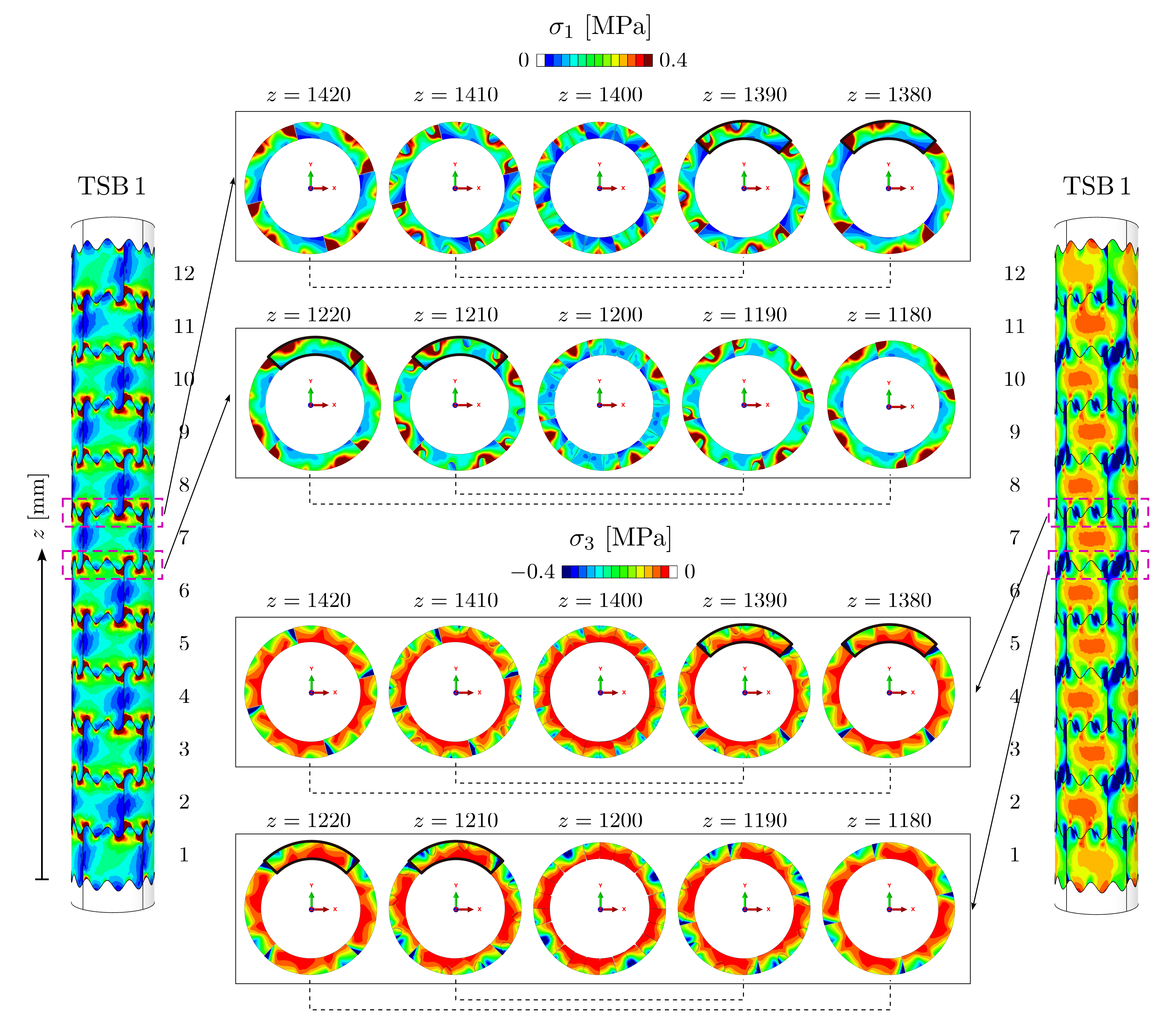}
    \caption{Principal stress distributions in the $\TSB\,1$ assembly obtained from cross-sections taken in five planes between the sixth and seventh layers and five planes between the seventh and eighth layers. The maximal $\sigma_1$ (top rows) and minimal $\sigma_2$ (bottom rows) principal stresses are shown. One representative block belonging to the seventh layer is highlighted by a black outline. Identical stress distributions are observed for the remaining blocks within the same layer, indicating that a single block is representative of the entire layer. Dashed lines connect layers exhibiting identical stress patterns, which are related through mirroring about the $y$-axis and an angular shift. Notably, the upper part of the seventh layer corresponds to the mirrored and rotated stress distribution of the lower part of the eighth layer, while an analogous relationship exists between the lower part of the seventh layer and the upper part of the sixth layer.}
    \label{fig:pipe-sin_sigma}
\end{figure}

\begin{figure}[h!]
    \centering
    \includegraphics[width=0.9\textwidth]{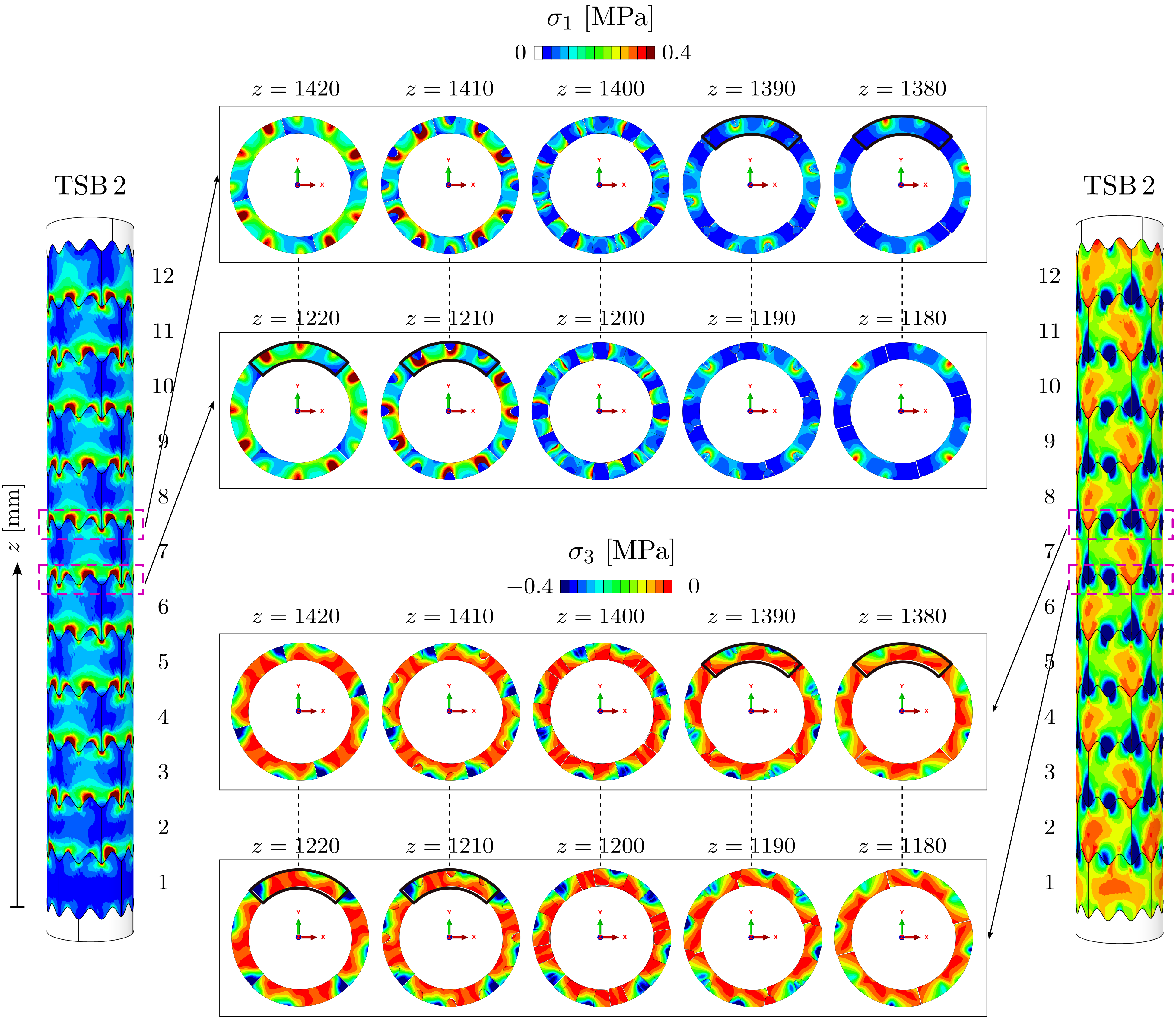}
    \caption{Principal stress distributions in the $\TSB\,2$ assembly obtained from cross-sections taken in five planes between the sixth and seventh layers and five planes between the seventh and eighth layers. The maximal $\sigma_1$ (top rows) and minimal $\sigma_2$ (bottom rows) principal stresses are shown. The block belonging to the seventh layer is highlighted by a black outline. Identical stress distributions are observed for the remaining blocks within the same layer, indicating that a single block is representative of the entire layer. Dashed lines connect layers exhibiting identical stress patterns, which are related through mirroring about the $y$-axis and an angular shift. Notably, the upper part of the seventh layer corresponds to the mirrored and rotated stress distribution of the upper part of the sixth layer, while an analogous relationship exists between the lower part of the seventh layer and the lower part of the eight layer.}
    \label{fig:pipe-cos_sigma}
\end{figure}

A more detailed investigation of the maximum and minimum principal stress distributions in the vicinity of the sinusoidal contact surfaces is presented in Figures~\ref{fig:pipe-sin_sigma} and \ref{fig:pipe-cos_sigma}. The analysis is restricted to the pipe configuration (exploding problem), as this case represents the most critical boundary value problem (BVP) among the investigated configurations.

The results demonstrate that, for both the $\TSB\,1$ and $\TSB\,2$ block assemblies, it is sufficient to examine a single layer in order to capture the characteristic stress distribution of the entire TIA. Moreover, due to the periodicity of the geometry and loading, the stress state of the whole assembly can effectively be represented by a single block. In the $\TSB\,1$ assembly, the blocks in every second layer exhibit a stress distribution similar to that of the preceding layer, but rotated by $180^\circ$ about the $y$-axis (see \Cref{fig:pipe-sin_sigma}). This behaviour is a direct consequence of the geometric asymmetry of the sine-shaped surface. In contrast, for the $\TSB\,2$ block TIA, the stress distributions in every second layer are mirrored with respect to the $yz$-plane.
For improved visualization, the figures depict the cross-section of the entire layer 7 rather than an individual block. Furthermore, only the top and bottom regions of the layer are shown, as these contain the sinusoidal surfaces responsible for the interlocking mechanism. The regions in between are of lesser interest, since no stress singularities or pronounced stress maxima occur there and the stress is distributed relatively uniformly.
Cross-sections are taken in five equally spaced planes within the interface region between the sixth and seventh layers, and in an additional five planes between the seventh and eighth layers: $z = 1210$, $z = 1220$, $z = 1380$, and $z = 1390$ mm. At these locations, the highest maximum principal stresses are observed. For each section, the distributions of the maximal (top row) and minimal (bottom row) principal stresses are shown. One representative block belonging to the seventh layer is highlighted by a black outline, serving as a reference for comparison within the assembly. 
As mentioned before, it can be observed that the remaining three blocks within the same layer exhibit identical principal stress patterns. This symmetry indicates that a single representative block is sufficient to characterize the stress state of the entire layer. Furthermore, dashed lines are used to connect layers that share identical stress distributions, which are related through mirroring with respect to the $y$-axis and a corresponding angular shift imposed by the assembly geometry.
%
% This comparison reveals a clear periodicity of the stress field across adjacent layers: for $\TSB\,1$ assembly the upper part of the seventh layer exhibits a stress distribution that is mirrored and rotated relative to the lower part of the eighth layer. An analogous relationship is observed between the lower part of the seventh layer and the upper part of the sixth layer; for $\TSB\,2$ assembly the upper part of the seventh layer exhibits a stress distribution that is mirrored and rotated relative to the upper part of the sixth layer and analogously for the lowe part.
% %
% These findings highlight the repeating and structured nature of stress transfer within the $\TSB$ TIA.
This comparison reveals a clear periodicity of the stress field across adjacent layers. For the $\TSB\,1$ assembly, the upper part of the seventh layer exhibits a stress distribution that is mirrored and rotated with respect to the lower part of the eighth layer. A corresponding relationship is observed between the lower part of the seventh layer and the upper part of the sixth layer. In contrast, for the $\TSB\,2$ assembly, the upper part of the seventh layer shows a stress distribution that is mirrored and rotated relative to the upper part of the sixth layer, with an analogous correspondence observed for the lower parts of the seventh and eight layer.
These observations highlight the repeating and highly structured nature of stress transfer within the $\TSB$ tubular interlocking.
%
% For both block geometries, the inner surface of the tube experiences lower levels of both maximum and minimum principal stresses compared to the outer surface.
%
A closer comparison reveals notable differences between the two block types. For the $\TSB\,1$ block, high peaks of maximum principal stress occur at the valleys of the sinusoidal surfaces in both the top and bottom parts of the layer. In contrast, for the $\TSB\,2$ block, such stress peaks are observed only in the bottom part of the layer. Additionally, among the three sinusoidal valleys present in each block, one valley consistently exhibits higher stress levels than the other two.
Overall, the results indicate that the $\TSB\,2$ block attains lower average maximum principal stresses than the $\TSB\,1$ block. Furthermore, the regions where peak tensile stresses occur are more localized in the $\TSB\,2$ block. On the other hand, the $\TSB\,2$ block exhibits larger regions subjected to minimum principal (compressive) stresses, highlighting a trade-off between tensile stress reduction and compressive stress distribution.

% #############################################################################################
%                                  DISPLACEMENTS
% #############################################################################################
\subsubsection{Deformed state and maximum deflection at a given load}
\begin{figure}[h!]
    \centering
    \includegraphics[width=0.9\textwidth]{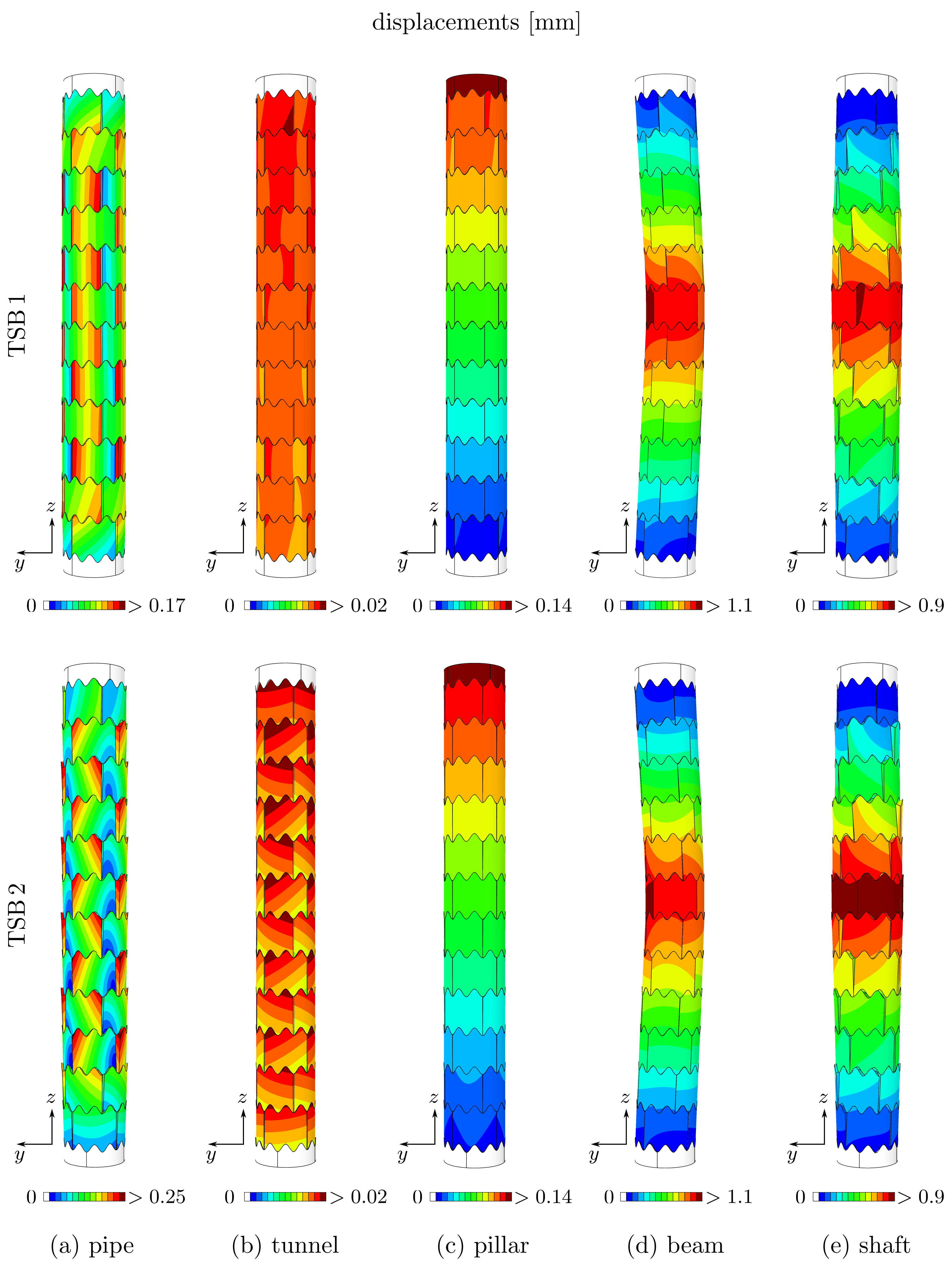}
    \caption{Magnitude of the displacement fields in the $zy$-plane for the five BVP configurations. The figure highlights the differences in deformation behaviour between assemblies composed of $\TSB\,1$ and $\TSB\,2$ blocks. The deformations are shown with different scaling factors: the pipe configuration is scaled by a factor of 80, the tunnel and pillar configurations by a factor of 200, the shaft configuration by 100, and the beam configuration by 50.}
    \label{fig:disp}
\end{figure}

Figure~\ref{fig:disp} illustrates the displacement magnitude fields of the TIA for the five investigated boundary value problem (BVP) configurations obtained from the FEM simulations. To enhance the visibility of the relative motion between individual blocks, the deformations are shown with different scaling factors: the pipe configuration is scaled by a factor of 80, the tunnel and pillar configurations by a factor of 200, the shaft configuration by 100, and the beam configuration by 50. The applied scaling allows a clear visualization of how the blocks within the TIA deform and move relative to one another. In the pipe configuration, the results reveal fundamentally different deformation patterns for the $\TSB\,1$ and $\TSB\,2$ block assemblies. In the pipe and tunnel configurations, the $\TSB\,1$ block assembly shows a tendency towards more homogeneous deformation patterns compared to the $\TSB\,2$ block assembly. This behaviour results in smaller average displacements for the $\TSB\,1$ block TIA under these loading conditions.
In the pillar configuration, both $\TSB\,1$ and $\TSB\,2$ block assemblies exhibit qualitatively a more stable load transfer compared to the pipe and tunnel configurations. In this case, the blocks do not tend to rotate or squeeze out of the assembly. Finally, for the beam and shaft configurations, the TIAs closely follow the deformation behaviour of a monolithic tubular structure.

%#############################################################################################
%                                  Computation Time Comparison
% #############################################################################################
\subsubsection{Comparison of computational cost}
In \Cref{tab:cpu-times} the computational efficiency of the proposed multibody dynamics framework is assessed in comparison to a high-fidelity finite element method model. Owing to its substantially lower computational cost, the MBD approach enabled the evaluation of 480 distinct assembly configurations within approximately 2.7 CPU hours on an 8-core AMD Ryzen 7 PRO 4750U processor. This corresponds to an average computational time of roughly 0.0056 CPU hours per configuration .
In contrast, a single high-fidelity FEM simulation of one assembly configuration required approximately 758 CPU hours on a 48-core Intel® Xeon® Platinum 8160 processor. Even without normalizing for hardware differences, the discrepancy in computational demand is several orders of magnitude.
\begin{table}[h!]
\centering
\caption{Comparison of computational cost between MBD and FEM approaches.}
\begin{tabular}{|l|c|c|c|}
\hline
\textbf{Method} & \textbf{Configurations} & \textbf{CPU Hours (Total)} & \textbf{Hardware} \\
\hline
MBD & 480 & 2.7 & AMD Ryzen 7 PRO 4750U (8 cores) \\
\hline
FEM & 1 & 758 & Intel® Xeon® Platinum 8160 (48 cores) \\
\hline
\end{tabular}
\label{tab:cpu-times}
\end{table}
This drastic difference in computational efficiency demonstrates the suitability of the MBD framework for large-scale parametric studies, preliminary design exploration, and the down-selection of promising block geometries prior to high-fidelity FEM verification.
% ############################################################################
\subsection{Hexagon-based block}\label{sec:Hex}
% ############################################################################
To assess the performance of the proposed Sine Block in comparison to the previously introduced hexagon-based block by \cite{HexBased}, a representative hexagonal configuration is selected as
\begin{equation}
    \THB \, {0} \coloneqq \THB(n=4, r_i=120, t=40),
\end{equation}
where the inner radius $r_i$ and the block thickness $t$ are given in millimetres. For consistency, the number of layers is set to $L = 12$, matching the configuration used for $\mathrm{TSB}\, {1}$ and $\mathrm{TSB}\, {2}$.

Using this setup, MBD simulations are conducted for the same set of boundary value problems defined in the preceding section. The corresponding results are summarized in Table~\ref{tab:hexagon-force-equilibrium}. 
\begin{table}[h!]
    \centering
    \begin{tabular}{|c|>{\centering\arraybackslash}m{3cm}|>{\centering\arraybackslash}m{3cm}|>{\centering\arraybackslash}m{3cm}|>{\centering\arraybackslash}m{3cm}|}
        \hline
        \textbf{BVP} & \textbf{Kinematic mechanism} & \textbf{Effective contact area} [\%] & $\bar{p}_{\mathrm{eff}}$ [MPa] & $\bar{p}_{\mathrm{max}}$ [MPa] \\
        \hline
        pipe & {activated} & - & - & -\\ 
        \hline
        tunnel & {suppressed} & $79.2$ & $0.039$ & $0.103$\\
        \hline
        pillar & {neutral} & - & - & -\\
        \hline
        beam & {activated} & - & - & - \\
        \hline
        shaft & {activated} & - & - & - \\
        \hline
    \end{tabular}
    \caption{Overview of the kinematic mechanism, effective contact area, effective mean contact pressure, and maximal mean contact pressure for the block configuration $\THB \, 0$ across all five boundary value problems (BVPs). The metrics indicate whether force equilibrium is achieved, quantify the extent of the effective contact area within the assembly, provide the mean contact pressure restricted to this area, and report the average of the maximum contact pressure per block.}
    \label{tab:hexagon-force-equilibrium}
\end{table}
The analysis reveals that the hexagon-based block, in this assembly has at least one feasible kinematic mechanism; only for the \emph{tunnel} BVP the kinematic mechanism is suppressed. This behaviour is attributable to the block geometry, which permits a kinematical explosion, as explained above.

This outcome highlights the substantially higher versatility of the proposed Sine Block when subjected to a broad range of loading scenarios. Although, in the \emph{tunnel} configuration, the hexagon-based block activates a larger effective contact area than $\TSB\,{2}$ (see Table~\ref{tab:cosin-force-equilibrium}), however, the effective mean contact pressure is approximately $30\,\%$ higher, and the maximal mean contact pressure is nearly twice as large compared to the Sine Block configuration.

\begin{figure}[h!]
    \centering
    \includegraphics[scale=1.0]{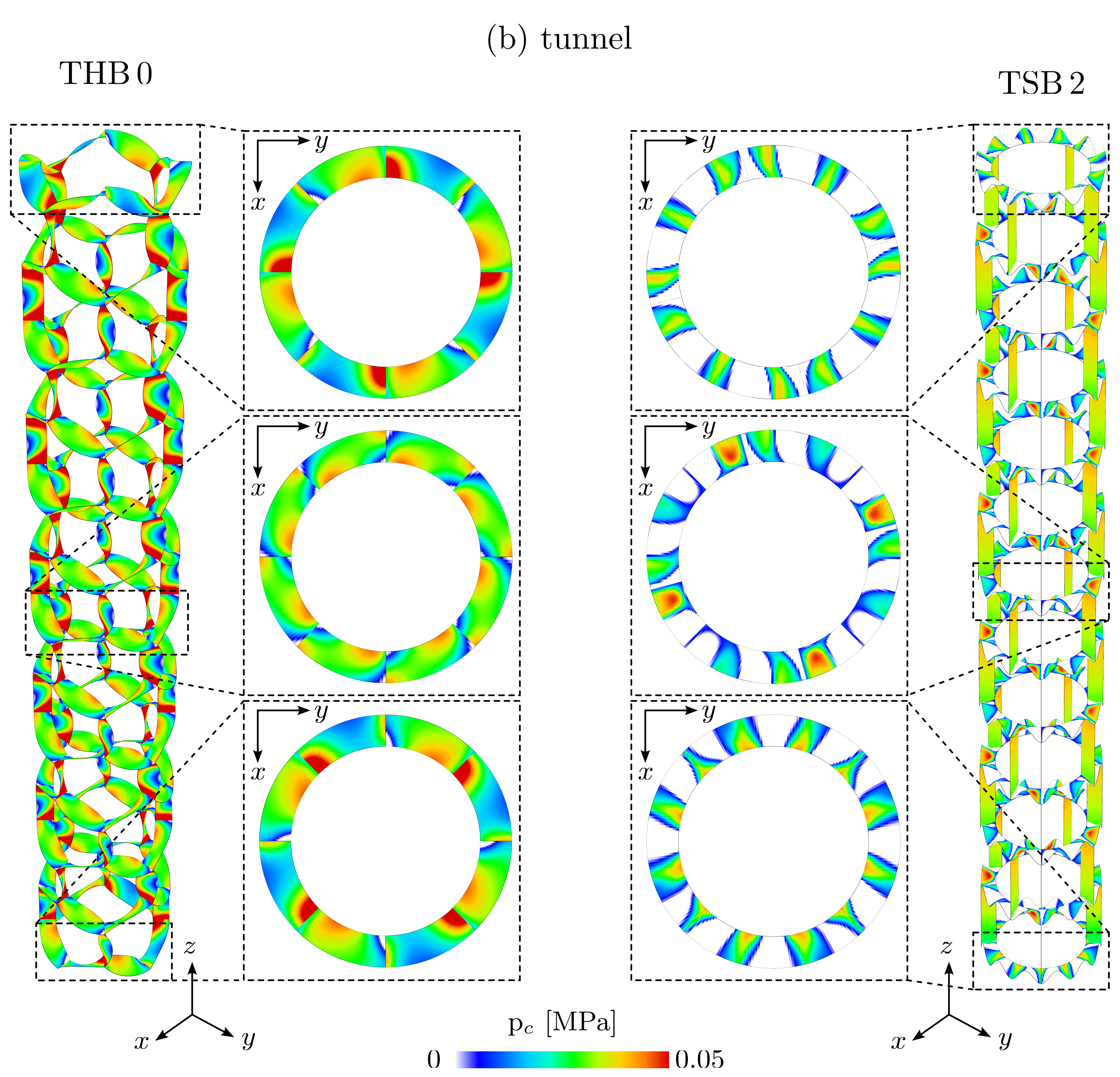}
    \caption{Comparison of contact pressure distributions obtained using the MBD approach for the \emph{tunnel} boundary value problem (BVP), highlighting differences in load transfer mechanisms between the $\THB\,0$ and $\TSB\,2$ block assemblies. The $\THB\,0$ assembly is shown on the left and the $\TSB\,2$ assembly on the right. Cross-sectional views in the top and bottom rows illustrate the transfer of the externally applied load from the assemblies to the bounding frame, while the middle row shows the inter-block load transfer between the sixth and seventh layers of the assemblies.}
    \label{fig:tunnel_MBD_hexagon_pc}
\end{figure}

A qualitative comparison of the resulting contact pressure distributions for both block types in the \emph{tunnel} BVP is provided in \Cref{fig:tunnel_MBD_hexagon_pc}, further illustrating the differences in load transfer mechanisms between the two block designs.

\section{Conclusions}
In this work, we introduced a novel tubular interlocking block, called the Sine Block, specifically designed for the construction of tubular interlocking structures. The defining characteristic of this block is its inherent resistance against kinematical explosion, i.e., simultaneously radially outward displacements of the blocks under radial loading, which typically represent a critical failure mode in such assemblies.
This property was rigorously demonstrated using mechanical equilibrium considerations. In particular, we showed that assemblies constructed from the Sine Block are capable of satisfying force equilibrium for the pipe boundary value problem (BVP). This BVP can be interpreted as an equivalent mechanical condition for kinematical explosion, as it directly probes the ability of the assembly to transfer radial loads to the bounding frame without loss of equilibrium. The successful fulfilment of this condition therefore provides strong evidence for the intrinsic interlocking capability of the proposed block geometry.

To systematically explore the design space and mechanical behaviour of the Sine Block, we developed a combined multibody dynamics (MBD) and finite element method (FEM) workflow. In this framework, the MBD approach functions as an efficient exploratory tool, enabling rapid assessment of numerous block configurations with respect to force equilibrium and contact force transmission.
The MBD framework provides a computational speed-up of approximately five orders of magnitude (on the order of $10^5$) compared to high-fidelity FEM simulations. This allows for the evaluation of hundreds of assembly configurations within just a few CPU hours, whereas a single FEM analysis can require several hundred CPU hours. This significant improvement in computational efficiency underscores the suitability of MBD for preliminary exploration and the down-selection of promising block designs.
The reliability of the proposed MBD framework was further assessed by a qualitative comparison of the resulting contact pressure distributions with those obtained from corresponding FEM simulations. The close agreement observed between both methods demonstrates that the MBD approach provides sufficiently accurate insight into the internal force transmission mechanisms, while maintaining a high level of computational efficiency. Consequently, the presented MBD-based workflow constitutes a robust and practical tool for the early-stage evaluation of novel interlocking systems, such as the Sine Block.

Based on the MBD-driven design space exploration, two prominent block configurations, denoted as $\TSB\,1$ and $\TSB\,2$, were identified and subsequently examined in greater detail. Beyond the pipe BVP, the assemblies of these were tested in additional boundary value problems, demonstrating their suitability for a wide range of engineering applications. Potential use cases include load-bearing tubular structures subjected to internal and external pressure, pillars supporting ceilings, beams carrying distributed loads, or shafts transmitting torque, for instance in vehicular applications. Owing to the modular nature of the assembly, the overall dimensions of such structures (e.g., the height of a pillar or the length of a beam) can be easily adapted by adding or removing blocks, providing a high degree of constructional flexibility. Thus, a single block geometry enables the realization of multiple functional structures with varying dimensions and mechanical roles.
Although the curves on the outer surface of $\TSB\,1$ and $\TSB\,2$ assemblies are identical, the Sine and Cosine Block assemblies exhibit markedly different mechanical responses. This highlights that even minor geometric variations can affect the mechanical behaviour of different BVPs in distinct ways. Thus, choosing an appropriate geometry for each BVP is essential, as seemingly minor differences can lead to non‑negligible effects in the overall behaviour.

All investigations in this study were conducted under frictionless contact assumptions. While such conditions cannot be fully realized in practice, they represent a conservative modelling choice that isolates the purely geometric interlocking effect of the block. Friction is known to significantly influence the global mechanical response of assembled systems and should therefore be incorporated in future studies. Nevertheless, the present results guarantee that, independent of friction, the blocks cannot kinematically slip out of the assembly as long as deformations remain within moderate limits, thereby confirming the robustness of the proposed interlocking concept.

%######################################################################################
%                              CONTRIBUTION ROLES, ACKNOWLEDGEMENTS
%######################################################################################
\section*{Contributor Roles}
\noindent\textbf{Domen Macek} reviewed and adapted MBD and FEM methodologies from the literature, implemented the MBD approach and developed the associated code, planned and performed the MBD and FEM analyses, validated the numerical results, and wrote the majority of the manuscript. \textbf{Meike Weiß} wrote the code, performed FEM analyses reviewed the relevant existing literature and wrote a part of this article. \textbf{Reymond Akpanya} developed the Sine Block and the assemblies, wrote the code, reviewed the relevant existing literature and wrote a part of this article.  \textbf{Tim Brepols} acquired funding, gave conceptual advice, contributed in the discussion of the results, read the article and gave valuable suggestions for improvement. \textbf{Hagen Holthusen} acquired funding, gave conceptual advice, contributed in the discussion of the results, read the article and gave valuable suggestions for improvement. \textbf{Alice C. Niemeyer} identified explosions problem, wrote a part of this article, acquired funding, gave conceptual advice and supervised.

\section*{Acknowledgements}
The authors gratefully acknowledge the funding by the Deutsche Forschungsgemeinschaft (DFG, German Research Foundation) in the framework of the Collaborative Research Centre CRC/TRR 280 “Design Strategies for Material-Minimized Carbon Reinforced Concrete Structures – Principles of a New Approach to Construction” (project ID 417002380). R.\ Akpanya was supported by a grant from the Simons Foundation (SFI-MPS-Infrastructure-00008650).
%
%The authors also thank the anonymous reviewers for their suggestions, many of which were very helpful.
%
Furthermore, Domen Macek also thanks Hannah Knobloch for her very valuable comments and suggestions. 

%######################################################################################
%                                 BIB
%######################################################################################
\bibliographystyle{plainnat_mod}
\bibliography{references.bib}

%######################################################################################
%                                 APPENDIX
%######################################################################################
\appendix
\section{Appendix}

\begin{figure}[H]
    \centering
    \includegraphics[width=0.8\linewidth]{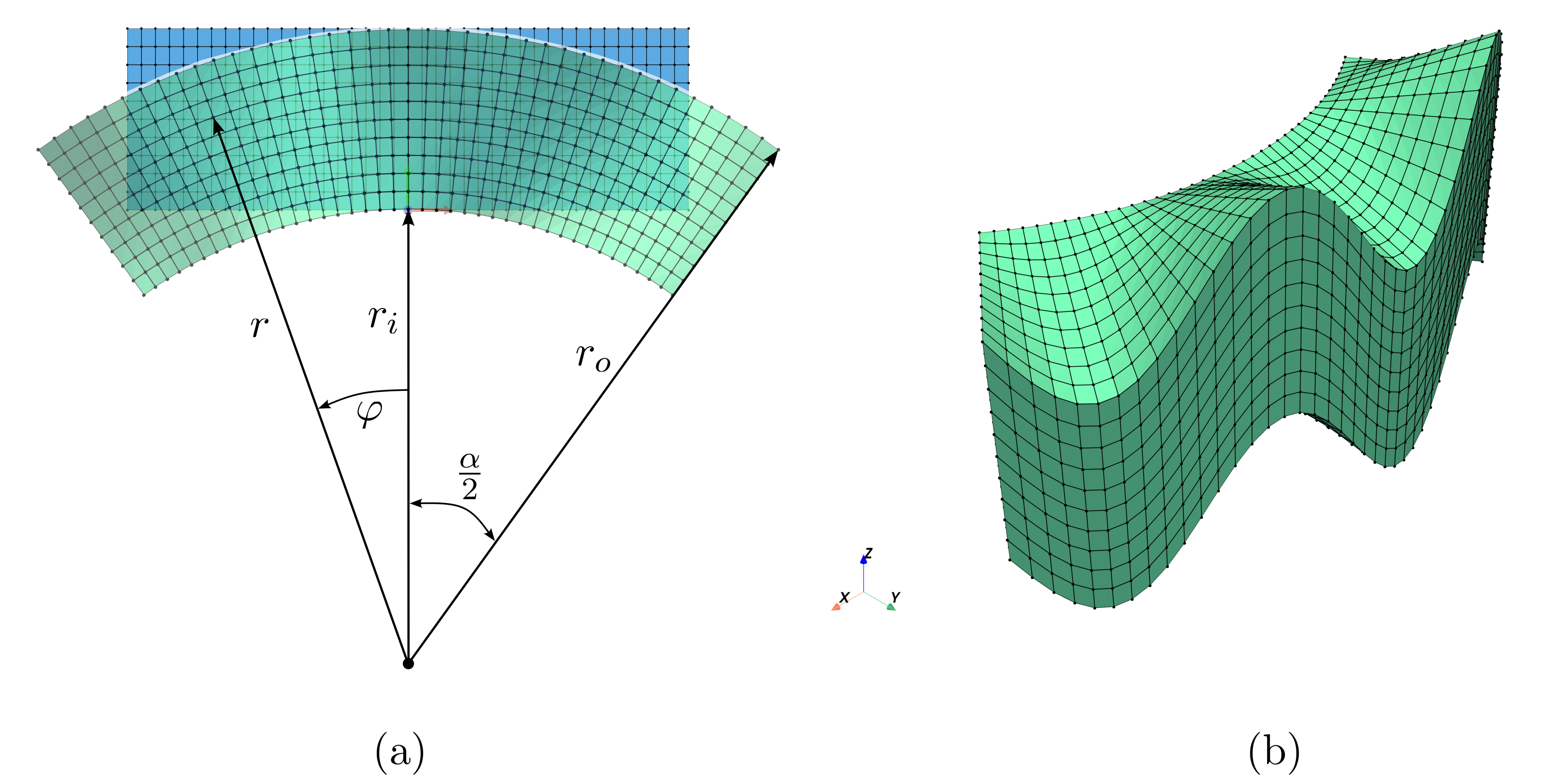}
    \caption{Cylindrical mapping of the planar block (a) and hexahedral mesh of $\TSB(h,a,f=1,s=0,n=5,r_i,t)$ (b)}
    \label{fig:cylindrical_mapping}
\end{figure}

\end{document}